%% file: main.tex
\documentclass[twocolumn]{aastex63}
\usepackage{booktabs}
\usepackage{natbib}

\received{\today}
\revised{--}
\accepted{--}

\shorttitle{ASASSN-14il}
\shortauthors{Dukiya et al.}

\newcommand{\il}{ASASSN-14il}
\newcommand{\csm}{\texttt{csm}}
\newcommand{\csmni}{\texttt{csmni}}
\newcommand{\mosfit}{\texttt{MOSFiT}}

\def \swift {\textit{Swift~}}

\begin{document}

\title{Probing the Circumstellar Environment of highly luminous type IIn SN ASASSN-14il}

\correspondingauthor{Naveen Dukiya}
\email{ndookia@gmail.com}

\author[0000-0002-0394-6745]{Naveen Dukiya}
\affiliation{Aryabhatta Research Institute of Observational Sciences, Manora Peak 263001, India}
\affiliation{Department of Applied Physics, Mahatma Jyotiba Phule Rohilkhand University, Bareilly, 243006, India}

\author[0000-0002-3884-5637]{Anjasha Gangopadhyay}
\affiliation{Oskar Klein Centre, Department of Astronomy, Stockholm University, AlbaNova, SE-106 91 Stockholm, Sweden}
\affiliation{Hiroshima Astrophysical Science Centre, Hiroshima University, 1-3-1 Kagamiyama, Higashi-Hiroshima, Hiroshima 739-8526, Japan}

\author[0000-0003-1637-267X]{Kuntal Misra}
\affiliation{Aryabhatta Research Institute of Observational Sciences, Manora Peak 263001, India}

\author[0000-0002-0832-2974]{Griffin Hosseinzadeh}
\affiliation{Steward Observatory, University of Arizona, 933 North Cherry Avenue, Tucson, AZ 85721-0065, USA}

\author[0000-0002-4924-444X]{K. Azalee Bostroem}
\affiliation{Steward Observatory, University of Arizona, 933 North Cherry Avenue, Tucson, AZ 85721-0065, USA}

\author[0009-0000-1020-9711]{Bhavya Ailawadhi}
\affiliation{Aryabhatta Research Institute of Observational Sciences, Manora Peak 263001, India}
\affiliation{Department of Physics, Deen Dayal Upadhyaya Gorakhpur University, Gorakhpur, 273009, India}

\author[0000-0003-4253-656X]{D. Andrew Howell}
\affiliation{Las Cumbres Observatory, 6740 Cortona Drive, Suite 102, Goleta, CA 93117-5575, USA}
\affiliation{Department of Physics, University of California, Santa Barbara, CA 93106-9530, USA}

\author[0000-0001-8818-0795]{Stefano Valenti}
\affiliation{Department of Physics and Astronomy, University of California, Davis, 1 Shields Avenue, Davis, CA 95616-5270, USA}

\author[0000-0001-7090-4898]{Iair Arcavi}
\affiliation{School of Physics and Astronomy, Tel Aviv University, Tel Aviv 69978, Israel}

\author[0000-0001-5807-7893]{Curtis McCully}
\affiliation{Las Cumbres Observatory, 6740 Cortona Drive, Suite 102, Goleta, CA 93117-5575, USA}
\affiliation{Department of Physics, University of California, Santa Barbara, CA 93106-9530, USA}

\author{Archana Gupta}
\affiliation{Department of Applied Physics, Mahatma Jyotiba Phule Rohilkhand University, Bareilly, 243006, India}

\begin{abstract}
We present long-term photometric and spectroscopic studies of Circumstellar Material (CSM)-Ejecta interacting supernova (SN) ASASSN-14il in the galaxy PGC 3093694. The SN reaches a peak $r$-band magnitude of $\sim$ $-20.3 \pm 0.2$ mag rivaling SN 2006tf and SN 2010jl. The multiband and the pseudo-bolometric lightcurve show a plateau lasting $\sim 50$ days. Semi-analytical CSM interaction models can match the high luminosity and decline rates of the lightcurves but fail to faithfully represent the plateau region and the bumps in the lightcurves. The spectral evolution resembles the typical SNe~IIn dominated by CSM interaction, showing blue-continuum and narrow Balmer lines. The lines are dominated by electron scattering at early epochs. The signatures of the underlying ejecta are visible as the broad component in the H$\alpha$ profile from as early as day 50, hinting at asymmetry in the CSM. A narrow component is persistent throughout the evolution. The SN shows remarkable photometric and spectroscopic similarity with SN~2015da. However, the different polarization in \il ~compared to SN~2015da suggests an alternative viewing angle. The late-time blueshift in the H$\alpha$ profiles supports dust formation in the post-shock CSM or ejecta. The mass-loss rate of 2-7 M$_{\odot} \mathrm{yr}^{-1}$ suggests a Luminous Blue Variable (LBV) progenitor in an eruptive phase for \il. 
\end{abstract}




\section{Introduction}
Core-collapse supernovae (CCSNe) that show narrow (10s to 1000s of km/s) Balmer lines in their spectra are typically classified as type IIn supernovae (hereafter SNe~IIn; \citealp{schlegel1990_IIn, filippenko97_review}). The narrow lines in their spectra arise due to the ejecta of the SN interacting with dense Circumstellar-Material (CSM) surrounding the progenitor \citep{chugai_danziger_1994, Chugai_2001_1998S, Chugai_1994w_2004}. Early spectra are dominated by a blue continuum, gradually cooling down as the SN evolves. The line profiles are usually multi-component and asymmetric based on the geometry of the explosion. Distinct narrow, intermediate-width, and broad components (corresponding to the unshocked CSM, Cold-Dense Shell (CDS), and fast-moving ejecta) can be seen simultaneously or at different times depending on the explosion geometry (see \citealp{smith_IIn_review} for a review on SNe~IIn).
In addition to Ultraviolet, Optical, and Infrared, interacting SNe can radiate in X-ray and radio wavelengths, providing independent constraints on the mass-loss of the progenitor \citep{Chevalier_1982_self-similar, Chevalier_1998}. Some SNe~IIn also form dust at the later stages of evolution, which leads to IR excess at late times or dust-echoes \citep{pozzo_1998s_dust, mattila_2006jc_dust, 2010jl_fransson, 2015da_tartaglia}.


SNe~IIn are relatively rare ($\sim$9\% intrinsic rate among all SNe II; \citealp{li_SNe_rate}) and form a heterogeneous and poorly understood class of transients due to their exceptionally diverse properties. These events have typical absolute magnitudes ranging from M$_{\mathrm{R}} \sim$ $-16$ to $-19$ mag but some unusually luminous events even reach M$_{\mathrm{R}} \sim$ $-22$ mag \citep{kiewe_iin_sample_2012, taddia_IIn_sample_2013, 2006gy_smith_mccray_shellshockeddiffusion}. Depending on their lightcurve shapes and luminosities, different physical scenarios have been invoked in the literature to explain the powering mechanism of these SNe~IIn. Many authors have developed analytical and numerical modeling frameworks to explain the explosion physics and observables of SNe~IIn \citep{balberg_loeb_2011, Svirski_2012, CW12, takashi_IIn_analytical, Ofek_2010jl_modelling, Dessart_IIn_simulation, Jiang_IIn_analytical_mosfit}. A lightcurve powered by ejecta interacting with optically thin CSM can explain moderately bright SNe~IIn, but superluminous SNe~IIn events e.g. SNe 2006gy, 2006tf, 2010jl, 2016aps \citep{2006gy_smith_mccray_shellshockeddiffusion, 2006gy_woosley, 2006tf_smith, 2010jl_fransson, 2016aps_nicholl} etc. require an extreme amount (more than 10 M$_{\odot}$ in some cases) of CSM to power their lightcurves. The dense CSM from the progenitors' mass-loss events may not extend far. In that case, transitioning from an interaction-dominated regime to a Ni-powered regime is similar to SNe~IIn-P \citep{2009kn_kankare, 2011ht_mauerhan}. However, all these events share the fact that a significant part of the total radiated energy during its lifetime comes from the CSM interaction of ejecta.

Different kinds of progenitor systems can give rise to SNe~IIn.
They have different CSM profiles (created by the mass-loss events of the progenitor or its binary companion) resulting from these different progenitor systems. SNe~IIn typically requires mass-loss rates higher than $10^{-3} \mathrm{M}_{\odot} \mathrm{ yr}^{-1}$ \citep{Mass_loss_IIn_moriya_2014} which is higher than what can be expected from the line-driven winds \citep{smith_owocki_linedrivenwinds}, although SNe~IIn falling in the lower range of luminosity can be explained by long-term strong winds \citep{1995N_fransson, smith_2009_vy_canis_majoris}. The enhanced mass-loss resulting from the great eruptions of Luminous Blue Variables (LBVs) in months to decades before an SN explosion \citep{Smith_mass_loss_2014} make them much more suitable progenitor candidates for most SNe~IIn. In the case of many SNe, archival images from Hubble Space Telescope ({\it HST}) and pre-SN outbursts have revealed their progenitors to be consistent with LBVs \citep{2005gl_gal-yam, 2009ip_precursor_smith_2010, 2010jl_smith, 1961V_progenitor}. Additionally, the large amount of CSM required to explain the superluminous supernovae (SLSNe) also favors a massive (40 - 100 M$_{\odot}$) LBV progenitor. 


In this context, we present the long-term observational photometric and spectroscopic analysis of ASASSN-14il. It was discovered by All Sky Automated Survey for SuperNovae (ASAS-SN) at UT 2014-10-01.11 at $V \sim 16.5$ mag \citep{14il_discovery}. The transient was detected in images from the double 14-cm ``Cassius'' telescope in Cerro Tololo, Chile at RA = 00h45m32.55s, Dec = -14d15m34.6s; approximately 0$''$.33 North and 0$''$.26 West from the centre of the galaxy 2MASX J00453260-1415328. No source was detected at the position of the transient down to the limiting magnitude of $V \sim 17.3$ mag in images taken on UT 2014-09-27.04 and before. Spectroscopic observation on UT 2014-10-03.55 with the Wide Field Spectrograph (WiFeS) mounted on the Australian National University 2.3-m telescope, using the B3000/R3000 gratings ($3500-9800$~\AA, $1$~\AA ~resolution) was used to classify the SN \citep{14il_classification}. \il ~was classified as a SN~IIn based on the narrow emission lines ($\sim$3000 km/s) in the Balmer series and the blue continuum. The redshift calculated from the SN spectrum is consistent with the redshift of the host at 0.022 \citep{host_galaxy_z}. 
The redshift of 0.022 corresponds to a distance of $88.5 \pm 6.2$ Mpc\footnote{\href{https://ned.ipac.caltech.edu/byname?objname=WISEA+J004532.52-141532.7&hconst=73&omegam=0.27&omegav=0.73&wmap=1&corr_z=4}{NED}}(corrected for Virgo + GA + Shapley) assuming H$_0 = 73$ km/s/Mpc, $\Omega_{\mathrm{matter}}=0.27$, and $\Omega_{\mathrm{vaccum}}=0.73$.
This distance is adopted throughout the paper for further analysis.

The paper is structured as follows. Section \ref{sec:observations} describes the observations and data reduction procedures. 
Estimation of extinction and explosion epoch is discussed in section \ref{sec:host_galaxy}. The lightcurve evolution and modeling in described in sections \ref{sec:lightcurve_evolution} and \ref{sec:Modelling}, respectively. The spectral evolution is described in section \ref{sec:spectral_evolution}. Finally, we conclude our paper in section \ref{sec:conclusions} and summarize the results in section \ref{sec:summary}.
 
\section{Observations and Data Reduction} \label{sec:observations}

\begin{table}
    \caption{General information about ASASSN-14il and its host galaxy}
    \label{tab:sn_and_host}
    \centering
    \input{sn_and_host.tex}
\end{table}

We observed \il ~with the Las Cumbres Observatory (LCO) network of telescopes as part of the Supernova Key Project (which eventually became the Global Supernova Project). The 1m class LCO telescopes were used for imaging observations. All these telescopes offer identical imaging systems. The imaging system uses a 4k $\times$ 4k CCD detector that covers a field-of-view of $15'.8 \times 15'.8$ \citep{LCO_telescopes_instruments}.
Observations started from $\sim 2$ days after discovery and continued for over a year with a high cadence in BVgri filters. The exposure times were varied from 60 to 300s based on the filter and the brightness of the SN to ensure good signal-to-noise (S/N) in the images. Two frames were taken in each filter to avoid spurious photometric measurements. The pre-processing (bias correction, flat correction, cosmic-ray correction, and astrometry for the frame) was conducted using the BANZAI pipeline \citep{Banzai}.

\begin{figure*}
    \centering
    \includegraphics[width=\textwidth]{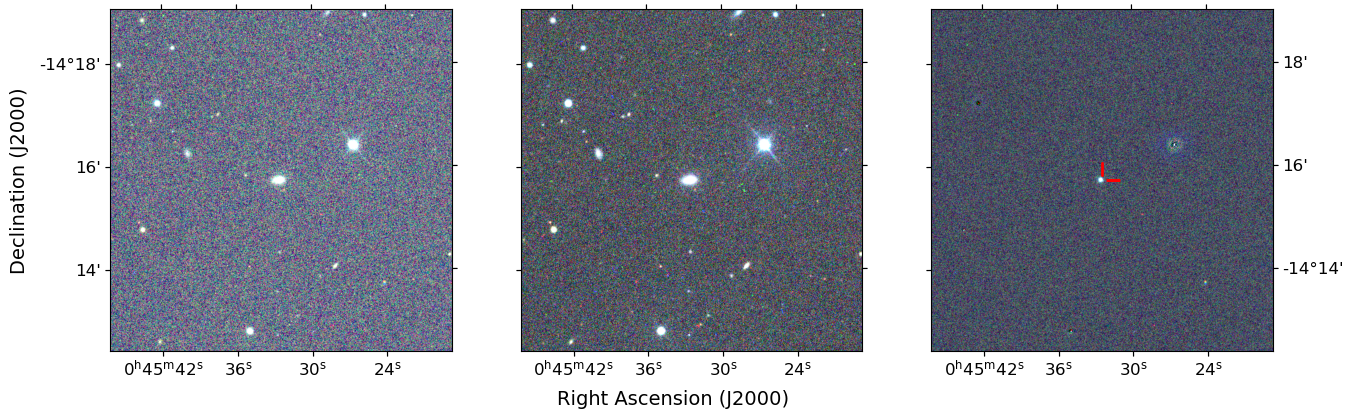}
    \caption{The left panel shows the {\it gri} color-composite image of ASASSN-14il with frames taken on 28 November 2014. The central panel shows the {\it gri} color-composite with the reference image taken on 27 December 2022, and the right panel shows the color-composite of the difference images with the position of SN marked.}
    \label{fig:14il_color}
\end{figure*}

Since the source was contaminated by the host galaxy,
we performed image subtraction using High Order Transform of PSF ANd Template Subtraction (HOTPANTS; \citealp{hotpants}) algorithm integrated in the \texttt{lcogtsnpipe}\footnote{\url{https://github.com/LCOGT/lcogtsnpipe/}} pipeline \citep{valenti_lcogtsnpipe}. The template images were taken in December 2022, long after the SN faded. \autoref{fig:14il_color} shows a {\it gri} composite image of the science frame, reference frame, and difference image of ASASSN-14il. The {\it BVgri} instrumental magnitudes were obtained from the difference images using PSF photometry. Nightly zero points and color terms were determined for each filter using the AAVSO Photometric All-Sky Survey (APASS) DR9 catalog\footnote{\href{https://vizier.cds.unistra.fr/viz-bin/VizieR?-source=II/336}{https://vizier.cds.unistra.fr/viz-bin/VizieR?-source=II/336}} \citep{apass_dr9}, which were later used to obtain the calibrated SN magnitudes. All detected stars in the frame crossmatched with APASS were used after sigma clipping to avoid potential variable stars. The photometry of \il ~is presented in \autoref{tab:photometry_log}.

\begin{figure}
    \centering
    \includegraphics[width=\columnwidth]{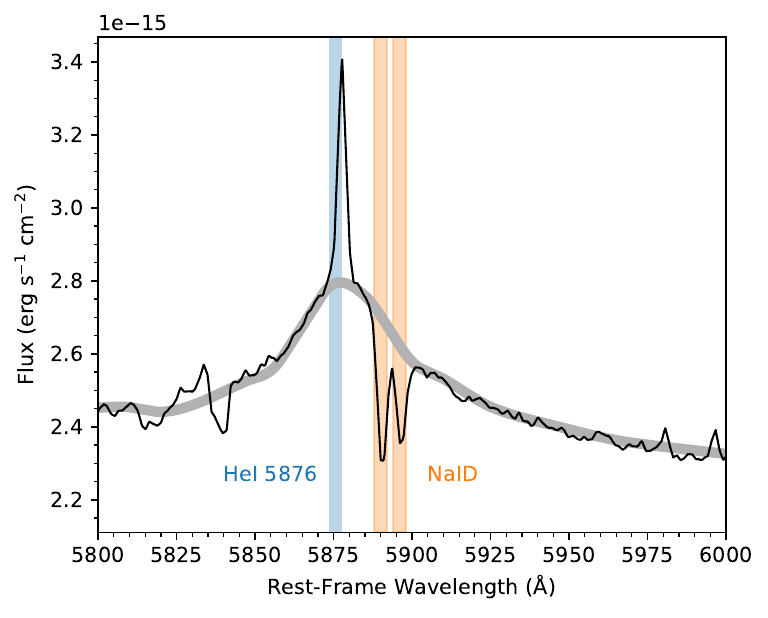}
    \caption{The spectrum of ASASSN-14il showing a prominent dip at 5890~\AA ~and 5896~\AA ~(marked as orange) in the rest-frame of the galaxy indicating high host galaxy extinction. The rest-frame wavelength of the nearby HeI emission line is marked in blue. The grey line shows the continuum used for normalization.}
    \label{fig:NaID}
\end{figure}

This supernova was also observed using the Ultra-Violet/Optical Telescope (UVOT; \citealp{uvot}) on {\it The Neil Gehrels Swift Observatory} (hereafter {\it Swift}; \citealt{swift}) in {\it UVW2, UVM2, UVW1, U, B,} and $V$ bands starting from day $\sim 8$ to $49$ after the explosion (c.f.r \autoref{sec:host_galaxy}). The photometry was obtained from the {\it Swift} Optical/Ultraviolet Supernova Archive (\href{https://archive.stsci.edu/prepds/sousa/}{SOUSA}; \citealp{sousa_brown_2014}). The reduction is based on that of \cite{sousa_reduction_brown_2009}, which includes subtraction of the host galaxy count rates (from templates taken $\sim$ 2 years after the explosion) using a 5\arcsec aperture at the source position, using the revised UV zero points and time-dependent sensitivity from \cite{sousa_zp}. Aperture photometry was performed using an aperture size of 3 or 5 arcsec based on the error. The final UVOT magnitudes are given in \autoref{tab:14il_swift_photometry}.

Low-resolution (R $\sim$ 400-700) optical spectroscopic observations were carried out with the FLOYDS spectrographs mounted on the LCO 2m telescopes using a 1\farcs6 slit placed at the parallactic angle. We used \texttt{floydsspec}\footnote{\url{https://github.com/LCOGT/floyds_pipeline}} pipeline \citep{Valenti_floyds} to extract the 1D spectrum, calibrate the wavelength to a HgAr lamp spectrum taken after the exposure, and calibrate the flux to a standard star observed on the same night. We include the publicly available data taken under the Public ESO Spectroscopic Survey of Transient Objects (PESSTO) program \citep{pessto_smartt_2015} using EFOSC2 mounted on the 3.6m ESO-NTT. The spectra at 4 epochs were taken with both Gr11 (3345 -- 7470 \AA; R $\sim$ 400) and Gr16 (6000 -- 9995 \AA; R $\sim$ 400) to cover the blue and red wavelengths. 1 spectra was taken with Gr13 (3650 -- 9250 \AA; R $\sim$ 350). A slit of width 1'' was used for these observations. The wavelength calibration was done with respect to HgAr lamp taken before and after the observation run, and the flux was calibrated to a spectrophotometric standard star.
A detailed description of data reduction can be found in \cite{pessto_smartt_2015}.

We also include the high resolution (R $\sim$ 3000) spectroscopic data, available in WISeREP \citep{wiserep}, which were taken with the Wide Field Spectrograph on the Australian National University 2.3-m telescope under the ANU WiFeS SuperNovA Programme (AWSNAP; \citealp{awsnap_2016}). The WiFeS instrument is an image-slicing integral field spectrograph with a wide 25'' $\times$ 38'' FoV. The WiFeS image slicer breaks the field of view into 25 ‘slitlets’ of width 1'', which are dispersed through VPH grating (B3000 or R3000).  The wavelength solution for WiFeS is derived using an optical model of the spectrograph, and the flux is calibrated to spectrophotometric standard stars. The detailed description of the data reduction can be found in \cite{awsnap_2016}.

Our optical spectroscopic observations span from day $\sim 8$ to $327$ after the explosion. NIR spectroscopy was acquired using SOFI mounted on the 3.6m ESO-New Technology Telescope (NTT) under the PESSTO program. The NIR observations span from day $\sim 19$ to $\sim 116$. The log of spectroscopic observations is given in \autoref{tab:spectra_log}. All the spectra were scaled to photometry by using \texttt{lightcurve-fitting}\footnote{\url{https://github.com/griffin-h/lightcurve_fitting}} module \citep{lk_fitting} to account for the slit loss corrections. Finally, all the spectra were corrected for the heliocentric redshift of the host galaxy. Despite careful extraction, some host-galaxy contamination will likely remain in our 1D spectra of this nuclear SN that may vary from night to night due to seeing, slit position, and instrumental effects. We keep this caveat in mind in our analysis below.
\section{Estimation of extinction and explosion epoch} 
\label{sec:host_galaxy}
The Milky Way extinction along the line-of-sight of ASASSN-14il is E(B-V)$_{\mathrm{MW}} = 0.0019$ \citep{milkyway_reddening}. To estimate the host galaxy contribution to the total reddening, we search for the NaID 5890, 5896~\AA ~doublet in the high-resolution ANU-WiFeS spectra. The NaID doublet is seen at the corresponding host-galaxy rest wavelength. We combine the ANU spectra taken on UT 2014-10-09.50, UT 2014-10-18.56, and UT 2014-10-27.47 to increase the S/N ratio. The NaID doublet is contaminated by the broad wings of a nearby emission line, as seen in figure \autoref{fig:NaID}. Additionally, we note that the HeI emission is blueshifted from its rest wavelength, the implications of which is discussed in \autoref{sec:spectral_evolution}. We model the wings of the emission line as a part of the continuum and estimate the equivalent width (EW) of Na D$_1$ and D$_2$ from the normalized spectra to be $0.44 \pm 0.03$~\AA, and $0.57 \pm 0.03$~\AA, respectively. We use the relation given by \cite{poznanski2012} between the Na1D features and $E(B-V)$. From this relation, we estimate host galaxy reddening of $0.22 \pm 0.09$ mag and $0.21 \pm 0.08$ mag, respectively, for D$_1$ and D$_2$. We also use the relation for the combined equivalent width of D$_1$ and D$_2$ to get the reddening value of $0.22 \pm 0.09$ mag, similar to the one estimated individually from D1 and D2. This estimate is consistent with \cite{14il_dickinson_thesis}, within errors. We multiply this reddening value by 0.86 to be consistent with the recalibration of Milkyway extinction by \cite{milkyway_reddening}. Therefore, the host-galaxy extinction is $0.19 \pm 0.08$~mag. The fitting errors have been propagated in quadrature. Thus, we adopt a total (galaxy+host) $E(B-V) = 0.21 \pm 0.08$ mag. We use this value of extinction throughout the paper.

To estimate the explosion epoch of \il, we perform parabolic fitting on the early time $V$-band lightcurve. The magnitudes are converted to fluxes up to 25 days post-discovery. The best-fit coefficients are used to find the roots of the equation, i.e., the value of time for which the flux equals zero. We perform the fit using MCMC simulations to constrain the associated errors. The explosion epoch is estimated to be JD $2456926.4 \pm 0.5$ (UT 2014-09-25.9) and adopted as the reference epoch throughout the paper. The general information about \il ~and its host galaxy is listed in \autoref{tab:sn_and_host}.

\section{Lightcurve Characteristics of asassn-14il}
\label{sec:lightcurve_evolution}

\begin{figure}
    \centering
    \includegraphics[width=\columnwidth]{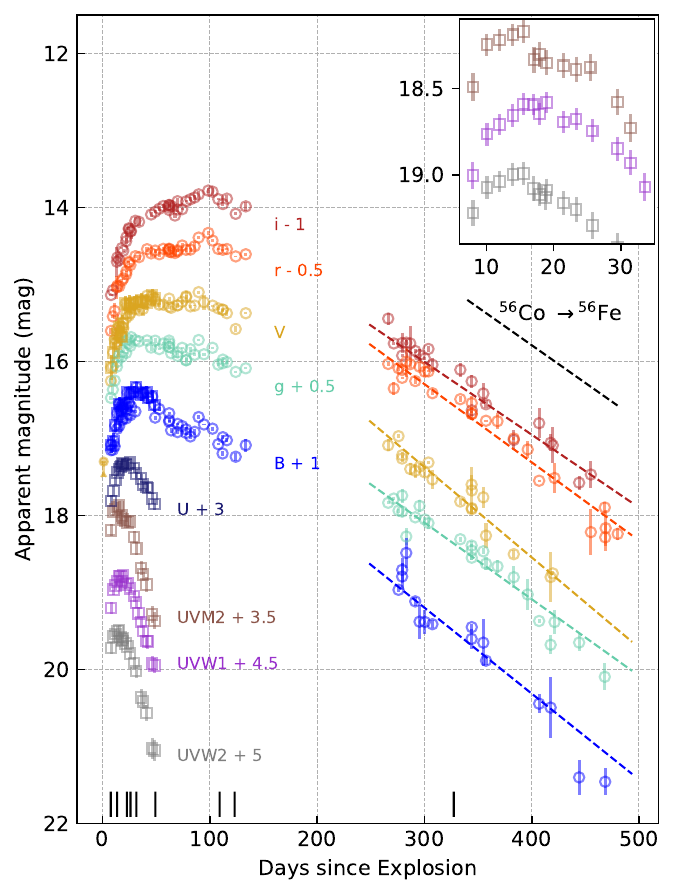}
    \caption{The evolution of optical and UV lightcurves of ASASSN-14il. Different bands are plotted with an offset for clarity. Circles represent the photometry from LCOGT, and squares represent the photometry from \swift UVOT. Epochs having accompanying spectra are represented by line symbols at the bottom. The inset shows a zoomed view of the UVM2, UVW1, and UVW2 bands. A plateau can be seen in the optical bands between days 40--80. A prominent bump in the lightcurve can be seen around day 90.}
    \label{fig:photometry_plot}
\end{figure}

\begin{table}
    \caption{The rise time and peak magnitudes of UV-Optical bands with a clear peak observed. The listed value and error of Rise Time and Observed Peak Magnitude are derived from the distribution resulting from the MC experiments.}
    \label{tab:peak_table}
    \centering
    \input{peak_table}
\end{table}
\begin{table}
    \caption{The lightcurve decline rates in all optical bands at various phases of evolution. The epochs are given with respect to the estimated explosion epoch.}
    \label{tab:slope_table}
    \centering
    \input{slope_table}
\end{table}

The UV and optical lightcurves of \il ~are shown in \autoref{fig:photometry_plot}. The first $\sim 40$ days after the explosion covers the initial rise in all UV and optical bands. We perform a spline fit on the lightcurves to estimate the peak magnitude and rise times. We have used a quartic spline implemented in the \texttt{UnivariateSpline}\footnote{\href{https://docs.scipy.org/doc/scipy/reference/generated/scipy.interpolate.UnivariateSpline.html}{https://docs.scipy.org/doc/scipy/reference/generated/scipy.interpolate.UnivariateSpline.html}} module of SciPy \citep{2020SciPy-NMeth}. The smoothing factor (between 0 and 1) was chosen by visual inspection such that it avoids overfitting. Monte-Carlo (MC) experiments were done to evaluate the associated uncertainties with the fit. For each MC experiment, we draw pseudo data points from the Gaussian distributions defined by the photometric magnitude and the error bars of the original data points and perform the fit.
The rise time and peak magnitudes in UV-Optical bands are listed in \autoref{tab:peak_table}. 
\cite{14il_dickinson_thesis} estimate the peak B magnitude to be much higher (based on visual inspection), which could be due to their higher adopted extinction value. We observe an increase in the rise time as we move towards the redder bands, which is expected as the photosphere cools down and moves towards redder wavelengths.
The lightcurve in \swift UV bands reaches a maximum around day 15-18 and declines very sharply thereafter. The {\it B} and {\it g} band attains a maximum at around day 36-38, much later than UV bands. The {\it V} and {\it r} bands do not have a distinct maximum as the initial rising phase directly transitions into a very flat plateau (day 40-100). The {\it i} band slowly rises in this duration.  The flattening of the optical lightcurves in optical bands between days 60-90 could be due to an ongoing CSM interaction. 

Both UV and optical lightcurves are bumpy in nature, with the optical bands showing a prominent bump around day 100. 
To check the significance of this lightcurve bump, we thoroughly check the diagnostic output from the \texttt{lcogtsnpipe} pipeline. No artefacts are seen in difference images before, during or after the bump, And the subsequent PSF extraction leaves behind just background noise. The images during this phase have high S/N ratio. Additionally, similar bumps are seen in the space based \swift UVOT data at early times. Therefore, we consider the bump to be a genuine feature of the lightcurve and not an artefact.
Similar bumps, attributed to the inhomogeneities in the CSM, have been seen in many SNe (c.f.r SN 2006jd - \citealp{2005ip_2006jd_stritzinger}; iPTF13z - \citealp{iptf13z_nyholm}; iPTF14hls - \citealp{iPTF14hls_Wang22}), including the 2012B eruption of SN~2009ip \citep{2009ip_graham}. \cite{martin_2009ip} argue that the lightcurve bumps in the 2012B outburst of SN~2009ip could be caused by the heating of the continuum-emitting region (caused by the density variations in the CSM), or due to line blanketing or a combination of both. Their detailed analysis of the detrended lightcurves in multiple bands makes a compelling case that while the first prominent bump can be explained by enhanced heating, the second and the third lightcurve bumps are influenced heavily by line blanketing.
The explosion mechanism of the 2012B outburst of SN~2009ip and whether it was a terminal explosion have been debated. \cite{2009ip_graham} note that the lightcurve bumps correspond to the fast ejecta catching up to the material ejected from the previous explosions. No pre-explosion activity has been seen in \il, but that does not necessarily negate the possibility that it may have just gone undetected. In addition, \cite{2009ip_graham} notes that the blackbody temperatures of the 2012B outburst show an uptick during this bump phase. A similar trend is seen in \il ~but not to the same prominence, likely due to the short-lived nature of the bump. In addition, the 2012B eruption of SN~2009ip experiences multiple bumps during its evolution; similarly, in \il, the bumps can be seen clearly in at least two epochs at around days 15 and 90. Based on the comparison, a similar explanation seems plausible for the bumps in the lightcurves of \il. Additionally, \cite{2009ip_graham} notes that the lightcurve bumps seen in SN~2009ip are more prominent in the bluer bands. In contrast, \cite{martin_2009ip} find that while it is true for the first bump, the second and third lightcurve bumps, influenced by line blanketing, do not follow the same trend. 
This behavior holds true in \il ~for the early bump at day 15, as the bump is more prominent in the UV bands. However, due to the scatter in the lightcurve, it is difficult to comment on the behavior of the bump at day 90. One distinction between the bumps seen in SN~2009ip and \il ~is that the bumps seen in SN~2009ip are longer lasting than \il, which can be attributed to the scale of inhomogeneities present.

All optical bands show a decline in late phases (day $> 250$). The initial decline of the lightcurves after the plateau is not well captured due to data gaps (days 120-260). Therefore, it is not possible to comment on the similarities with the transitional phase of SNe~IIn-P. The lightcurve settles into the decline of 1.0-1.2 mag (100d)$^{-1}$ at late phases, which is close to the expected $^{56}$Co decay rate of 0.98 mag (100d)$^{-1}$. It is possible to explain the observed decay rate as a result of the ejecta-CSM interaction as well, as this mechanism can give rise to a wide diversity of decay rates in SNe~IIn. The lightcurve decline rate in all optical bands at different phases of the evolution is given in \autoref{tab:slope_table}.  

\subsection{Comparison Sample}

\begin{table*}
\caption{General properties of the SNe~IIn comparison sample. We have taken redshift dependent distances assuming H$_0$ = 73 km sec$^{-1}$ Mpc$^{-1}$, $\Omega_{\mathrm{matter}}$=0.27, and $\Omega_{\mathrm{vaccum}}$=0.73. The redshifts are corrected for the effects of GA, Virgo, and Shapley unless mentioned otherwise. We take the reference epoch to be the explosion epoch if available in the literature; otherwise, the discovery epoch is considered. The peak magnitudes are as the brightest absolute magnitude showed by the SN based on the distance and reddening considered here.}
\centering
\smallskip
\label{tab:comparison_sample}      
\input{comparison_sample.tex}
\end{table*}

\begin{figure}
    \centering
    \includegraphics[width=\columnwidth]{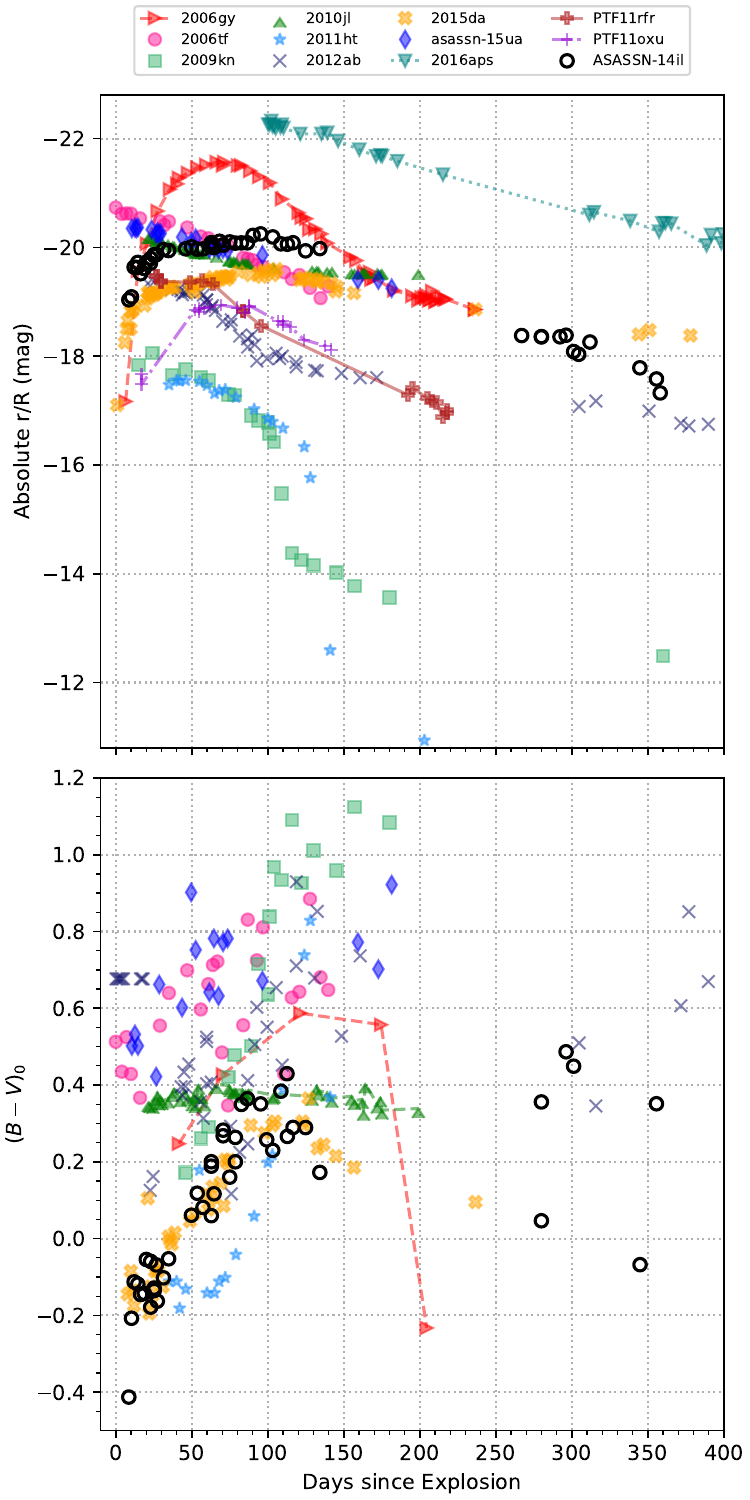}
    \caption{\textit{top:} Absolute r/R-band lightcurve of \il ~compared with those of the SNe~IIn in the comparison sample. \il ~is very luminous only dwarfed by SNe~2006gy and 2016aps. The lightcurve is very similar to SN~2015da, but the decline rate of \il ~is higher than SN~2015da and similar to SNe~2006tf, 2016aps, and ASASSN-15ua. \textit{bottom:} Intrinsic B-V color of \il ~compared with the reference sample. The gradual reward evolution of B-V color evolution is similar to SNe~IIn.}
    \label{fig:V_abs}
\end{figure}

We have chosen a fairly diverse comparison sample of eleven other well-researched SNe. The sample consists of the following objects: SNe~ 2005ip, 2006gy, 2006tf, 2009kn, 2010jl, 2011ht, PTF11rfr, PTF11oxu, 2012ab, ASASSN-15ua, and 2016aps. These SNe~IIn cover a diverse range in luminosity and show interaction signatures up to late times.
The basic parameters of the comparison SNe are laid out in \autoref{tab:comparison_sample}. SN 2012ab represents the normal luminosity SNe~IIn. SNe~2006gy, 2006tf, 2010jl, ASASSN-15ua, and 2016aps are all very luminous type SNe~IIn having peak M$_{r/R} < -20$ 
that result from various CSM configurations. 
SNe~IIn-P 2009kn and 2011ht, with a distinct plateau in their lightcurves, are also included in this comparison sample. As noted by \cite{nyholm_IIn_survey_2020}, PTF11rfr and PTF11oxu also show a plateau in their lightcurve, but unlike the SNe~IIn-P, they don't show a sharp drop from the plateau into a $^{56}$Ni-dominated tail. The lightcurve modeling of these SNe reveals a diversity in their CSM configuration. SN~2006gy results from a massive compact CSM shell \citep{2006gy_smith_mccray_shellshockeddiffusion}, while more extended CSM structures power the comparatively long-lived lightcurves of SNe 2006tf, 2010jl, and ASASSN-15ua \citep{2006tf_smith, Ofek_2010jl_modelling, 15ua_dickinson}. 

We have compiled a sample containing the normal luminosity and very luminous SNe~IIn from the literature. Also, the different lightcurve shapes resulting from different physical scenarios enable us to highlight the lightcurve heterogeneity of SNe~IIn and the role of CSM interaction in their photometric evolution. The sample allows us to put the luminosity of \il ~in perspective and infer the possible physical scenario by comparing and contrasting their lightcurve and color evolution. 

\subsection{Absolute lightcurve and Color Evolution}
Adopting the distance modulus $\mu = 34.74 \pm 0.15$ mag and total reddening $E(B-V) = 0.21 \pm 0.08$ mag, we generate the absolute magnitude lightcurves of \il. The $r$-band magnitude of \il ~plateaus at $\sim -20.2$ mag over many tens of days. The $V$-band also shows similar behavior. \autoref{fig:V_abs} shows the comparison of the absolute $r$-band lightcurve of \il ~with other SNe from the comparison sample. 
\il ~stands out as one of the most luminous SN in the comparison sample, which is in concordance with SNe~2006tf, ASASSN-15ua, 2010jl and is only dwarfed by the extremely luminous SNe~2006gy and 2016aps. The lightcurve shape of \il ~is very similar to SN~2015da except at the late epochs when SN~2015da declines much more slowly. However, the decline rate of \il ~is similar to those of SNe~2006tf, ASASSN-15ua, and 2016aps. 
The lightcurves of all these SNe have been successfully modeled by interaction of SN ejecta with massive amounts (in order of tens of M$_{\odot}$) of dense extended CSM structures that allow the ejecta-CSM interaction to persist until late times \citep{2006tf_smith, 15ua_dickinson, 2015da_tartaglia, 2015dalate_nathansmith_2023, 2016aps_nicholl}. Based on the required mass-loss, the progenitors for these events are speculated to have undergone an LBV giant eruption.
A similar progenitor and physical scenario seems plausible for explaining the observed luminosities and late-time decline rate of the lightcurves of \il ~as well.

SNe~IIn-P 2009kn and 2011ht show a plateau in their lightcurve, but afterward, they steeply drop into a radioactivity-dominated tail. On the other hand, SNe like PTF11rfr and PTF11oxu show a plateau in their lightcurve, but their lightcurves smoothly decline afterward in the $r$-band \citep{nyholm_IIn_survey_2020}. Due to the data gaps between days 120-260, it is not possible to establish parallels with either of these classes of objects. However, we note that the SNe~IIn-P in our sample are generally much fainter than \il.


The {\it B-V} color evolution of \il ~is in agreement with the other SNe~IIn from the comparison sample. All the SNe show a redward evolution in the \textit{B-V} color as the photosphere cools down, \il ~maintains an overall bluer color. The overall color evolution of \il ~is strikingly similar to SN 2015da.


After correcting the multi-band magnitudes obtained for extinction, the flux integration (in {\it UVW2, UVM2, UVW1, U, B, g, V, r, i} bands) was performed with blackbody corrections to attain the bolometric luminosities using {\sc SuperBol} \citep{2018Nicholl}. The lightcurve peaks at day $\sim$16 attaining a luminosity of $1.6 \times 10^{44}$ erg s$^{-1}$, radiating a total energy of $0.9 \times 10^{51}$ erg during the observational campaign. This estimate is significantly higher than that of \cite{14il_dickinson_thesis} due to the fact that their calculation was done with V-band lightcurve without any bolometric correction and accounts for only the first 120 days.
\cite{slsn_gal-yam_2012} suggested that SNe showing peak magnitude less than --21 in any band or peak luminosity higher than $7 \times 10^{43}$ erg s$^{-1}$ may be considered as SLSNe. According to both these criteria, \il ~can be considered as a SLSNe; however, this threshold is somewhat arbitrary, as mentioned in \cite{slsn_gal-yam_2019}, and it is unclear whether two separate luminosity populations exist for SNe~IIn.
%


\input{mosfit_param.tex}
\section{multi-band lightcurve modeling using MOSFiT} \label{sec:Modelling}
\begin{figure}
    \includegraphics[width=\columnwidth]{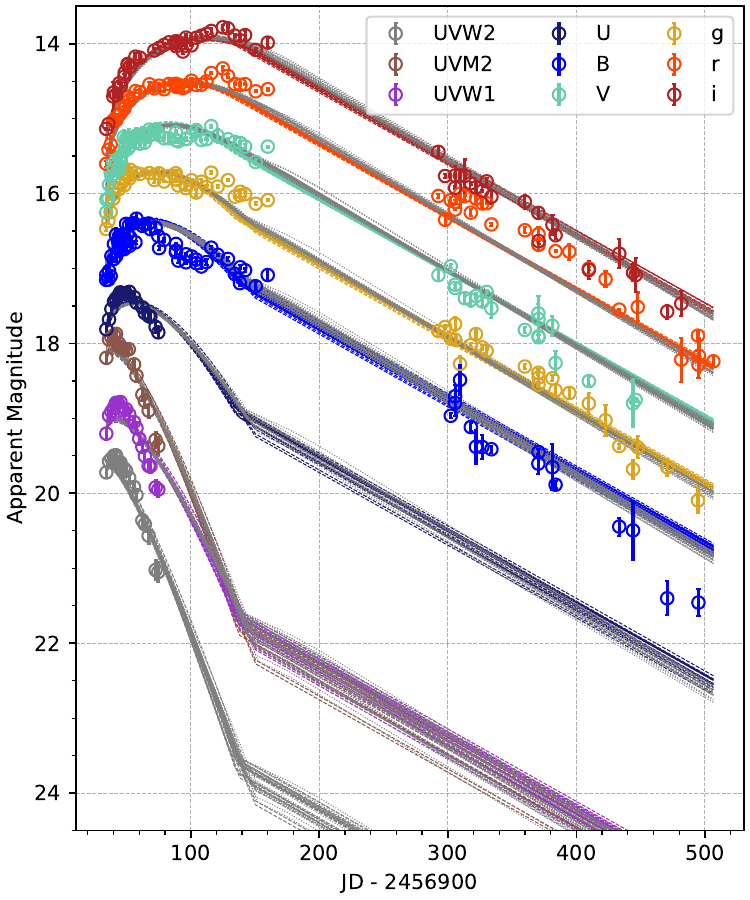}
    \caption{Results of \csm ~models from \mosfit. The lightcurves representative of the evaluated posterior are plotted on top of the observational data. The colored and gray lines represent the posterior of wind and shell \csm ~models, respectively. The lightcurves of \il ~are well reproduced by the CSM models with M$_{\mathrm{csm}}$ between 5-9 M$_{\odot}$.}
    \label{fig:mosfit_lc}
\end{figure}

To estimate the physical parameters associated with the explosion, we also performed analytical multi-band lightcurve modeling of ASASSN-14il. \texttt{MOSFiT} is an open-source lightcurve fitter for transients \citep{mosfit}. It fits the multiband lightcurves with a list of built-in models using Markov Chain Monte Carlo (MCMC) methods. 

Since narrow and intermediate width lines, which are indicative of interaction, are seen in the spectra of \il ~throughout its evolution, we attempted to fit its lightcurves with the models which include CSM as a powering mechanism.
\cite{Chevalier_1982_self-similar, Chevalier_1998} lays the analytical groundwork for the radiation expected from ejecta interacting with CSM. To achieve a self-similar solution, it is assumed that the ejecta is expanding gas into a stationary CSM. Both the expanding outer-ejecta and the stationary CSM are assumed to follow a power-law distribution ($\rho_{\mathrm{ejecta}} \propto r^{-n}$ and $\rho_{\mathrm{CSM}} \propto r^{-s}$).
\cite{CW12} presented generalized semi-analytical models that take into account shock power from interaction, magnetar spin-down, and $^{56}$Ni radioactivity. This work has been used in \cite{CW13} with $\chi^2$-minimization algorithm to find optimal parameters for a sample of interacting SNe. 
\cite{Chevalier_1982_self-similar} presented the solutions for $s=0$ and $s=2$ and later
\cite{Jiang_IIn_analytical_mosfit} extended these solution to $0 \leq s \leq 2$.
This model, along with a generalized framework to easily fit multiband lightcurves, is implemented in \texttt{MOSFiT}. 

In the \mosfit ~\texttt{csm} model, it is assumed that the SN luminosity results from the conversion of kinetic energy from both the forward and reverse shock into heating. While in the \texttt{csmni} model, $^{56}$Ni is considered as an additional power source along with the csm interaction. We fitted the lightcurves of \il ~with both the \texttt{csm} and \texttt{csmni} models. We fix the power law index of the outer ejecta $n=12$, which is expected for the explosions of RSGs \citep{Matzner_Mckee_1999, takashi_IIn_analytical}. We fix the opacity $k=0.34$~cm$^2$ g$^{-1}$ for fully ionized hydrogen-rich material. We take the rest of the parameters as fitting parameters. The fitting was performed using the dynamic nested sampler \texttt{dynesty} \citep{dynesty}. For ejecta mass and csm mass, we consider uniform priors from 1-150~M$_{\odot}$. For the inner radius of CSM (R$_0$), we consider a uniform prior from $1-500$~AU, and for the CSM density at the inner radius ($\rho_0$), we consider a log-uniform prior from $10^{-14} - 10^{-9}$ g cm$^{-3}$. From both \texttt{csm} and \texttt{csmni} cases, we fitted with a shell csm ($s=0$) and a wind csm ($s=2$) models. For the \texttt{low\_Ni} and \texttt{high\_Ni} models we consider a uniform prior from 0.1-10\% and 0.1-100\% for the nickel fraction in ejecta (f$_{\mathrm{Ni}}$ = M$_{\mathrm{Ni}}$ / M$_{\mathrm{ej}}$), respectively.
The estimated parameters from fitting various models are listed in \autoref{tab:mosfit_param}. The $1\sigma$ errors listed represent the statistical uncertainty associated with fitting and hence are not indicative of the model uncertainties.

Both the \csm ~and \csmni ~models (with either a wind or shell CSM configuration) provide a very similar fit to the observed lightcurves, despite the latter being the more complex. Both are able to reproduce the broad lightcurve features such as peak luminosity, rise times, duration etc. reasonably well. However, these models have a smooth CSM distribution, and therefore they are unable to reproduce both the lightcurve bumps and the decline of lightcurve in the UV bands after the peak. We tried to model the lightcurve after removing the bump seen around day 90, in a manner similar to \cite{Bumpy_lk_typeI_slsn_Hosseinzadeh}, but this approach doesn't improve the fit to the lightcurves in our case. \autoref{fig:mosfit_lc} shows the sample lightcurves for the \csm ~(shell and wind) models, which are representative of the posterior for these models.
The late-time luminosity is powered by the reverse shock running through the ejecta after the forward shock terminates.
The models having an extremely high fraction of $^{56}$Ni in the ejecta can reproduce the steep decline of UV lightcurves. But due to their unphysical nature, we neglect these models.
The variation of the estimated parameters for different models dominates over the fitting uncertainties. However, the requirement of a high ejecta mass (10--100 $M_{\odot}$) and a very dense CSM ($\sim 10^{-11}$ g cm$^{-3}$) are consistent. 

The mass-loss rate for the wind-like csm models is calculated as 
$ \dot{M} = 4 \pi v_w \rho_0 R_0^2 $,
and the time of ejection of CSM (t$_{\mathrm{csm}}$) is calculated as $R_0 / v_w$, where we have assumed a typical LBV-like wind velocity ($v_w$) of 100 km sec$^{-1}$.
We estimate a mass-loss rate of 2.8 M$_{\odot}$ yr$^{-1}$ (for the csm\_s2 model), and the CSM shell/wind expelled about 1-5 years prior to the explosion.

\section{Spectroscopic Evolution} \label{sec:spectral_evolution}
\begin{figure*}
    \centering
    \includegraphics[width=\textwidth]{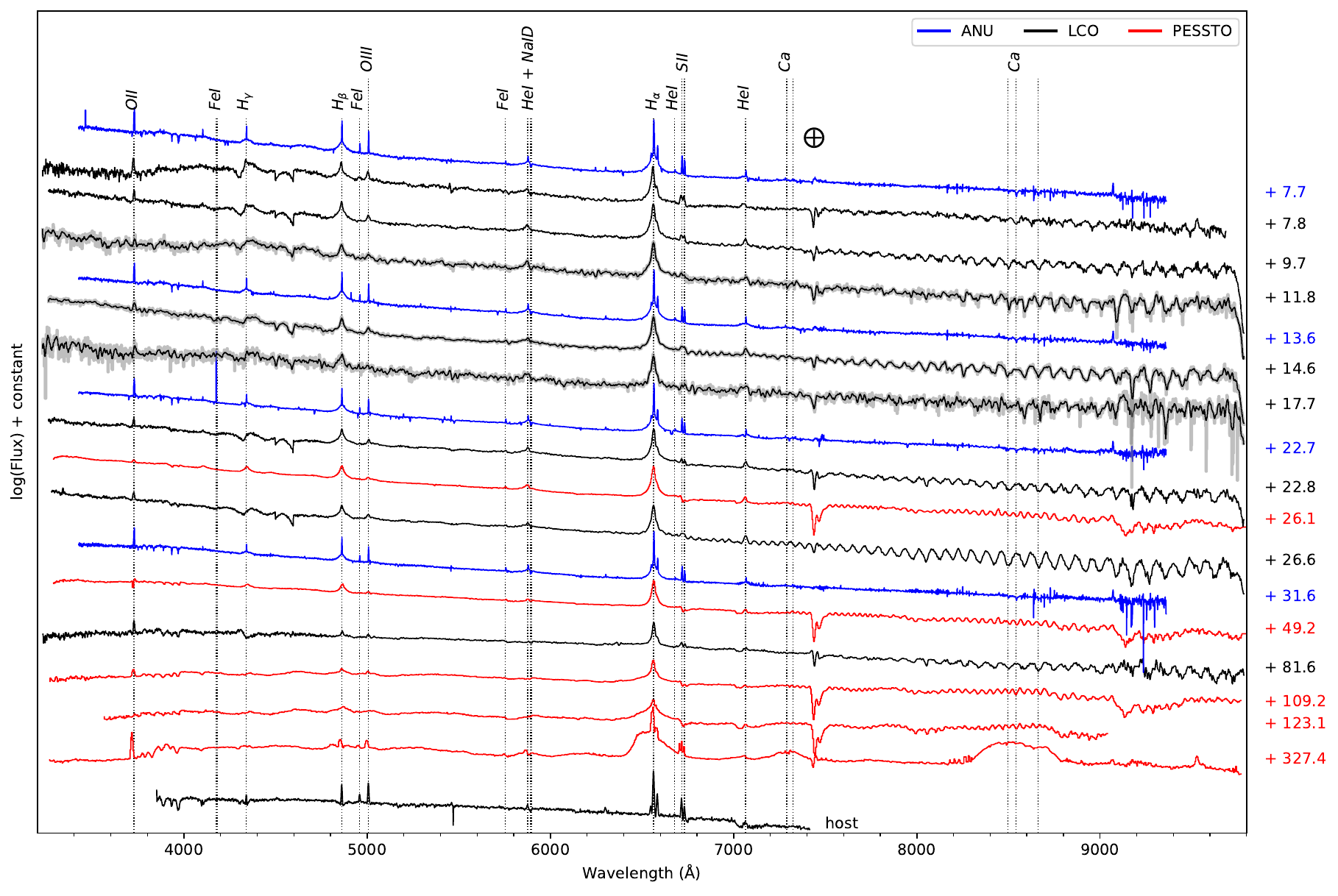}
    \caption{Spectral Evolution of \il ~from day 8--327. Spectra are displayed in the source rest frame and are corrected for extinction. The grey color represents the original spectra, and the colored lines represent smoothed spectra. The blue, black, and red spectra correspond to ANU-WiFeS, LCO-FLOYDS, and NOT-EFOSC spectra from PESSTO, respectively. Early spectra have a blue continuum and narrow H and He lines with Lorentzian wings. Broad ejecta components can be seen in H$\alpha$ and H$\beta$ lines along with intermediate width lines at later epochs. FLOYDS spectra suffer from imperfect fringing correction at $\gtrsim$8000 \AA. The host spectrum is plotted at the bottom.}
    \label{fig:spectra}
\end{figure*}

\begin{figure*}
    \centering
    \includegraphics[width=0.335\linewidth]{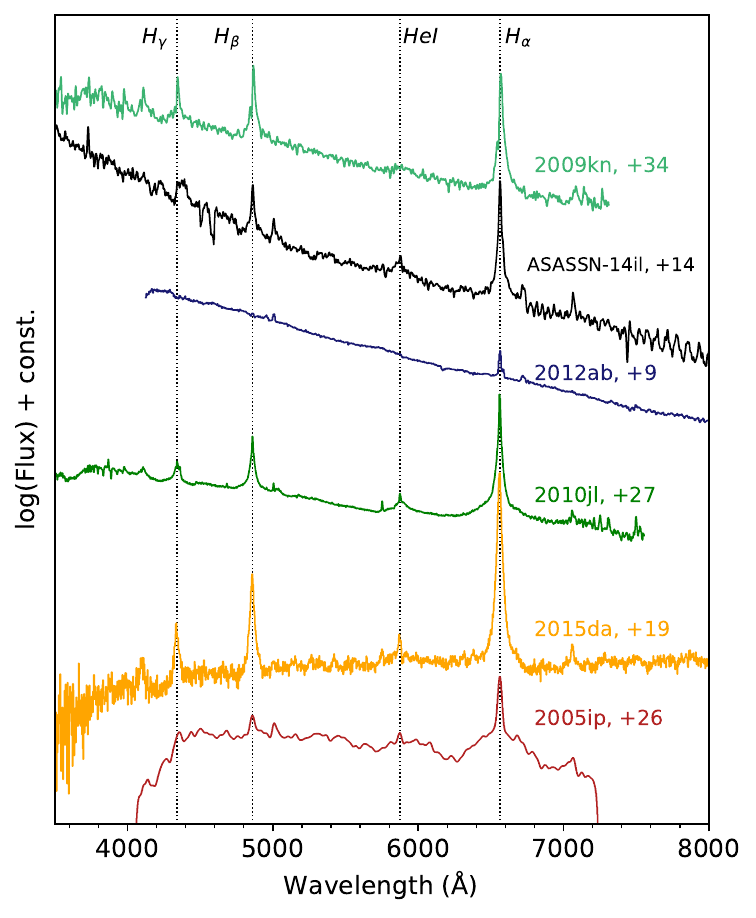}
    \includegraphics[width=0.32\linewidth]{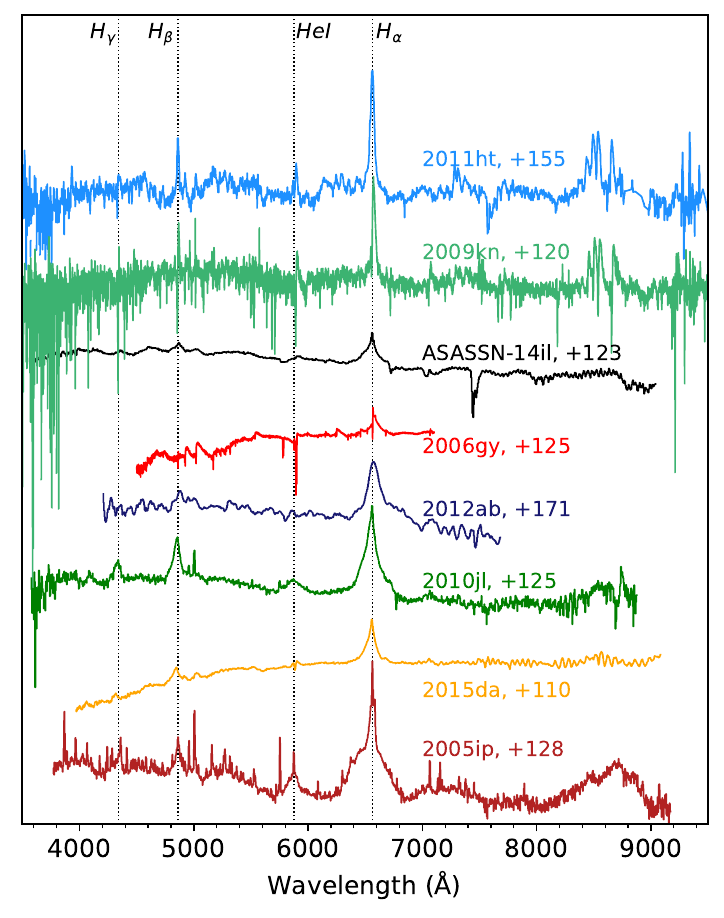}
    \includegraphics[width=0.32\linewidth]{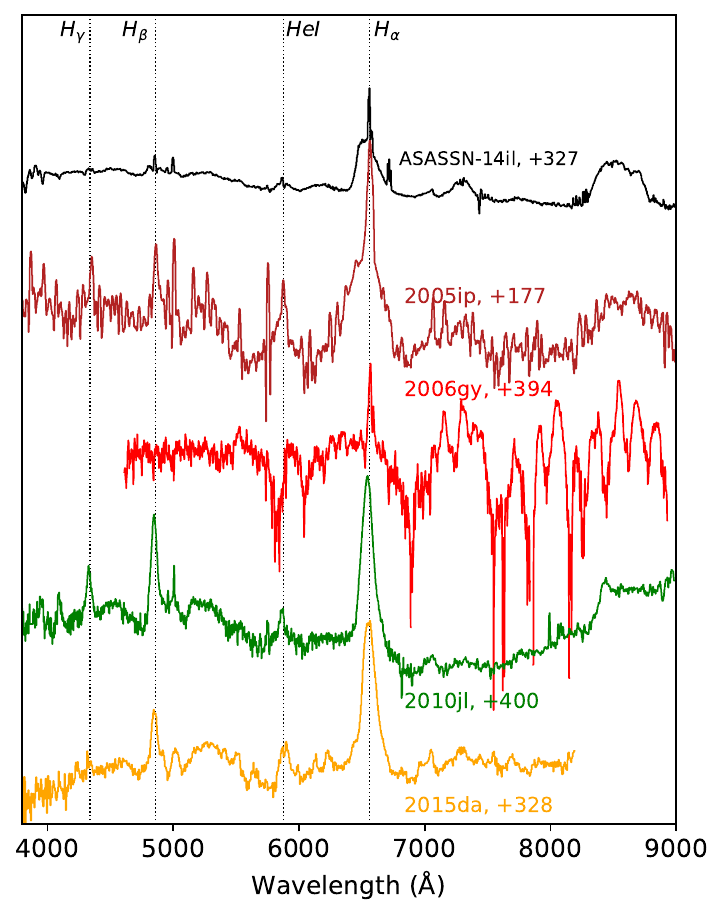}
\caption{Comparison of ASASSN-14il spectra with the comparison sample at distinct stages of its evolution. SNe~IIn-P and SNe~IIn are plotted above and below the spectra of \il ~(black). \il ~shows prominent Balmer features from early times similar to other SNe~IIn and SNe~IIn-P. At mid epochs, the Balmer lines are much more prominent in the SNe~IIn-P owing to their weaker continuum, in contrast to the SLSNe. At late epochs, many of the SNe develop a noticeable blueshift in their H$\alpha$ profiles.}
\label{fig:spectra_comparison}
\end{figure*}

\begin{figure*}
    \centering
    \includegraphics[width=\linewidth]{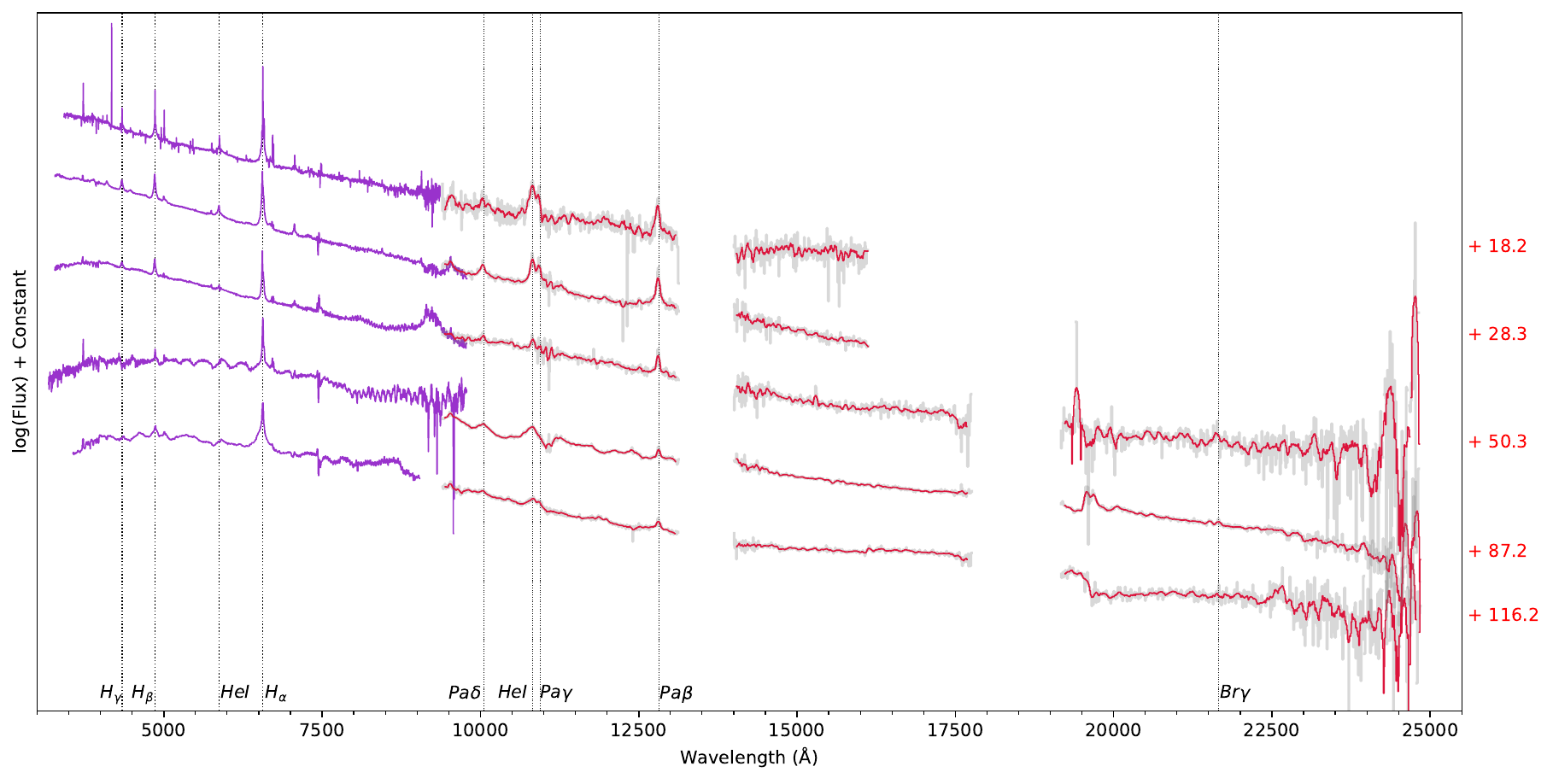}
    \caption{NIR spectral evolution of ASASSN-14il. The NIR spectra are plotted along with optical spectra from nearby epochs. The spectra are displayed in the source rest frame and are not corrected for reddening.
    The gray color represents the original spectra, and the colored lines (violet - Optical, red - NIR) represent smoothed spectra. The evolution of Pashcen lines is similar to the optical Balmer lines.}
    \label{fig:nir_spectra}
\end{figure*}

The spectral evolution of ASASSN-14il from day $\sim$8--327 is shown in \autoref{fig:spectra}. From the first spectra, narrow line features of Hydrogen are visible, which contrasts with some SNe~IIn, where initial spectra are almost featureless. The overall spectral evolution is very slow. Multicomponent H$\alpha$ line is visible throughout its spectral evolution. Other Balmer lines, as well as weak He lines (5876, 6678, and 7065~\AA) are also present since the beginning. The evolution of H$\alpha$ is very smooth, and a detailed description is given in subsection~\ref{Halpha}. The lines originating from SN ejecta like Ba {\sc II}, Sc {\sc II}, Mg {\sc I}, Ti {\sc II}, and Ca {\sc II} that are usually seen in a normal CCSNe without predominant interaction are not visible in the early phase of spectral evolution. Some lines like O {\sc III} and S {\sc II} are seen in the spectral evolution, indicating contamination by the host galaxy. For comparison, we have also shown the host-galaxy spectrum in \autoref{fig:spectra}. The host galaxy spectrum was taken on UT 2004-09-06.61 under the 6dF galaxy survey \citep{6dF2004, host_galaxy_z}. 
The 6dF spectrum is acquired using two VPH gratings, V(5400-7500\AA) and R(3900-5600\AA). Jones et al. (2004) note that the combined spectra provide a resolution of R$\sim$1000.
The aperture is fed using a 6”.7 fiber. The fiber was positioned 0”.2 south of the position of ASASSN-14il. So, the region sampled by the spectrograph should be a good approximation of the host galaxy.
The host-galaxy spectra closely resemble the 1442 day spectra presented in \cite{14il_dickinson_thesis}, and the H-to-N line ratio seems similar upon visual inspection. The unavailability of the spectrum presented by \cite{14il_dickinson_thesis}, hinders any further quantitative comparison.


The first spectrum was taken on day 8, featuring a blackbody continuum at a temperature of 11,000~K and a radius of 100~AU. This spectrum shows narrow H$\alpha$ profiles on top of smooth Lorentzian wings. Similar Lorentzian wings are present in other Balmer and He lines. The N {\sc II} contamination from the host galaxy is also seen in this spectrum.

The spectra are dominated by Lorentzian line profiles and a smooth blue continuum up to day 30. At day 50, the line profile slowly shifts from Lorentzian into more complex multi-component profiles. During this time, the temperature of the continuum decreases to 6000 K and becomes constant thereafter. After day 50, ejecta signatures are visible as a broad component in the H$\alpha$ line.

After day 100, a clear blueshift can be noticed in the H$\alpha$ profile, the extent of which increases with time. At day 327, the Balmer lines are dominated by the broad ejecta component, and Ca{\sc II} (both forbidden and NIR) lines are also visible. The blueshift seen earlier is even more prominent at this stage. The nebular phase spectral lines typical for the CCSNe are not visible at these later stages either.
The systematic blueshift in the H$\alpha$ profile can arise due to the asymmetric CSM or dust formation in the post-shock gas \citep{2006tf_smith,2010jl_jencson}.

The narrow emission feature of the HeI 5876~\AA ~line is noticeably redshifted compared to the rest-wavelength in the higher resolution ANU spectra (taken between day 8-32; \autoref{fig:NaID}). Such a redshift can be caused by an unresolved P-cygni profile characteristic of a CSM outflow \citep{2015dalate_nathansmith_2023}. The blue-edge of the narrow P-Cygni in SN~2017hcc lies at -70 to -60 km s$^{-1}$, which is completely washed out in low to moderate resolution spectra \citep{2017hcc_smith_andrews}. \il ~maybe a similar case where the P-Cygni profile in HeI and corresponding narrow P-cygni profiles in Balmer lines are unresolved due to the limited resolution of the spectra.
It is also possible that these narrow P-Cygni features are masked due to contamination from the host.

Comparison of ASASSN-14il spectra with other SNe~IIn from the comparison sample at different stages of its evolution is shown in \autoref{fig:spectra_comparison}. All the SNe have a blue continuum in the early spectrum (day $\sim 20$). All interacting SNe show narrow emission lines superimposed on the continuum. The H$\alpha$ of \il ~is very prominent, similar to most of the SNe in the comparison sample signifying strong ongoing interaction. However, SN~2012ab exhibits relatively weaker H$\alpha$, which results from directly seeing the high-velocity jet-like ejecta with relatively lesser CSM interaction at this phase \citep{2012ab_gangopadhyay}. The early spectra of \il ~bear an overall resemblance to the other events in the sample except for SN~2005ip, which shows a spectrum similar to normal type II, with narrow lines from preshock CSM on top \citep{2005ip_smith}.

The spectral evolution of \il ~at mid-phases closely resembles SNe~2006gy and 2015da around day $\sim$ 120. SN~2010jl has similar H$\alpha$ line morphology but with more prominent Balmer lines than \il. This spectral synergy between the three SNe is also evident in their photometric behavior, showing high luminosities and long-lasting lightcurves. In contrast, for the SNe~IIn-P 2009kn and 2011ht, H$\alpha$ line emission is more prominent with respect to their weaker continuum. In addition to the narrow component, a broad component can be clearly seen in \il ~ and all the other SNe~IIn. This component arises due to the ejecta contribution in the H$\alpha$ profile. Additionally, the broad component of \il ~shows a deficit in the red side flux of the H$\alpha$ line similar to SNe~2005ip, 2010jl, and 2015da.

After $\sim 1$ year of the explosion,  
\il ~shows a flat-topped broad component in H$\alpha$, with the narrow component on top. The observed blueshift in \il ~and SNe 2005ip, 2010jl, and 2015da is much more prominent compared to the mid-phases. The overall blueshift in the line profile of SN~2010jl was interpreted as the broadening due to electron scattering \citep{2010jl_fransson}. However, \cite{2010jl_smith_dust, 2015dalate_nathansmith_2023} argue that broadening by electron scattering should be symmetric about the original source of the narrow-line photons. The narrow lines in SNe~2010jl and 2015da are found at rest-frame velocities even at late phases where a blueshift is seen in the H$\alpha$ profile. Instead, this blueshift is explained as a result of dust formation in the post-shock CSM/ejecta, which preferentially blocks the emission from receding material, which is consistent with the other evidence of dust-formation \citep{2010jl_maeda, gall2014}. The H$\alpha$ profile of \il ~also shows an overall blueshift and a rest-frame emission component. The H$\alpha$ profile of \il ~at this epoch is similar to SN~2005ip at day $\sim 177$, which shows a similar blueshifted flat top profile. This blueshifted profile appears in SN~2005ip as early as day $\sim 120$ and is attributed to the post-shock dust formation in the CDS \citep{2005ip_orifox_dust, 2005ip_bak_neilsen_earlydust}. Based on these comparisons, the blueshifted H$\alpha$ profile of \il ~is likely the result of dust formation in the post-shock CSM/ejecta.


\subsection{Infrared Spectrum}
Five NIR spectra were obtained for ASASSN-14il between days 19--116 after the explosion. They are displayed in \autoref{fig:nir_spectra} along with the closest optical spectra. Prominent lines of Pa$\alpha$, Pa$\beta$, and Pa$\delta$ can be seen in the spectral sequence of ASASSN-14il. Pa$\gamma$ is also visible but is usually blended with He {\sc i} 10830~\AA ~and Pa$\epsilon$ line. The FWHM of the Pa$\beta$ and Pa$\delta$ typically varies between 8,000 - 13,000 km sec$^{-1}$. 
The lines here are isolated, and FWHMs concord with H$\alpha$ within error bars. Overall, the NIR spectra of \il ~are consistent with optical spectra regarding interaction features, mostly implied from the FWHM's being consistent in both the wavelength regimes.

\subsection{H-alpha Evolution}
\label{Halpha}
ASASSN-14il suffers significantly from host galaxy lines of O{\sc II}, N{\sc II}, and S{\sc II} (seen clearly in the higher resolution WiFeS spectral sequence). The host-galaxy spectrum taken under the 6dF survey (see \autoref{sec:spectral_evolution}) has an FWHM resolution of 9--12~\AA ~in the H$\alpha$ region. The SN spectra have lower resolution compared to the host spectrum. The SN spectrum was subtracted after degrading the resolution of the host galaxy spectrum, taking the S{\sc II} 6717-6731~\AA ~line as a reference for host contribution. In most low to medium-resolution spectra, the host galaxy contribution is almost removed, leaving an unresolved narrow residual component behind. For the ANU-WIFES spectra, the analysis was performed as is, without performing the host-galaxy subtraction, as they are much higher in resolution than the available host spectra.

\begin{figure*}
    \centering
    \includegraphics[width=0.32\linewidth]{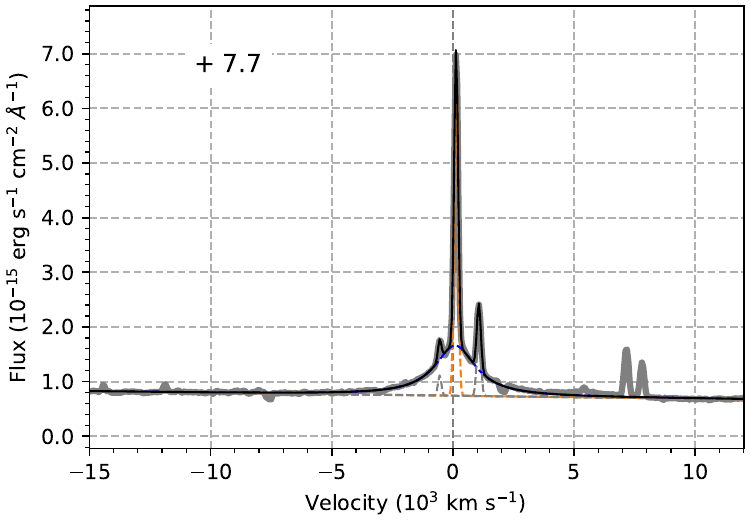}%
    \includegraphics[width=0.32\linewidth]{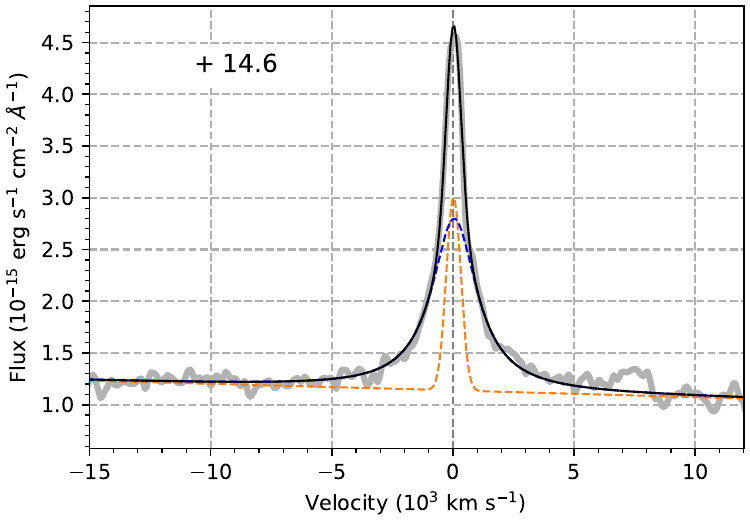}%
    \includegraphics[width=0.32\linewidth]{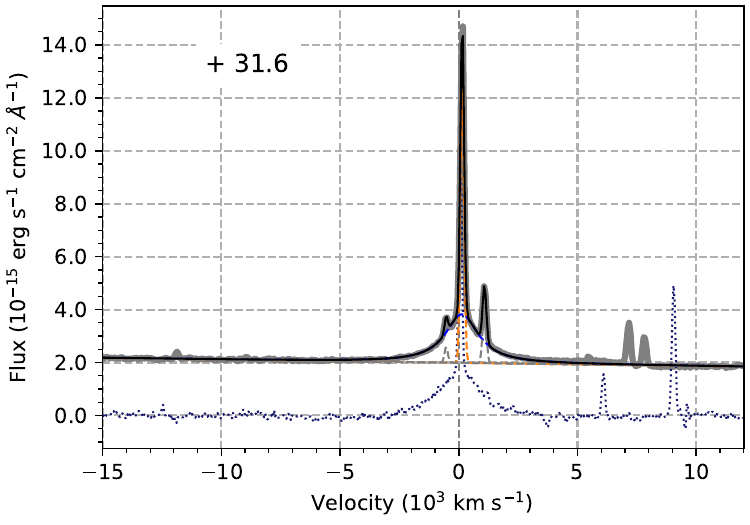}
    \includegraphics[width=0.32\linewidth]{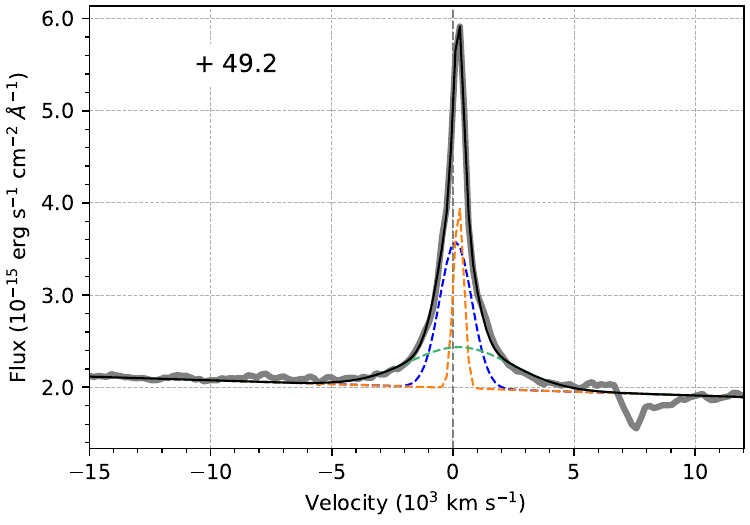}%
    \includegraphics[width=0.32\linewidth]{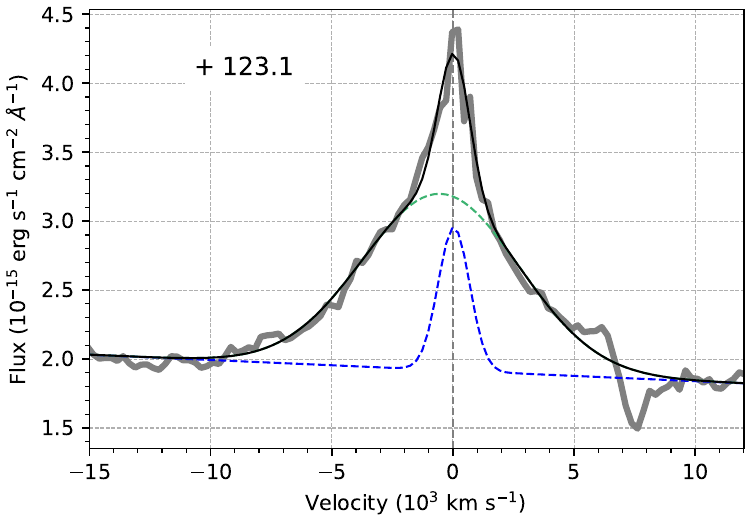}%
    \includegraphics[width=0.32\linewidth]{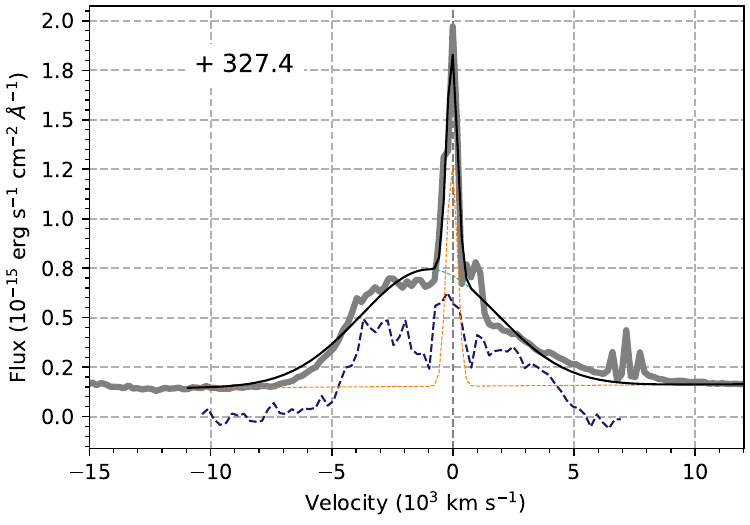}
    \caption{The H$\alpha$ evolution of ASASSN-14il being fitted with multiple Lorentzian and Gaussian components. The host galaxy contribution is subtracted, and the continuum is selected by masking the line regions. ASASSN-14il shows a very complex geometry of the spectral profiles. The orange, blue, and green components represent the narrow, intermediate, and broad components, respectively. For day 31.6 and 327.4 The H-beta (violet) from is plotted in violet.}
    \label{fig:spectral_decomp}
\end{figure*}

\begin{table*}
    \centering
    \caption{This table describes the spectral decomposition of H$\alpha$ profile of ASASSN-14il. The line centers and FWHMs of the narrow, intermediate-width, and broad components are reported. We have expressed the fitting error with the line centers and fwhm measurements. The total error in measurement is the sum in quadrature of the fitting error and the resolution uncertainty of the instrument.}
    \label{tab:spectral_decomposition}
    \input{14il_spectral_components}
\end{table*}

\begin{figure}
    \centering
    \includegraphics[width=\linewidth]{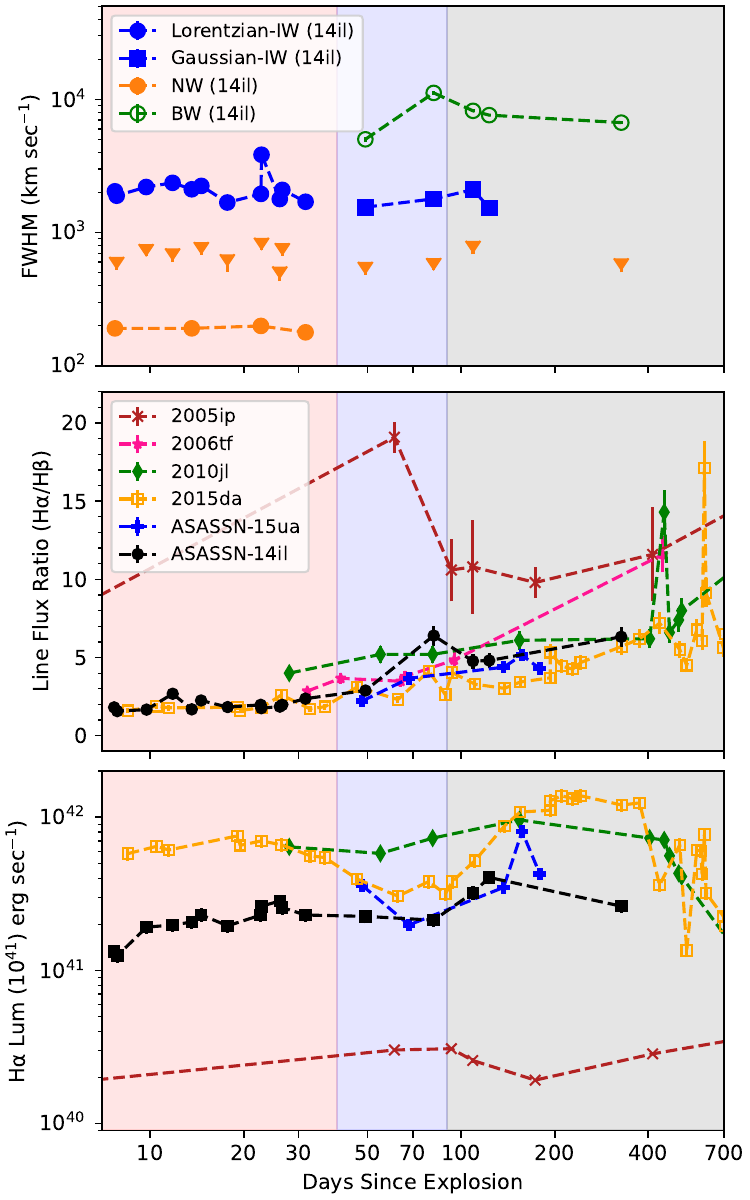}
    \caption{Top Panel: The FWHMs of the different components of the H$\alpha$ obtained by fitting Gaussian/Lorentzian to line profiles. The three color-coded zones in the panels demarcate the regions where \il ~enters into three phases of its evolution as described in \autoref{Halpha}; Middle Panel: the line flux ratios of H$\alpha$ and H$\beta$ for long-lasting SNe~IIn. 10$\%$ error has been included in the flux-ratios; Bottom Panel: the line-luminosity variations of H$\alpha$ as compared with other members of the sub-class.}  
    \label{fig:eqw_halpha_hbeta}
\end{figure}

In the host-galaxy subtracted spectra, the H$\alpha$ profile was fitted with combinations of the following to reproduce the overall line profile - i) A narrow Gaussian component, ii) A Lorentzian/Gaussian intermediate width component, iii) A broad Gaussian component. At early times, the overall H$\alpha$ profile is better represented by a Lorentzian intermediate width component as the emission lines are dominated by electron scattering. At late times (day $>$50), a Gaussian intermediate width component better reproduces the overall line profile as the lines evolve into a more complex multi-component structure. The choice of the continuum is very critical for performing these fits. The continuum is selected to be far from the line region by at least 50~\AA. The continuum selection was performed repetitively to optimize the consistency of the fits. The spectral evolution and the corresponding fits at selected epochs are shown in \autoref{fig:spectral_decomp}. The parameters of components obtained from the fitting at all epochs are given in \autoref{tab:spectral_decomposition}. The errors listed represent only fitting errors, and other uncertainties like resolution matching, subtraction of host-spectra, and imperfections in wavelength calibration have not been included. The evolution of the line centers and FWHMs of the various components is plotted in the top panel of Figure~\ref{fig:eqw_halpha_hbeta}.

Three components can describe the line profiles: i) A narrow component (unresolved in the lower resolution spectra) centered around the rest wavelength, ii) An intermediate width component centered around the rest wavelength, with FWHM typically varying between 1600--2000 km sec$^{-1}$, and iii) A broad emission varying in FWHM between 5000--8500 km sec$^{-1}$ seen after day 50. This broad component progressively develops a blueshift of $\sim 600-1000$ km~s$^{-1}$. The variance shown by the centers of the narrow and intermediate components is statistically insignificant. The minimum FWHM of the narrow component is 178 km sec$^{-1}$ seen in the ANU spectra (day 31.6), however it still may be unresolved. To check whether the narrow component seen in the high-resolution ANU spectrum is due to the host or intrinsic to \il, the host-galaxy spectrum was subtracted by degrading the resolution of the ANU spectrum. A residual narrow component persists, indicating narrow line emission from the SN as seen in our spectral sequence. \cite{14il_dickinson_thesis} performed a similar analysis with the high-resolution spectrum taken after the SN had faded and arrived at a similar conclusion. It should be noted that this residual narrow component may still be from the host galaxy as the H to N emission ratio is variable depending on the slit positions and orientation. However, a strong narrow component is seen in the HeI 5876~\AA line as well, in spite of the host galaxy having very low HeI emission.

The H$\alpha$ evolution of \il ~can be divided into three distinct phases. From day 8--50, the H$\alpha$ is dominated by Lorentzian wings caused by the electron scattering of narrow line photons. A narrow component is also visible in the spectra even after subtracting the host galaxy contribution. This indicates that the photosphere lies in the unshocked CSM at this phase. The UV (and bolometric) lightcurves peak during this phase (around day 15), which can ionize the unshocked CSM. During this phase, no signatures of underlying ejecta are seen in the H$\alpha$ profile, unlike some SNe where broad emission or broad P-Cygni profiles are seen from early times, which would indicate direct line-of-sights to the ejecta. In SN~2012ab, this is achieved by disk-like CSM \citep{2012ab_gangopadhyay} and in SN~2005ip, a clumpy CSM structure is present \citep{2005ip_smith}.

After day 50, the combination of a narrow Gaussian and a Lorentzian profile can no longer accurately reproduce the line profiles. Instead, the line profiles are more faithfully represented by combining three Gaussian (narrow, intermediate width, and broad) components. The broad wings of high-velocity ejecta are distinguishable in the line profiles. The Gaussian intermediate width and broad with components have FWHMs of 1550--2000 km s$^{-1}$ and 5000--8000 km s$^{-1}$, respectively. No significant P-Cygni absorption associated with the broad H$\alpha$ component is visible, suggesting that the underlying ejecta is already entering the optically thin regime. 

After day 81.5, both the intermediate width and broad component are visible up to day 123.0, along with the narrow component present throughout the evolution. During the evolution, the broad component changes in FWHM between 5000---8800 km s$^{-1}$. However, after day 109, a prominent blueshift can be noticed in the broad component of H$\alpha$, now centered between $\sim$ -500 to -900~km~s$^{-1}$. The H$\alpha$ profile in the day 327.3 spectrum can not be described as a combination of Gaussians with shifted centroids. Instead, it can be better described as missing flux at velocities higher than -4000~km~s$^{-1}$.
\autoref{fig:spectral_decomp} also shows the H$\beta$ line plotted (with offset for clarity) together with H$\alpha$ at days 31.6 and 327.4. The H$\beta$ line is scaled such that it matches the blue-side profile of the H$\alpha$ line. With this comparison, it is clear that the blueshift seen at late epochs is wavelength-dependent, more prominent at shorter wavelengths. This wavelength dependence is also noted by \cite{14il_dickinson_thesis}, based on their higher resolution spectra. The implications of this is further discussed in \autoref{sec:dust_formation}.

Figure~\ref{fig:eqw_halpha_hbeta} shows the total integrated line flux ratio of H$\alpha$ to H$\beta$, line luminosities of H$\alpha$ as compared with some other superluminous and long-lasting interacting SNe~IIn (SN 2005ip -- \citealp{2005ip_smith}; SN 2006tf -- \citealp{2006tf_smith}; SN 2010jl --\citealp{2010jl_jencson}; SN 2015da --\citealp{2015da_tartaglia} and ASASSN-15ua -- \citealp{15ua_dickinson}) along with FWHMs of the H$\alpha$ components of \il. The H$\alpha$ luminosity includes the contribution from the narrow, intermediate-width, and broad components of the H$\alpha$ profile. The H$\alpha$ line fluxes and H$\alpha$ to H$\beta$ flux ratios were estimated from the dereddened spectra, including a 10$\%$ error.
As seen in the middle panel of \autoref{fig:eqw_halpha_hbeta}, the H$\alpha$ to H$\beta$ flux ratio of ASASSN-14il steadily increases throughout its evolution. \cite{15ua_dickinson} have proposed that the H$\alpha$ to H$\beta$ flux ratio at early epochs is dominated by recombination after photoionization, where the line ratio typically approaches 3. Then, at later times, it is dominated by heating by the post-shock gas dominated by collisional excitation. A similar trend is noticed in \il ~and other objects where a transition is seen from photoionization to collisional excitation. The flux ratio of \il ~also closely resembles SN~2015da throughout the evolution.

The bottom panel of Figure~\ref{fig:eqw_halpha_hbeta} compares the H$\alpha$ line luminosities of \il ~with other SNe~IIn. For \il, the luminosity initially increases from day 8--23 by a factor of 7; the luminosity is almost constant from day 23--82, and from day 82--327, it increases again by a factor of 3. The H$\alpha$ luminosity of SNe~2010jl, 2015da and ASASSN-15ua shows an increasing trend when the interaction becomes maximum and decreases later. \il ~shows a similar trend, except for the flattening seen between days 23--82, which could be due to persistent interaction with shell/clump.

\begin{figure*}
    \centering
    \includegraphics[height=5.4cm]{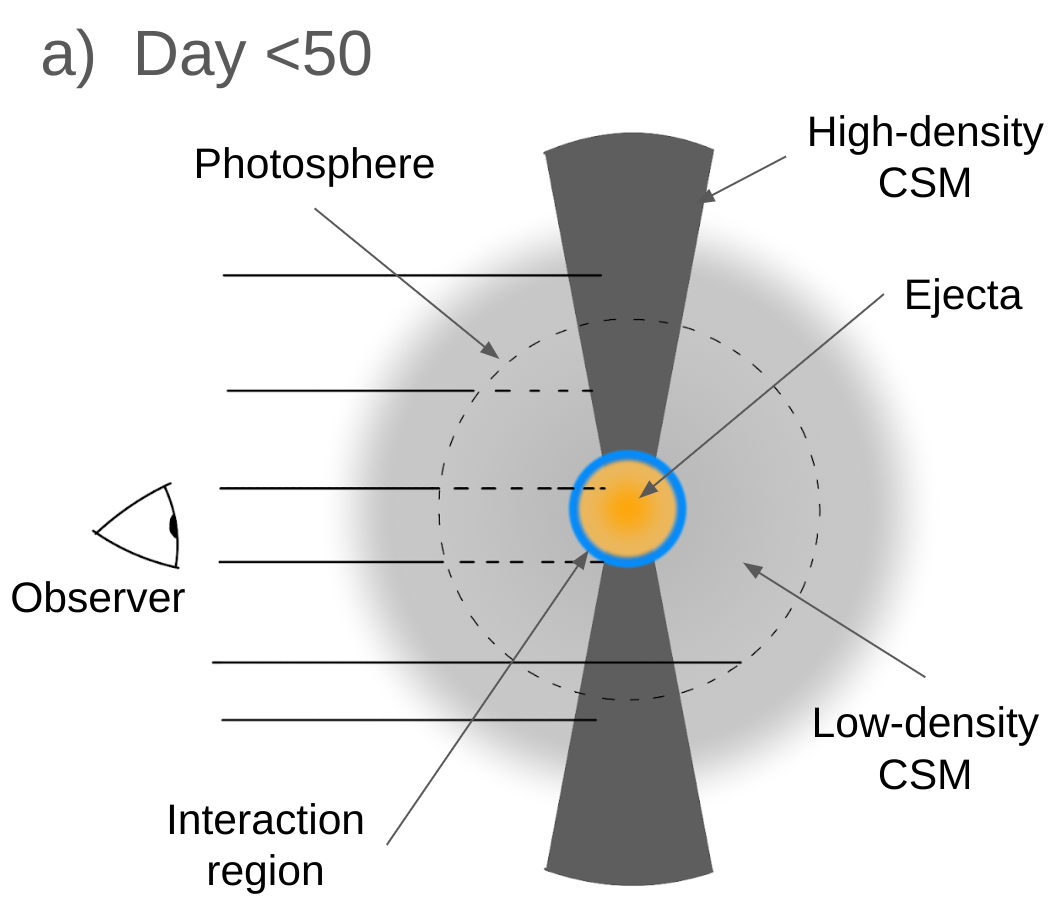}%
    \includegraphics[height=5.4cm]{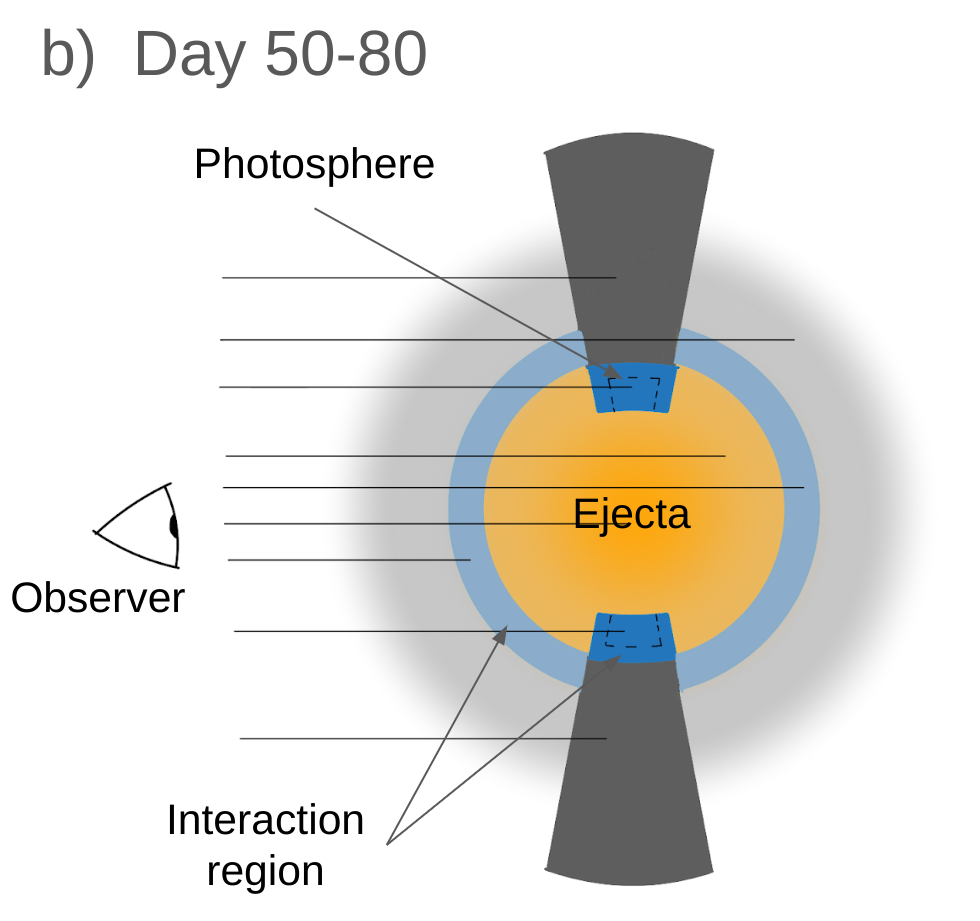}%
    \includegraphics[height=5.4cm]{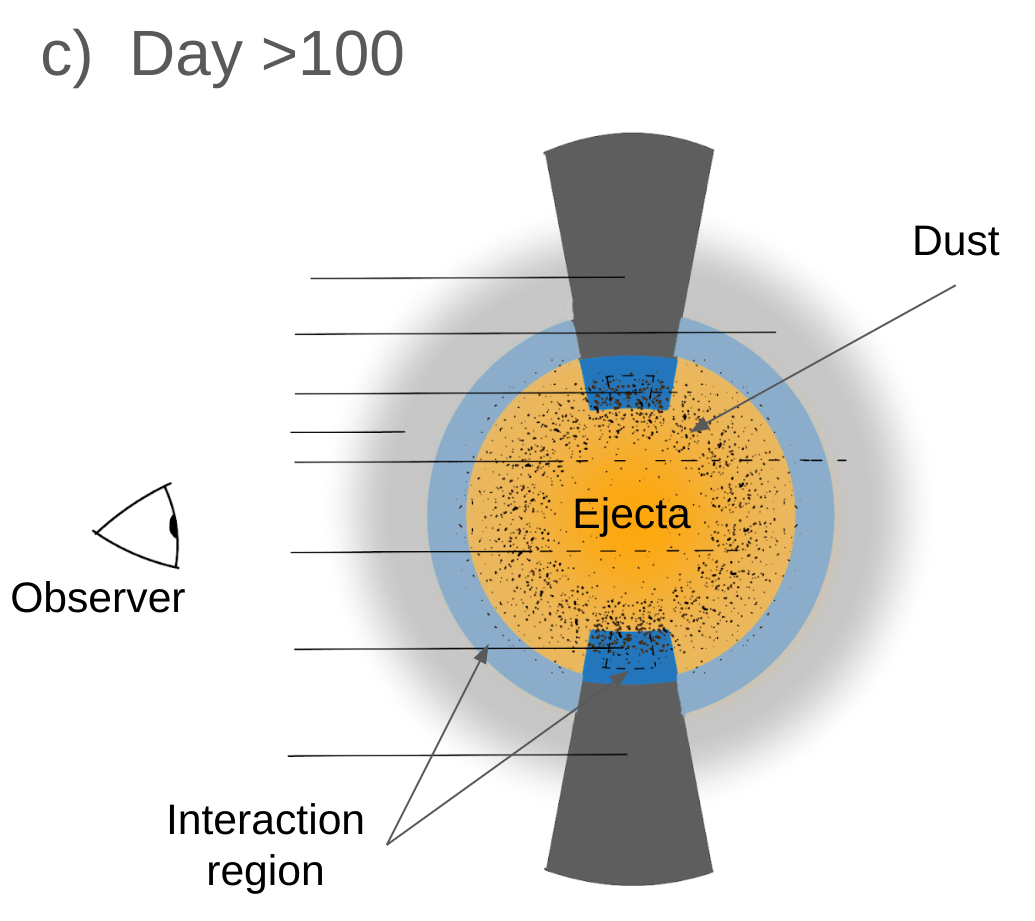}
    \caption{Cartoon diagram representing the evolution of ASASSN-14il through various distinct phases. Representative line-of-sights (LOSs) are shown by solid lines, where dashed lines indicate hindered LOSs. A dense disk-like CSM is located face-on to the observer, and A low-density CSM is present elsewhere. Interaction with dense CSM gives rise to most of the luminosity. At mid-phases, the photosphere recedes from the pre-shock CSM and signatures from the post-shock region and ejecta are visible. At late times the formation of dust obscures the tradition from receding material.}
    \label{fig:cartoonpic}
\end{figure*}

\section{Results and Discussion} \label{sec:conclusions}
This paper presents the long-term photometric (up to day 480) and spectroscopic (up to day 327) monitoring campaign of a SN~IIn ASASSN-14il. The lightcurves show bumpy behavior at different stages, which can be attributed to ejecta interacting with regions of enhanced CSM density. The multi-band lightcurves also display a plateau-like behavior due to ongoing interaction. A similar plateau is seen in SNe~IIn-P \citep{2011ht_mauerhan} followed by a steep drop (2-4 mag in optical) from the plateau into a radioactivity-dominated tail. In contrast, some SNe~IIn show a plateau followed by a linear decline \citep{nyholm_IIn_survey_2020}. However, we can not ascertain the nature of \il ~in the context of these two classes due to gaps in the observations. \il, with a peak absolute magnitude M$_{r}$ $\sim -20.3 \pm 0.2$ mag, is one of the brightest SN~IIn.
In UV bands, it is brighter than $-21$ mag, which satisfies the criteria for SLSN category \citep{slsn_gal-yam_2012}. However, whether such bright interacting transients form a separate population or fall in the tail end of normal SNe~IIn is unclear. The color curves are concordant with a typical SN~IIn. Overall, it bears a stark resemblance with SN~2015da in terms of the lightcurve and color evolution which indicates similar powering mechanism(s).

Both wind and shell type CSM models in MOSFiT can reproduce the broad lightcurve features of \il. To reproduce the observed luminosities of \il, a high amount of ejecta mass (a few tens to hundred M$_{\odot}$) is required to interact with a high-density CSM ($\sim$ 10$^{-11}$ g cm$^{-3}$). These one-zone models, however, can not reproduce the steep decline of UV lightcurves and the small bump seen in the optical lightcurves around day 90. The assumptions of homologous expansion, constant opacity, and spherical symmetry are simply not representative of the actual explosion scenario. More sophisticated modeling techniques are required to adequately model the lightcurve, which is beyond the scope of this work.

The spectral evolution of \il ~resembles typical SNe~IIn dominated by smooth blue continuum and Balmer lines. The Balmer lines are dominated by Lorentzian wings in the early epochs. The H$\alpha$ line profile evolves to be more complex and multicomponent after day 50. The signs of underlying high-velocity ejecta are visible after this epoch. The emergence of ejecta signatures at the time of peak optical luminosity indicates asymmetries in the explosion/CSM geometry, which is quite common in SNe~IIn \citep{smith_IIn_review, IIn_specpol_bilinski}. This is further discussed in \autoref{sec:asymmetry}.

After day 123.0, a prominent overall blueshift can be observed in the H$\alpha$ profile, which becomes more pronounced in the final spectra at day 327. The spectra of \il ~show similarity with SNe~2006gy, 2010jl, and 2015da. The late-time flat-topped profile is reminiscent of SN~2005ip. Also, the late-time blueshift can be explained by dust formation in the post-shock CSM or ejecta (similar to SNe 2005ip, 2010jl, and 2015da; \citealp{2005ip_smith, 2010jl_smith_dust, 2015dalate_nathansmith_2023}). The case of dust formation and alternative explanation are discussed in \autoref{sec:dust_formation}. The evolution of H$\alpha$ to H$\beta$ flux ratio supports the transition scenario from recombination to collisional excitation. The line luminosity of H$\alpha$ shows an increasing trend due to persistent ongoing interaction. 

\subsection{Narrow lines} \label{sec:narrow_lines}
\il ~shows narrow lines throughout its spectral evolution. However, the narrow lines do not show a P-Cygni profile which is expected from a CSM outflow. The lack of a narrow-width P-Cygni feature is very uncommon among SNe IIn. In the case of \il, there could be several reasons for the lack of this feature. First, it could be just due to the spectral resolution of the available spectra. P-Cygni profiles corresponding to outflow velocities of $< 100$~km~s$^{-1}$ would be unresolved in the presented spectra. An unresolved P-Cygni profile will appear redshifted as explained by \cite{2015dalate_nathansmith_2023}, and there are indeed some hints of it as the narrow HeI 5876 feature is redshifted from its expected rest wavelength. It is possible that \il ~is surrounded by an asymmetric CSM structure such that the CSM along the line-of-sight has a relatively lower density (as well as slow-moving) which will give rise to a weaker P-Cygni feature that is hard to resolve under limiting conditions. Secondly, it could be due to the contamination of the narrow lines by the host-galaxy emission. 

Additionally, a recent study by \cite{Ishii_2024_narrow_lines} explores the relations between the line shapes and CSM structure by Monte Carlo radiative transfer codes. They find that a narrow line exhibits a P-Cygni profile only when an eruptive mass-loss event forms the CSM. The CSM structure from a steady mass loss will have a negative velocity gradient after the SN event due to radiative acceleration. Therefore, an H$\alpha$ photon emitted at the deeper CSM layers, traveling outwards, will never be able to undergo another H$\alpha$ transition. Therefore, the CSM residing in the direction of the observer may be created by steady-mass loss (e.g., LBV winds) in the case of \il.

\subsection{Asymmetry} \label{sec:asymmetry}
In a spherical symmetric model for interacting SNe at early times the continuum photosphere lies in the CSM ahead of the shock. After some time, the photosphere moves into the post-shock CDS, and the uninterrupted ejecta is visible only at late times when the interaction has slowed down enough \citep{smith_IIn_review}.  The line profile of H$\alpha$ earlier spectra (up to day 31) of \il ~show a symmetric Lorentzian profile expected from the continuum being in the pre-shock CSM. However, the broad component from the SN ejecta is visible as soon as day $50$, along with the intermediate component from the post-shock CDS. The optical luminosity is near-peak at this epoch, indicating the presence of strong CSM interaction. This is not explained from a spherically symmetric CSM geometry. Instead, the emergence of broad ejecta suggests significantly lower CSM along the LOS compared to the CSM needed to sustain the peak luminosity. This can be explained by an asymmetric CSM configuration such that dense CSM, at angles alternate to the LOS, giving rise to the majority of the observed luminosity. Such a configuration is not uncommon in SNe~IIn. SNe 2010jl, 2012ab, and 2015da all reveal the underlying ejecta earlier than expected from a spherical CSM distribution and are suggested to feature a disk-like/torus-like CSM geometry \citep{2010jl_xray_asymmetry_katsuda, 2014ab_bilinksi, 2012ab_gangopadhyay, 2015dalate_nathansmith_2023}. 

A recent spectropolarimetry study of SNe~IIn (including \il) by \cite{IIn_specpol_bilinski} reveals a significant amount of polarization in the majority of events in their sample, indicating asymmetries in the CSM. They study \il ~at 3 epochs (day 31, 63, 119 from the explosion date estimated in this work). There is a change in continuum polarization going from day 31 to 63, which coincides with the transition of the continuum photosphere from the pre-shock CSM to CDS/ejecta. The continuum polarization of \il ~is one of the lowest (along with SN~2014ab) in the sample. Additionally, there is no significant line depolarization noticed in \il. This polarization doesn't necessarily eliminate the possibility of asymmetric CSM; instead, they can be explained in a complimentary manner. A disk-like CSM viewed from the polar region will have a face-on symmetry for the continuum photosphere and, therefore, low-polarization \citep{2014ab_bilinksi}.

\subsection{Mass-loss Rate}
\label{mass-lossrate}
The intermediate-width component seen throughout the evolution of ASASSN-14il indicates the persistent ejecta-CSM interaction, as also observed in the case of SN~2012ab \citep{2012ab_gangopadhyay}. Assuming that the luminosity of the ejecta-CSM interaction is fed by energy at the shock front, the progenitor mass loss rate $\dot{M}$ can be calculated using the relation of \cite{chugai_danziger_1994}: 

\begin{equation}
\dot{M}=\frac{2L}{\epsilon}\frac{v_w}{v_{SN}^{3}}
\end{equation}

\noindent where $\epsilon$ ($<1$) is the efficiency of conversion of the shock's kinetic energy into optical radiation (an uncertain quantity), $v_w$ is the velocity of the pre-explosion stellar wind, $v_{\mathrm{SN}}$ is the velocity of the post-shock shell, and $L$ is the bolometric luminosity of \il. 

The shock velocity is inferred from the intermediate width component after the emission lines are no longer dominated by electron scattering (day $>=$ 50). We take the shock velocity to be 1750 km s$^{-1}$, similar to the value used by \cite{14il_dickinson_thesis}. Using the bolometric luminosity at day $\sim 50$ (L=4.7 x 10$^{43}$ erg sec$^{-1}$) and assuming 50$\%$ conversion efficiency with $\epsilon$=0.5, the mass-loss rate in terms of wind velocity will be:
\begin{equation}
      \dot{M} = 5.6 \times \frac{v_w}{100\mathrm{~km~s}^{-1}} \times \mathrm{M}_{\odot} ~\mathrm{yr}^{-1}
\end{equation}

If we assume a typical unshocked wind velocity observed for LBV winds $v_{w}$$\sim$100 km sec$^{-1}$ \citep{smith_IIn_review}, the estimated mass-loss rate for \il ~is 5.6 M$_{\odot}$ yr$^{-1}$. However, we find that the estimated mass-loss rate is 1.0~M$_{\odot}$ yr$^{-1}$ for only the integrated observed luminosity with no bolometric corrections. \cite{14il_dickinson_thesis} found the mass-loss rate to be 1 M$_{\odot}$ yr$^{-1}$ following similar analysis based on the V-band luminosity. These values are also comparable to the mass-loss value for the \texttt{csm\_s2} model derived from lightcurve fitting.

These values are comparable, albeit slightly higher, than the estimates for SN~2015da \citep{2015dalate_nathansmith_2023} at similar epochs. This can be attributed to higher bolometric luminosity and slower shock speed in \il. The estimated value of mass-loss rate is much higher than the typical LBV winds \citep{hillier_2001_eta_carinae, LBV_winds_Vink_deKoter, groh_hillier_lbv_2009, Smith_mass_loss_2014} and higher than the values often attributed to some SNe~IIn, which are of the order of 0.1 M$_{\odot}$ yr$^{-1}$ as observed in some great eruptions of LBVs \citep{Chugai_1994w_2004, kiewe_iin_sample_2012}. These values are also much higher than those of normal-luminosity SNe~IIn like SN~2005ip ($2-4\times10^{-4}$ M$_{\odot}$ yr$^{-1}$; \citealp{2005ip_smith}). It is also much larger than the typical values of RSG and yellow hypergiants ($10^{-4}-10^{-3}$ M$_{\odot}$ yr$^{-1}$; \citealp{mass_loss_rsg_de_jager, ysg_de_jager, mass_loss_rsg_van_loon, Smith_mass_loss_2014}), and quiescent winds of LBV (10$^{-5}-10^{-4}$ M$_{\odot}$ yr$^{-1}$, \citealp{Vink2018}). The obtained mass-loss rate in \il ~indicates the progenitor underwent an LBV giant eruptive prior to the SN event. The CSM may be the result of interaction with a binary companion, which in would explain the expected asymmetry in the geometry.

\subsection{A case for Dust Formation} \label{sec:dust_formation}
The H$\alpha$ profile of \il ~shows a deficit in the flux of the redside wing. This deficit is clearly visible as an overall blueshift of the line after day 100 and increases with time. At day 327, even a combination of blueshifted Gaussian components can not appropriately reproduce the H$\alpha$ line profile. At this epoch, the line profile can be best described as missing flux from the redside wing and near rest velocity. The flux deficit phenomenon is wavelength dependent as well, affecting the shorter wavelength H$\beta$ more severely compared to H$\alpha$.

An overall blueshift could be caused by many reasons, such as radiative acceleration, lopsided SN explosion or asymmetric CSM, obscuration of the receding material by the continuum photosphere, and dust formation. However, except for dust formation, none of the other mechanisms can explain the observed time dependence and wavelength dependence.

In the case of a blueshift caused by the radiative acceleration of the CSM, the blueshift should decrease as the luminosity drops, and we expect no significant wavelength dependence for this case.  Additionally, in the case of a blueshift caused by the radiative acceleration scenario, the original narrow line photon source should be blueshifted as well, which is not the case for \il ~\citep{Dessart_IIn_simulation}.
Similarly, in the case of obscuration by the continuum photosphere, the expected blueshift will be strongest in early times and decrease later on as the continuum optical depth drops. 
For a lopsided explosion/CSM any blueshift present should be present from early times as well, remaining consistent throughout the evolution. Thus none of these scenarios are consistent with the observed evolution of \il.

However, all these requirements are readily explained by a scenario where new dust grains form in the ejecta or the post-shock CSM. The dust formation will increase at late times as the ejecta and the post-shock CSM cool down, explaining the time evolution of the observed deficit of flux in the side wing of H$\alpha$. Also, the wavelength dependence of the flux deficit is a natural consequence of the extinction from dust affecting shorter wavelengths more prominently. Dust formation has other signatures, such as an increase in the NIR flux and an increased rate of fading the optical flux. However, due to data gaps between days 120--250 and the lack of late-time NIR data, we are unable to comment on it. Dust formation is very common in SLSNe~IIn as seen in the case of SNe 2006tf \citep{2006tf_smith}, 2010jl \cite{2010jl_smith_dust, 2010jl_maeda, gall2014}, 2015da \cite{2015dalate_nathansmith_2023}, ASASSN-15ua \cite{15ua_dickinson},  and 2017hcc \cite{2017hcc_smith_andrews}, so expecting dust formation in \il ~would not be unreasonable.

\subsection{Physical Scenario} \label{sec:physical_scenario}
Figure~\ref{fig:cartoonpic} describes a physical scenario that, although not unique, explains the observables of \il ~through various stages of its evolution. In this scenario, the ejecta from the SN explosion interacts with a disk-like CSM. The viewing angle of the observer is such that the CSM disk is face-on to the observer. This configuration is surrounded by distant CSM that may result from steady mass loss from the progenitor.

\autoref{fig:cartoonpic}a describes a scenario where the photosphere lies in the pre-shock CSM surrounding the interaction region. The line profiles are dominated by narrow lines and Lorentzian wings resulting from electron scattering in the ionized CSM. The signatures from the SN ejecta are masked by the interaction region. This figure explains the observed line profile in \il ~at early phases ($<$ day 50).

Between day 50--80, the photosphere recedes into the interaction region. Due to the asymmetric CSM geometry, both uninterrupted SN ejecta and post-shock region are visible simultaneously (\autoref{fig:cartoonpic}b). The line profiles see contributions from the SN ejecta, shock region, and distant uninterrupted CSM, which manifests as broad, intermediate, and narrow-width Gaussian components, respectively.

\autoref{fig:cartoonpic}c depicts the evolution of \il ~post day 80. 
The contributions from the distant CSM, post-shock region, and the ejecta are still seen. Dust formation is seen in the post-shock gas, an efficient location for dust formation in SNe~IIn. This dust preferentially obscures the receding SN ejecta and shock, causing a deficit in the redside wing of line profiles. This effect manifests as an overall blueshift of the broad component. Such blueshift is observed in the late time ($>$ day 100) H$\alpha$ profile of \il ~as well as many other SNe~IIn (e.g. SNe~2005ip, 20006tf, 2010jl, ASASSN-15ua, 2015da).

Similar explosion/CSM geometry have been proposed for SNe~2010jl, 2014ab, 2015da, and 2017hcc \citep{2010jl_smith, 2014ab_bilinksi, 2015dalate_nathansmith_2023, 2017hcc_smith_andrews} with different viewing angles. However, the viewing angle in SN~2014ab and \il ~are proposed to be similar, which is further emphasized by both having remarkably low polarization \citep{IIn_specpol_bilinski}.


\section{Summary} \label{sec:summary}
\begin{enumerate}
    \item \il ~was observed with an extensive follow-up campaign spanning $\sim 480$ days in photometric and $\sim 327$ days in spectroscopy observations.
    \item \il ~is a very luminous SN~IIn showing long-term interaction signatures and a plateau in the optical light curves around the maximum. It peaks at $\sim - 20.3$ mag in r-band ($< -21$ mag in UV bands), comparable to SLSNe~2006tf, 2010jl, ASASSN-15ua. The lightcurve shape is very similar to SN~2015da, but the late time (after $\sim 250$ days) decline rate of $\sim$0.01 mag/day in optical bands is faster than SN~2015da but comparable to SNe~2006tf, 2010jl, ASASSN-15ua. The slow reward evolution of the intrinsic B-V resembles the typical SNe~IIn.
    \item The multi-band lightcurve modeling of \il ~indicates a CSM-driven explosion. Both wind/shell CSM model generates our observed lightcurves with a CSM mass between 4.7--9.1 M$_{\odot}$ and the CSM shell/wind being expelled about 1-5 years before the explosion. 
    \item Spectroscopically, \il ~shows long-term predominant interaction signatures with narrow H and He lines on top of otherwise featureless spectra similar to other SLSNe~IIn 2006gy, 2010jl, and 2015da.
    \item The H$\alpha$ profile of \il ~can be described by the following components- (i) An unresolved component from the pre-shock ionized CSM that stays consistent throughout the evolution, (ii) An intermediate width component initially dominated by Lorentzian electron-scattering wings but later (day $>50$) show Gaussian profile (FWHM $\sim$ 1750~km s$^{-1}$) representative of the post-shock region.
    (iii) A broad Gaussian emission component (FWHM $\sim$ 7000~km s$^{-1}$) representative of the uninterrupted ejecta, which becomes visible (day $>50$).
    \item The emergence of the ejecta component when the lightcurves are near peak luminosity indicates asymmetry in the CSM structure.
    \item The blueshift of H$\alpha$ profile at late phases (day $\geq 81$) indicates dust formation in the post-shock CSM/ejecta.
    \item The mass-loss rate of upto 2-7 M$_{\odot} \mathrm{yr}^{-1}$ indicates that the progenitor star underwent an LBV giant eruption prior to the SN explosion.
\end{enumerate}

\section*{Acknowledgements}
We thank the anonymous referee for providing us with valuable suggestions and scientific insights that enhanced the quality of the paper.
This work uses data from the Las Cumbres Observatory Global Telescope network. The LCO group is supported by NSF grant AST-1911151. This work uses {\it Swift}/UVOT data reduced by P. J. Brown and released in the {\it Swift} Optical/Ultraviolet Supernova Archive (SOUSA). SOUSA is supported by NASA's Astrophysics Data Analysis Program through grant NNX13AF35G. This research has made use of the APASS database, located on the AAVSO website. Funding for APASS has been provided by the Robert Martin Ayers Sciences Fund. This work is based (in part) on observations collected at the European Organisation for Astronomical Research in the Southern Hemisphere, Chile as part of PESSTO, (the Public ESO Spectroscopic Survey for Transient Objects Survey) ESO program 188.D-3003, 191.D-0935, 197.D-1075. ND, KM and BA acknowledge the support from BRICS grant DST/ICD/BRICS/Call-5/CoNMuTraMO/2023 (G) funded by the Department of Science and Technology (DST), India. BA acknowledges the Council of Scientific $\&$ Industrial Research (CSIR) fellowship award (09/948(0005)/2020-EMR-I) for this work.



\bibliography{references}
\bibliographystyle{aasjournal}

\appendix
\counterwithin{figure}{section}
\counterwithin{table}{section}
\section{Observation log}
\input{14il_photometry_apj.tex}

\begin{table*}[h]
    \centering
    \caption{Log of {\it Swift} UV and optical photometric observations of ASASSN-14il.}
    \label{tab:14il_swift_photometry}
    \input{14il_swift_photometry}
\end{table*}

\begin{table*}[h]
    \centering
    \caption{Log of spectroscopic observations of ASASSN-14il.}
    \label{tab:spectra_log}
    \input{spectra_log}
\end{table*}

\newpage
\section{MOSFiT Corner Plot}

\begin{figure*}
    \centering
    \includegraphics[width=\textwidth]{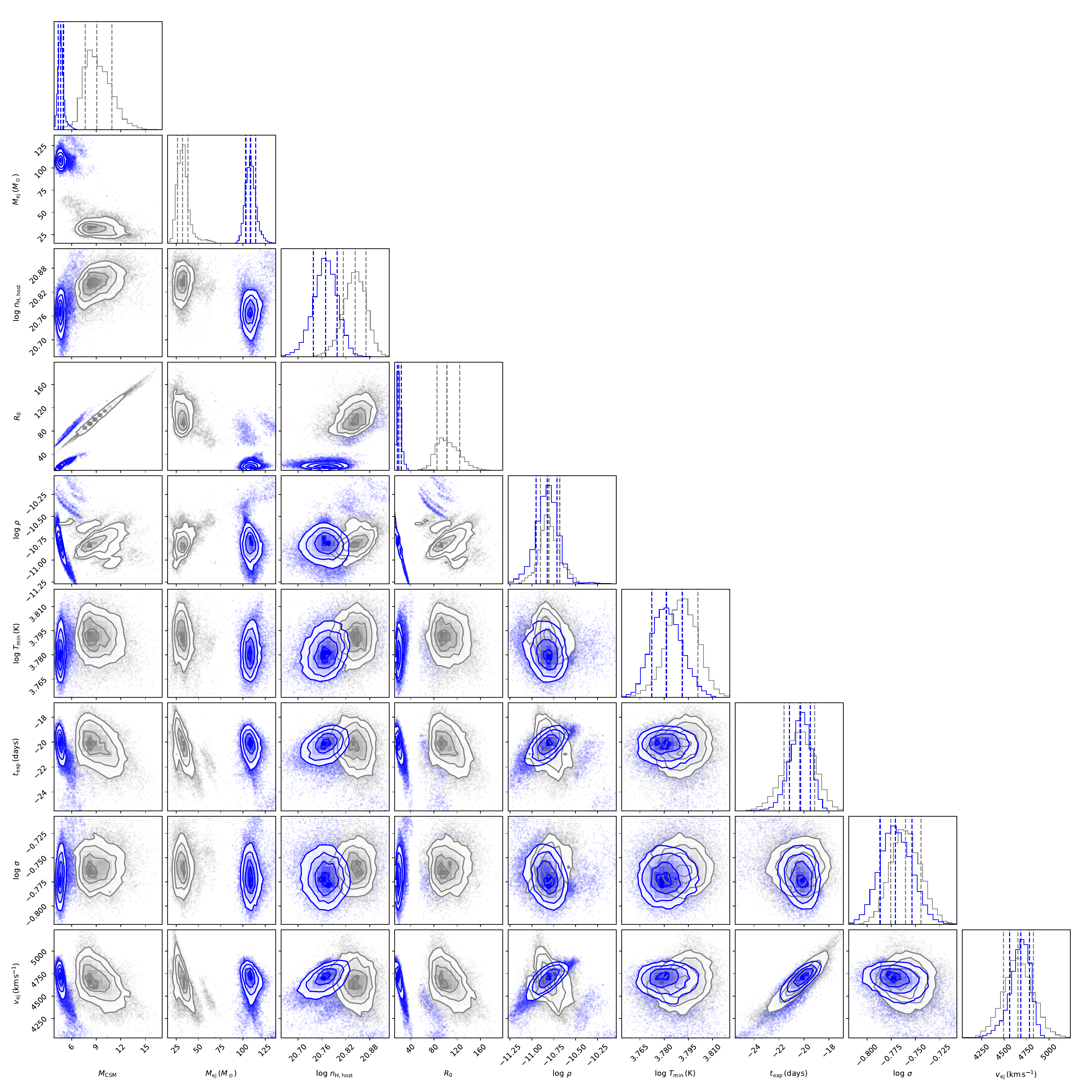}
    \caption{Corner plot showing the best fit parameters of the \csm ~wind (blue) and \csm ~shell (gray) models for ASASSN-14il.}
    \label{fig:mosfit_corner}
\end{figure*}

\label{lastpage}
\end{document}

%% file: sn_and_host.tex
\begin{tabular}{c|c}
\hline \hline
\multicolumn{2}{c}{\textbf{ASASSN-14il}} \\ \hline
Discovery Date$^{(1)}$ & UT 2014-10-01.11 \\
Explosion Date & UT 2014-09-25.9 \\
SN Type$^{(2)}$ & IIn \\
RA (J2000) & 00h45m32.55s \\
Dec (J2000) & -14d15m34.6s \\
Discovery Magnitude & 16.5 mag (V-band) \\
E(B-V)$_{\mathrm{total}}$ & 0.21 ± 0.08 mag \\
\hline \multicolumn{2}{c}{\textbf{Host Galaxy}$^{(3)}$} \\ \hline
Galaxy Names & PGC 3093694\\
& WISEA J004532.52-141532.7 \\
& 6dF J0045326-141533 \\
Morphology Type & Dwarf Galaxy \\
RA (J2000) & 00h45m32.601s \\
Dec (J2000) & -14d15m32.50s \\
redshift (z) & 0.022 \\
v$_{\mathrm{Virgo+Shapley+GA}}$ & 6462 ± 47 km/s \\
D & $88.52 \pm 6.2$ Mpc \\
$\mu$ & 34.74 ± 0.15 mag \\ \hline
\end{tabular}
\newline
References - (1) \cite{14il_discovery}, (2) \cite{14il_classification}, \,\,\,\,\,\,\,\,
(3) \href{https://ned.ipac.caltech.edu/byname?objname=WISEA+J004532.52-141532.7&hconst=73&omegam=0.27&omegav=0.73&wmap=1&corr_z=4}{NED}

%% file: peak_table.tex
\begin{tabular}{cccc}
\hline \hline 
 & Rise Time & \multicolumn{2}{c}{Peak Magnitude}  \\
 &  & Observed & Absolute \\
 & (day) & (mag) & (mag) \\
\hline
UVW2 & $14.7 \pm 0.9$ & $14.54 \pm 0.03$ &  $-22.0 \pm 0.7$\\
UVM2 & $15.2 \pm 0.9$ & $14.43 \pm 0.03$ &  $-22.2 \pm 0.7$\\
UVW1 & $17.8 \pm 0.8$ & $14.30 \pm 0.02$ &  $-21.8 \pm 0.5$\\
U    & $21.6 \pm 1.2$ & $14.32 \pm 0.02$ &  $-21.4 \pm 0.4$\\
B    & $36.3 \pm 0.5$ & $15.57 \pm 0.01$ &  $-20.0 \pm 0.3$\\
g    & $38.3 \pm 0.3$ & $15.18 \pm 0.01$ &  $-20.3 \pm 0.3$\\

\hline
\end{tabular}

%% file: slope_table.tex
\begin{tabular}{cccc}
\hline \hline 
 & \multicolumn{2}{c}{Time Interval} \\
 & 40-90d & 250-500d \\
 & (mag d$^{-1}$) & (mag d$^{-1}$) \\
\hline
B & 0.009$\pm$0.002 & 0.011$\pm$0.001 \\
g & 0.004$\pm$0.001 & 0.010$\pm$0.001 \\
V & 0.002$\pm$0.001 & 0.012$\pm$0.000 \\
r & -0.001$\pm$0.001 & 0.010$\pm$0.001 \\
i & -0.004$\pm$0.001 & 0.009$\pm$0.001 \\
\hline
\end{tabular}

%% file: comparison_sample.tex
\setlength{\tabcolsep}{3pt}
\begin{tabular}{l c c c c c c c}
\hline \hline
SN       & Host Galaxy     & Distance    & Extinction & Peak magnitude (r/R)  & Explosion & Reference$^\dagger$  \\
         &                  & (Mpc)                    &$E(B-V)$ (mag) & (mag)    &   
(mjd)    &                       \\
\hline
SN 2005ip & NGC 290 & 33.7 & 0.047 & -17.7 & 53679.16$^{\ddagger}$ & 
1 \\
SN 2006gy & NGC 1260  & 77.7 & 0.56 & -21.6 & 53967.0 & 2, 3 \\
SN 2006tf & SDSS J124615.80+112555.5 & 330 & 0.027 & -20.7 & 54081.0$^{\ddagger}$ & 4 \\
SN 2009kn & MCG -3-21-6              & 68.9 & 0.114 & -18.0 & 55115.5 & 5 \\
SN 2010jl & UGC 5189A                & 49.2 & 0.058 & -20.2 & 55478.6$^{\#}$ & 6, 7    \\
SN 2011ht      &  UGC 5460        & 20.5    & 0.061 & -17.6 & 55833.0$^{\ddagger}$ & 8 \\
PTF11oxu & WISEA J033834.32+223242.7 & 376 & 0.176 & -18.9 & 55837.3 & 9, NED\\
PTF11rfr & WISEA J014216.97+291625.6 & 286 & 0.042 & -19.5 & 55895.4 & 9, NED\\
SN 2012ab  & SDSS J122247.61+053624.2 &  83.6  & 0.079 & -19.4 &  55955.3 & 10 \\             
SN 2015da & NGC 5337 & 37.2 & 0.98 & -19.6 & 57030.4 & 11, 12\\
ASASSN-15ua & GALEXASC J133454.49+105906.7 & 244.5* & 0.0578 & -20.4 & 57368$^{\ddagger}$ & 13\\
SN 2016aps & - & 1302* & 0.0263 & -22.3 & 57440.0$^{\ddagger}$ & 14\\
ASASSN-14il & WISEA J004532.52-141532.7 & 88.5 & 0.24 & -20.3 & 56925.9 & This work\\

\hline                                                                              
\end{tabular}
\newline
$^\dagger$ References:
(1) \cite{2005ip_2006jd_stritzinger}
(2) \cite{2006gy_smith_mccray_shellshockeddiffusion}; 
(3) \cite{2006gy_agnoletto}; 
(4) \cite{2006tf_smith};
(5) \cite{2009kn_kankare};  
(6) \cite{2010jl_jencson}; 
(7) \cite{2010jl_zhang}; 
(8) \cite{2011ht_mauerhan}; 
(9) \cite{nyholm_IIn_survey_2020};
(10) \cite{2012ab_gangopadhyay}; 
(11) \cite{2015da_tartaglia};
(12) \cite{2015dalate_nathansmith_2023};
(13) \cite{15ua_dickinson};
(14) \cite{2016aps_nicholl}\\
$*$ Distances are not corrected for the effects of GA, Virgo and Shapley.\\
$^{\#}$ The adopted explosion date is that of the first detection.\\
$^{\ddagger}$ The adopted explosion date is the discovery date.

%% file: mosfit_param.tex
\begin{table*}
    \caption{Fitting parameters for MOSFiT \csm ~and \csmni ~models. The shell and wind type CSM are represented by `s0' and `s2' suffixes, respectively.}
    \label{tab:mosfit_param}
    \centering
    \begin{tabular}{cccccccccc}
    \hline\hline
        Parameter & M$_{\mathrm{ej}}$ & v$_{\mathrm{ej}}$ & f$_{\mathrm{Ni}}$ & M$_{\mathrm{Ni}}$ & M$_{\mathrm{csm}}$ & R$_{0}$ & log $\rho_{0}$ & $\dot{M}$ & t$_{\mathrm{csm}}$\\ 
        Unit & (M$_{\odot}$) & (km s$^{-1}$) & (\%) & (M$_{\odot}$) & (M$_{\odot}$) & (AU) & (g cm$^{-3}$) & (M$_{\odot}$ yr$^{-1}$) & (yr) \\ 
        Prior & 1--150 & $10^3$ -- $10^5$ & 0--100 & - & 0--150 & 1--500 & -14 -- -9 & - & - \\ 
        \hline
        \hline \\
        csm\_s0 & $32.2^{+6.1}_{-5.7}$ & $4654.4^{+171}_{-162}$ & - & - & $9.1^{+1.9}_{-1.4}$ & $102.4^{+22.3}_{-17.2}$ & $-10.8^{+0.1}_{-0.1}$ & - & 4.8 \\
        \\
        csm\_s2 & $108.4^{+5.8}_{-5.3}$ & $4685^{+97}_{-124}$ & - & - & $4.7^{+0.3}_{-0.3}$ & $19.8^{+4.2}_{-2.9}$ & $-10.8^{+0.1}_{-0.1}$ & 2.8 & 0.9 \\
        \\
        low\_Ni+csm\_s0 & $20.9^{+16.7}_{-12.8}$ & $4670^{+210}_{-200}$ & $0.4^{+0.3}_{-0.2}$ & $0.08^{+0.18}_{-0.07}$ & $15.87^{+12.53}_{-5.73}$ & $187.11^{+145.4}_{-68.2}$ & $-10.43^{+0.7}_{-0.3}$ & - & 8.9 \\
        \\
        low\_Ni+csm\_s2 & $14.6^{+2.7}_{-2.6}$ & $4897^{+172}_{-111}$ & $8^{+0.5}_{-0.5}$ & $1.2^{+0.3}_{-0.3}$ & $5.4^{+0.8}_{-1.0}$ & $80.2^{+13}_{-16}$ & $-10.36^{+0.11}_{-0.16}$ & 125.3 & 3.8\\
        \\
        \hline \\
        high\_Ni+csm\_s0 & $6.93^{+0.6}_{-0.5}$ & $5370^{+253}_{-241}$ & $66^{+5}_{-6}$ & $4.6^{+0.8}_{-0.7}$ & $1.9^{+0.7}_{-0.5}$ & $87.76^{+22.1}_{-14.5}$ & $-11.87^{+0.03}_{-0.02}$ & - & 4.2\\
        \\
        high\_Ni+csm\_s2 & $7.6^{+0.6}_{-0.4}$ & $6025^{+284}_{-271}$ & $65^{+4}_{-5}$ & $4.9^{+0.7}_{-0.6}$ & $3.2^{+0.4}_{-0.3}$ & $126^{+16}_{-15}$ & $-12.1^{+0.1}_{-0.!}$ & 5.63 & 5.97\\
        \\
    \hline
    \end{tabular}
\end{table*}

%% file: 14il_spectral_components.tex
\begin{center}

\begin{tabular}{| c | c |  c | l l | l l | l l |}
\hline \hline
Phase & Source & Resolution & \multicolumn{2}{c|}{Narrow Component} & \multicolumn{2}{c|}{Intermediate-width Component} & \multicolumn{2}{c|}{Broad Component} \\ 
 & & & center & fwhm & 
center & fwhm & 
center & fwhm\\ 
(d) & & (km s$^{-1}$) & (km s$^{-1}$) & (km s$^{-1}$) & (km s$^{-1}$) & (km s$^{-1}$) & (km s$^{-1}$) & (km s$^{-1}$)\\ 
\hline
7.7 & ANU & 100 & \,\,130$\pm$0.3 & 190$\pm$0.9 &  100\,\,$\pm$9.6 & 2040$\pm$28 & \,\,\,\,\,\,\,\,- & \,\,\,\,\,\,\,\,- \\
7.8 & LCO & 500 & -100$\pm$5.3 & 500$\pm$18 & -100$\pm$10 & 1900$\pm$56 & \,\,\,\,\,\,\,\,- & \,\,\,\,\,\,\,\,- \\
9.7 & LCO & 500 & \,\,\,\,\,\,\,\,0$\pm$4.7 & 700$\pm$17 & \,\,\,\,\,\,\,\,0$\pm$10 & 2200$\pm$61 & \,\,\,\,\,\,\,\,- & \,\,\,\,\,\,\,\,- \\
11.8 & LCO & 500 & \,\,\,\,\,\,\,\,0$\pm$8.4 & 600$\pm$29 & \,\,100$\pm$24 & 2400$\pm$130 & \,\,\,\,\,\,\,\,- & \,\,\,\,\,\,\,\,- \\
13.6 & ANU & 100 & \,\,150$\pm$0.3 & 190$\pm$0.8 & \,\,140$\pm$8.0 & 2110$\pm$23 & \,\,\,\,\,\,\,\,- & \,\,\,\,\,\,\,\,- \\
14.6 & LCO & 500 & \,\,\,\,\,\,\,\,0$\pm$6.1 & 700$\pm$22 & \,\,100$\pm$14 & 2200$\pm$90 & \,\,\,\,\,\,\,\,- & \,\,\,\,\,\,\,\,- \\
17.7 & LCO & 500 & \,\,100$\pm$18 & 600$\pm$59 & \,\,\,\,\,\,\,\,0$\pm$27 & 1700$\pm$150 & \,\,\,\,\,\,\,\,- & \,\,\,\,\,\,\,\,- \\
22.7 & ANU & 100 & \,\,140$\pm$0.3 & 200$\pm$0.9 & \,\,130$\pm$8.3 & 1950$\pm$24 & \,\,\,\,\,\,\,\,- & \,\,\,\,\,\,\,\,- \\
22.8 & LCO & 500 & \,\,\,\,\,\,\,\,0$\pm$4.1 & 800$\pm$12 & \,\,200$\pm$27 & 3800$\pm$80 & \,\,\,\,\,\,\,\,- & \,\,\,\,\,\,\,\,- \\
26.1 & PESSTO & 500 & \,\,200$\pm$8.5 & 500$\pm$29 & \,\,100$\pm$11 & 1800$\pm$52 & \,\,\,\,\,\,\,\,- & \,\,\,\,\,\,\,\,- \\
26.6 & LCO & 500 & \,\,\,\,\,\,\,\,0$\pm$5.0 & 700$\pm$18 & \,\,100$\pm$13 & 2100$\pm$82 & \,\,\,\,\,\,\,\,- & \,\,\,\,\,\,\,\,- \\
31.6 & ANU & 100 & \,\,140$\pm$0.3 & 180$\pm$0.9 & \,\,130$\pm$8.2 & 1700$\pm$25 & \,\,\,\,\,\,\,\,- & \,\,\,\,\,\,\,\,- \\
49.2 & PESSTO & 500 & \,\,200$\pm$4.8 & 500$\pm$17 & \,\,100$\pm$13 & 1500$\pm$57 & \,\,\,\,\,300$\pm$70 & 5000$\pm$240 \\
81.6 & LCO & 500 & \,\,200$\pm$2.5 & 500$\pm$9.4 & \,\,200$\pm$9.4 & 1800$\pm$47 & -200$\pm$280 & 11200$\pm$400 \\
109.2 & PESSTO & 500 & \,\,100$\pm$7.6 & 700$\pm$29 & -100$\pm$29 & 2100$\pm$130 & \,\,\,-600$\pm$57 & 8200$\pm$190 \\
123.1 & PESSTO & 500 & \,\,\,\,\,\,\,\,- & - & \,\,\,\,\,\,\,\,0$\pm$28 & 1500$\pm$83 & \,\,\,-500$\pm$52 & 7600$\pm$160 \\
327.4 & PESSTO & 500 & \,\,\,\,\,\,\,\,0$\pm$9.7 & 500$\pm$24 & \,\,\,\,\,\,\,\,- & \,\,\,\,\,\,\,\,- & -1000$\pm$69 & 6500$\pm$160 \\
\hline
\end{tabular}
\end{center}     


%% file: 14il_photometry_apj.tex
\startlongtable
\begin{deluxetable*}{cccccccc}
\tabletypesize{\footnotesize}
\tablecaption{Log of photomteric observations of ASASSN-14il obtained from LCOGT network of telescopes. \label{tab:photometry_log}}
\tablehead{
\colhead{Date} & \colhead{Phase} & \colhead{B} & \colhead{V} & \colhead{g} & \colhead{r} & \colhead{i} & \colhead{telescope}\\
\colhead{(yyyy-mm-dd)} & \colhead{(d)} & \colhead{(mag)} & \colhead{(mag)} & \colhead{(mag)} & \colhead{(mag)} & \colhead{(mag)} & \colhead{}
}
\startdata
2014-10-03 & 7.8 & $16.15 \pm 0.03$ & $16.25 \pm 0.04$ & $15.97 \pm 0.02$ & $16.10 \pm 0.04$ & $16.14 \pm 0.01$ & 1m0-12 \\
2014-10-05 & 9.5 & $16.15 \pm 0.02$ & $16.07 \pm 0.01$ & $15.87 \pm 0.04$ & $15.91 \pm 0.02$ & $16.07 \pm 0.01$ & 1m0-11 \\
2014-10-07 & 11.3 & $16.10 \pm 0.08$ & $15.71 \pm 0.06$ & $15.75 \pm 0.05$ & $15.84 \pm 0.11$ & - & 1m0-08 \\
2014-10-09 & 13.3 & - & $15.58 \pm 0.16$ & $15.57 \pm 0.05$ & $15.54 \pm 0.15$ & $15.70 \pm 0.20$ & 1m0-08 \\
2014-10-09 & 13.7 & $15.87 \pm 0.02$ & $15.74 \pm 0.03$ & $15.50 \pm 0.02$ & $15.52 \pm 0.04$ & $15.65 \pm 0.05$ & 1m0-11 \\
2014-10-11 & 15.5 & $15.84 \pm 0.06$ & $15.63 \pm 0.01$ & $15.42 \pm 0.01$ & $15.52 \pm 0.03$ & $15.68 \pm 0.05$ & 1m0-03 \\
2014-10-13 & 17.9 & $15.76 \pm 0.04$ & $15.55 \pm 0.01$ & $15.44 \pm 0.01$ & $15.45 \pm 0.02$ & $15.53 \pm 0.01$ & 1m0-10 \\
2014-10-15 & 19.3 & $15.74 \pm 0.05$ & $15.57 \pm 0.03$ & $15.37 \pm 0.02$ & $15.42 \pm 0.02$ & $15.56 \pm 0.02$ & 1m0-08 \\
2014-10-17 & 22.0 & $15.74 \pm 0.03$ & $15.44 \pm 0.02$ & $15.39 \pm 0.01$ & $15.37 \pm 0.04$ & $15.44 \pm 0.04$ & 1m0-13 \\
2014-10-18 & 22.2 & $15.63 \pm 0.01$ & $15.38 \pm 0.03$ & $15.23 \pm 0.01$ & $15.33 \pm 0.01$ & $15.39 \pm 0.05$ & 1m0-08 \\
2014-10-19 & 23.9 & $15.68 \pm 0.02$ & - & - & - & - & 1m0-12 \\
2014-10-20 & 24.5 & $15.54 \pm 0.06$ & $15.40 \pm 0.04$ & $15.27 \pm 0.02$ & $15.22 \pm 0.02$ & $15.28 \pm 0.03$ & 1m0-03 \\
2014-10-20 & 24.4 & $15.54 \pm 0.02$ & $15.30 \pm 0.03$ & $15.21 \pm 0.01$ & $15.21 \pm 0.01$ & $15.27 \pm 0.02$ & 1m0-11 \\
2014-10-21 & 25.9 & $15.70 \pm 0.02$ & $15.37 \pm 0.02$ & $15.30 \pm 0.01$ & $15.25 \pm 0.03$ & $15.31 \pm 0.02$ & 1m0-10 \\
2014-10-22 & 26.5 & $15.59 \pm 0.02$ & $15.30 \pm 0.04$ & $15.18 \pm 0.03$ & - & $15.31 \pm 0.10$ & 1m0-03 \\
2014-10-26 & 30.5 & $15.65 \pm 0.02$ & $15.32 \pm 0.07$ & $15.22 \pm 0.02$ & $15.13 \pm 0.02$ & $15.17 \pm 0.05$ & 1m0-11 \\
2014-10-29 & 34.0 & - & $15.33 \pm 0.02$ & $15.24 \pm 0.01$ & $15.14 \pm 0.03$ & $15.19 \pm 0.05$ & 1m0-12 \\
2014-11-02 & 37.7 & - & $15.32 \pm 0.09$ & - & - & - & 1m0-03 \\
2014-11-06 & 41.4 & - & $15.15 \pm 0.03$ & $15.23 \pm 0.02$ & $15.04 \pm 0.03$ & - & 1m0-11 \\
2014-11-10 & 45.2 & - & $15.20 \pm 0.04$ & $15.32 \pm 0.04$ & $15.10 \pm 0.04$ & $15.09 \pm 0.03$ & 1m0-08 \\
2014-11-13 & 49.0 & $15.73 \pm 0.03$ & $15.22 \pm 0.02$ & $15.23 \pm 0.03$ & $15.06 \pm 0.03$ & $15.06 \pm 0.02$ & 1m0-12 \\
2014-11-17 & 52.8 & $15.62 \pm 0.02$ & $15.26 \pm 0.02$ & $15.34 \pm 0.01$ & $15.11 \pm 0.04$ & $15.02 \pm 0.03$ & 1m0-12 \\
2014-11-21 & 56.6 & - & $15.26 \pm 0.04$ & $15.27 \pm 0.02$ & $15.04 \pm 0.03$ & $14.98 \pm 0.03$ & 1m0-03 \\
2014-11-26 & 61.9 & $15.78 \pm 0.02$ & - & $15.23 \pm 0.02$ & $15.04 \pm 0.02$ & $14.97 \pm 0.02$ & 1m0-10 \\
2014-11-26 & 62.0 & - & - & $15.31 \pm 0.03$ & $15.05 \pm 0.01$ & $14.97 \pm 0.04$ & 1m0-13 \\
2014-11-27 & 62.2 & $15.76 \pm 0.03$ & $15.29 \pm 0.04$ & $15.31 \pm 0.01$ & $15.08 \pm 0.02$ & $15.03 \pm 0.02$ & 1m0-05 \\
2014-11-27 & 62.0 & $15.67 \pm 0.01$ & $15.18 \pm 0.03$ & $15.26 \pm 0.01$ & $15.04 \pm 0.04$ & $15.00 \pm 0.04$ & 1m0-08 \\
2014-11-28 & 63.8 & $15.89 \pm 0.04$ & $15.30 \pm 0.02$ & $15.39 \pm 0.01$ & $15.08 \pm 0.03$ & $14.98 \pm 0.05$ & 1m0-12 \\
2014-12-02 & 67.4 & - & - & - & $15.09 \pm 0.05$ & $15.10 \pm 0.04$ & 1m0-03 \\
2014-12-04 & 69.8 & $15.92 \pm 0.02$ & $15.27 \pm 0.03$ & $15.43 \pm 0.02$ & $15.06 \pm 0.02$ & $15.03 \pm 0.04$ & 1m0-10 \\
2014-12-04 & 69.9 & $15.81 \pm 0.01$ & $15.27 \pm 0.02$ & $15.38 \pm 0.01$ & $15.07 \pm 0.01$ & $14.92 \pm 0.04$ & 1m0-12 \\
2014-12-09 & 74.2 & $15.84 \pm 0.02$ & $15.23 \pm 0.03$ & $15.32 \pm 0.02$ & $14.99 \pm 0.03$ & $15.02 \pm 0.01$ & 1m0-05 \\
2014-12-12 & 77.8 & $15.93 \pm 0.03$ & $15.17 \pm 0.03$ & $15.48 \pm 0.02$ & - & - & 1m0-12 \\
2014-12-12 & 77.9 & $15.88 \pm 0.03$ & $15.15 \pm 0.06$ & $15.48 \pm 0.05$ & $15.05 \pm 0.03$ & $14.92 \pm 0.01$ & 1m0-13 \\
2014-12-16 & 81.8 & $15.97 \pm 0.03$ & $15.28 \pm 0.02$ & $15.34 \pm 0.03$ & $15.05 \pm 0.02$ & $14.89 \pm 0.03$ & 1m0-12 \\
2014-12-20 & 85.8 & $15.91 \pm 0.04$ & $15.27 \pm 0.01$ & $15.41 \pm 0.02$ & $15.00 \pm 0.05$ & $14.87 \pm 0.03$ & 1m0-12 \\
2014-12-24 & 89.5 & $15.72 \pm 0.02$ & $15.10 \pm 0.02$ & $15.22 \pm 0.02$ & $14.89 \pm 0.02$ & $14.83 \pm 0.04$ & 1m0-03 \\
2014-12-29 & 94.5 & $15.82 \pm 0.01$ & $15.23 \pm 0.10$ & $15.41 \pm 0.02$ & - & - & 1m0-11 \\
2015-01-02 & 98.5 & - & - & - & $14.83 \pm 0.03$ & $14.78 \pm 0.03$ & 1m0-11 \\
2015-01-06 & 102.4 & $15.87 \pm 0.03$ & $15.27 \pm 0.01$ & $15.32 \pm 0.01$ & $14.93 \pm 0.02$ & $14.79 \pm 0.05$ & 1m0-03 \\
2015-01-11 & 107.8 & $16.07 \pm 0.01$ & $15.31 \pm 0.02$ & $15.54 \pm 0.02$ & $15.04 \pm 0.01$ & $14.89 \pm 0.02$ & 1m0-13 \\
2015-01-15 & 111.8 & $16.19 \pm 0.04$ & $15.33 \pm 0.01$ & $15.52 \pm 0.02$ & - & - & 1m0-13 \\
2015-01-16 & 112.1 & $15.98 \pm 0.02$ & $15.33 \pm 0.03$ & $15.48 \pm 0.01$ & $15.05 \pm 0.02$ & $14.96 \pm 0.02$ & 1m0-05 \\
2015-01-19 & 115.8 & $16.02 \pm 0.02$ & $15.37 \pm 0.04$ & $15.51 \pm 0.02$ & $15.04 \pm 0.02$ & $14.89 \pm 0.05$ & 1m0-10 \\
2015-01-28 & 124.0 & $16.24 \pm 0.05$ & $15.58 \pm 0.04$ & $15.63 \pm 0.02$ & $15.14 \pm 0.01$ & $15.08 \pm 0.02$ & 1m0-05 \\
2015-02-06 & 133.4 & $16.09 \pm 0.08$ & $15.37 \pm 0.01$ & $15.59 \pm 0.01$ & $15.11 \pm 0.02$ & $14.99 \pm 0.06$ & 1m0-03 \\
2015-06-19 & 266.2 & - & $17.09 \pm 0.08$ & $17.33 \pm 0.02$ & $16.53 \pm 0.02$ & $16.44 \pm 0.08$ & 1m0-12 \\
2015-06-24 & 271.4 & - & - & $17.28 \pm 0.02$ & $16.85 \pm 0.09$ & $16.77 \pm 0.02$ & 1m0-05 \\
2015-06-28 & 275.8 & $17.97 \pm 0.03$ & $16.97 \pm 0.03$ & $17.44 \pm 0.03$ & $16.59 \pm 0.04$ & - & 1m0-11 \\
2015-07-02 & 279.2 & $17.71 \pm 0.15$ & $17.22 \pm 0.04$ & $17.45 \pm 0.07$ & $16.71 \pm 0.01$ & $16.92 \pm 0.02$ & 1m0-12 \\
2015-07-02 & 279.2 & $17.79 \pm 0.17$ & $17.26 \pm 0.04$ & $17.24 \pm 0.09$ & $16.61 \pm 0.03$ & $16.76 \pm 0.13$ & 1m0-13 \\
2015-07-06 & 283.0 & $17.49 \pm 0.20$ & - & $17.77 \pm 0.11$ & $16.50 \pm 0.19$ & $16.76 \pm 0.03$ & 1m0-12 \\
2015-07-09 & 286.7 & - & $17.40 \pm 0.08$ & - & $16.53 \pm 0.05$ & $16.76 \pm 0.22$ & 1m0-11 \\
2015-07-14 & 291.4 & $18.12 \pm 0.05$ & $17.42 \pm 0.09$ & $17.52 \pm 0.03$ & $16.75 \pm 0.04$ & $16.86 \pm 0.02$ & 1m0-05 \\
2015-07-18 & 295.4 & $18.38 \pm 0.22$ & $17.37 \pm 0.04$ & $17.38 \pm 0.08$ & $16.57 \pm 0.06$ & $16.92 \pm 0.08$ & 1m0-08 \\
2015-07-23 & 300.4 & $18.38 \pm 0.16$ & $17.44 \pm 0.06$ & $17.55 \pm 0.04$ & $16.63 \pm 0.06$ & $16.96 \pm 0.01$ & 1m0-05 \\
2015-07-26 & 303.7 & - & $17.31 \pm 0.02$ & $17.60 \pm 0.07$ & $16.63 \pm 0.01$ & $16.84 \pm 0.04$ & 1m0-03 \\
2015-07-30 & 307.4 & $18.41 \pm 0.05$ & $17.53 \pm 0.13$ & - & $16.91 \pm 0.02$ & $17.04 \pm 0.09$ & 1m0-05 \\
2015-08-25 & 333.4 & - & $17.82 \pm 0.04$ & $17.81 \pm 0.03$ & $16.99 \pm 0.04$ & $17.11 \pm 0.09$ & 1m0-08 \\
2015-09-04 & 343.9 & $18.60 \pm 0.13$ & $17.60 \pm 0.12$ & $17.89 \pm 0.02$ & $17.18 \pm 0.03$ & $17.26 \pm 0.08$ & 1m0-10 \\
2015-09-05 & 344.1 & $18.45 \pm 0.05$ & $17.92 \pm 0.03$ & - & - & - & 1m0-10 \\
2015-09-05 & 344.1 & - & $17.68 \pm 0.30$ & $17.96 \pm 0.12$ & $17.18 \pm 0.07$ & $17.63 \pm 0.04$ & 1m0-12 \\
2015-09-05 & 344.0 & - & $17.90 \pm 0.07$ & $18.05 \pm 0.04$ & $17.04 \pm 0.09$ & - & 1m0-13 \\
2015-09-15 & 354.8 & $18.65 \pm 0.30$ & $17.76 \pm 0.13$ & $17.97 \pm 0.04$ & - & $17.42 \pm 0.13$ & 1m0-12 \\
2015-09-18 & 357.3 & $18.89 \pm 0.05$ & $18.26 \pm 0.16$ & $18.13 \pm 0.09$ & $17.27 \pm 0.01$ & $17.56 \pm 0.05$ & 1m0-05 \\
2015-09-28 & 368.0 & - & - & $18.16 \pm 0.08$ & - & - & 1m0-10 \\
2015-09-29 & 368.0 & - & - & - & $17.28 \pm 0.11$ & - & 1m0-10 \\
2015-10-14 & 383.2 & - & $18.50 \pm 0.06$ & $18.30 \pm 0.14$ & $17.50 \pm 0.02$ & $18.02 \pm 0.12$ & 1m0-08 \\
2015-10-27 & 396.2 & - & - & $18.53 \pm 0.21$ & $17.64 \pm 0.11$ & - & 1m0-05 \\
2015-11-06 & 406.9 & $19.44 \pm 0.12$ & - & $18.87 \pm 0.03$ & $18.05 \pm 0.01$ & $17.80 \pm 0.19$ & 1m0-10 \\
2015-11-17 & 417.6 & $19.49 \pm 0.40$ & $18.80 \pm 0.32$ & $19.18 \pm 0.12$ & - & $18.06 \pm 0.11$ & 1m0-03 \\
2015-11-19 & 419.6 & - & $18.75 \pm 0.03$ & - & - & $18.09 \pm 0.23$ & 1m0-03 \\
2015-11-21 & 421.2 & - & - & $18.88 \pm 0.15$ & $18.01 \pm 0.19$ & - & 1m0-05 \\
2015-12-14 & 444.1 & $20.40 \pm 0.23$ & - & $19.15 \pm 0.11$ & - & $18.57 \pm 0.09$ & 1m0-05 \\
2015-12-25 & 455.1 & - & - & - & $18.72 \pm 0.29$ & $18.47 \pm 0.17$ & 1m0-05 \\
2016-01-07 & 468.4 & $20.46 \pm 0.18$ & - & - & $18.78 \pm 0.18$ & - & 1m0-03 \\
2016-01-07 & 468.1 & - & - & $19.59 \pm 0.17$ & $18.40 \pm 0.08$ & - & 1m0-05 \\
2016-01-07 & 468.8 & - & - & - & $18.66 \pm 0.23$ & - & 1m0-10 \\
2016-01-19 & 480.1 & - & - & - & $18.74 \pm 0.08$ & - & 1m0-05 \\
\enddata
\end{deluxetable*}

%% file: 14il_swift_photometry.tex
\begin{tabular}{ccccccccc}
\hline \hline
Phase & MJD & UVW1 & UVW2 & UVM2 & U & B & V & Instrument \\ 
(d) &   & (mag) & (mag) & (mag) & (mag) & (mag) & (mag) & \\
\hline
7.90 & 56933.90 & $14.70 \pm 0.08$ & $14.72 \pm 0.07$ & $14.69 \pm 0.08$ & $14.81 \pm 0.07$ & $16.08 \pm 0.11$ & $16.08 \pm 0.16$ & UVOT \\
9.99 & 56935.99 & $14.46 \pm 0.07$ & $14.57 \pm 0.07$ & $14.45 \pm 0.06$ & $14.68 \pm 0.07$ & $16.03 \pm 0.10$ & $15.88 \pm 0.14$ & UVOT \\
11.84 & 56937.84 & $14.41 \pm 0.06$ & $14.54 \pm 0.06$ & $14.42 \pm 0.06$ & $14.54 \pm 0.06$ & $15.83 \pm 0.09$ & $15.77 \pm 0.11$ & UVOT \\
13.83 & 56939.83 & $14.36 \pm 0.07$ & $14.50 \pm 0.07$ & $14.39 \pm 0.06$ & $14.44 \pm 0.06$ & $15.67 \pm 0.09$ & $15.64 \pm 0.12$ & UVOT \\
15.43 & 56941.43 & $14.29 \pm 0.06$ & $14.49 \pm 0.07$ & $14.37 \pm 0.06$ & $14.40 \pm 0.06$ & $15.55 \pm 0.08$ & $15.55 \pm 0.11$ & UVOT \\
16.88 & 56942.88 & $14.29 \pm 0.06$ & $14.58 \pm 0.07$ & - & $14.37 \pm 0.06$ & $15.64 \pm 0.09$ & - & UVOT \\
16.94 & 56942.94 & - & $14.58 \pm 0.07$ & $14.53 \pm 0.08$ & - & - & $15.66 \pm 0.14$ & UVOT \\
17.83 & 56943.83 & $14.35 \pm 0.06$ & $14.61 \pm 0.07$ & $14.50 \pm 0.07$ & $14.35 \pm 0.06$ & $15.57 \pm 0.08$ & $15.68 \pm 0.12$ & UVOT \\
18.69 & 56944.69 & - & $14.63 \pm 0.07$ & - & $14.32 \pm 0.06$ & - & - & UVOT \\
18.87 & 56944.87 & $14.28 \pm 0.06$ & $14.59 \pm 0.07$ & $14.55 \pm 0.07$ & $14.33 \pm 0.06$ & $15.60 \pm 0.09$ & $15.50 \pm 0.11$ & UVOT \\
21.37 & 56947.37 & $14.39 \pm 0.06$ & $14.66 \pm 0.07$ & $14.57 \pm 0.07$ & $14.34 \pm 0.06$ & $15.55 \pm 0.09$ & $15.25 \pm 0.10$ & UVOT \\
23.25 & 56949.25 & $14.38 \pm 0.06$ & $14.70 \pm 0.07$ & $14.59 \pm 0.07$ & $14.33 \pm 0.06$ & $15.40 \pm 0.08$ & $15.27 \pm 0.10$ & UVOT \\
25.37 & 56951.37 & - & - & $14.58 \pm 0.08$ & - & - & $15.25 \pm 0.12$ & UVOT \\
25.67 & 56951.67 & $14.45 \pm 0.07$ & $14.79 \pm 0.08$ & - & $14.31 \pm 0.06$ & $15.42 \pm 0.08$ & - & UVOT \\
29.42 & 56955.42 & $14.55 \pm 0.07$ & $14.92 \pm 0.08$ & $14.78 \pm 0.08$ & $14.38 \pm 0.06$ & $15.40 \pm 0.08$ & $15.25 \pm 0.10$ & UVOT \\
31.34 & 56957.34 & $14.63 \pm 0.07$ & $15.02 \pm 0.08$ & $14.93 \pm 0.08$ & $14.46 \pm 0.06$ & $15.34 \pm 0.07$ & $15.23 \pm 0.10$ & UVOT \\
33.47 & 56959.47 & $14.77 \pm 0.08$ & - & - & - & - & - & UVOT \\
35.92 & 56961.92 & $14.88 \pm 0.08$ & $15.36 \pm 0.10$ & $15.18 \pm 0.09$ & $14.52 \pm 0.06$ & $15.43 \pm 0.08$ & $15.21 \pm 0.09$ & UVOT \\
37.85 & 56963.85 & $15.00 \pm 0.08$ & $15.41 \pm 0.10$ & $15.27 \pm 0.09$ & $14.60 \pm 0.06$ & $15.41 \pm 0.08$ & $15.23 \pm 0.09$ & UVOT \\
40.78 & 56966.78 & $15.13 \pm 0.09$ & $15.56 \pm 0.11$ & $15.40 \pm 0.09$ & $14.63 \pm 0.07$ & $15.40 \pm 0.08$ & $15.23 \pm 0.09$ & UVOT \\
42.21 & 56968.21 & $15.14 \pm 0.09$ & - & - & $14.64 \pm 0.07$ & $15.47 \pm 0.08$ & - & UVOT \\
46.23 & 56972.23 & $15.42 \pm 0.10$ & $16.02 \pm 0.14$ & $15.79 \pm 0.11$ & $14.78 \pm 0.07$ & $15.46 \pm 0.08$ & $15.15 \pm 0.09$ & UVOT \\
48.31 & 56974.31 & $15.45 \pm 0.10$ & $16.05 \pm 0.14$ & $15.87 \pm 0.11$ & $14.85 \pm 0.07$ & $15.57 \pm 0.08$ & $15.18 \pm 0.09$ & UVOT \\
\hline
\end{tabular}

%% file: spectra_log.tex
\begin{tabular}{ccccccc}
\hline \hline
Date & MJD & phase & Telescope & Range & Slit Width & Resolution \\ 
(yy-mm-dd) & (day) & & (\AA) & (\AA) & (arcsec) & ($\lambda / \Delta\lambda$) \\\hline
2014-10-03 & 56933.55 & 7.6 & ANU+WiFeS & 3050--11000 & - & 3000 \\
2014-10-03 & 56933.73 & 7.7 & LCO+FLOYDS & 3200--10000 & 1.6 & 400--700 \\
2014-10-05 & 56935.63 & 9.6 & LCO+FLOYDS & 3200--10000 & 1.6 & 400--700 \\
2014-10-07 & 56937.71 & 11.7 & LCO+FLOYDS & 3200--10000 & 1.6 & 400--700 \\
2014-10-09 & 56939.50 & 13.5 & ANU+WiFeS & 3050--11000 & - & 3000 \\
2014-10-10 & 56940.47 & 14.5 & LCO+FLOYDS & 3200--10000 & 1.6 & 400--700 \\
2014-10-13 & 56943.64 & 17.6 & LCO+FLOYDS & 3200--10000 & 1.6 & 400--700 \\
2014-10-14 & 56944.06 & 18.1 & ESO-NTT + SOFI & 9000--25000 & 1.0 & 600--2200 \\
2014-10-18 & 56948.56 & 22.6 & ANU+WiFeS & 3050--11000 & - & 3000 \\
2014-10-18 & 56948.67 & 22.7 & LCO+FLOYDS & 3200--10000 & 1.6 & 400--700 \\
2014-10-22 & 56952.02 & 26.0 & ESO-NTT + EFOSC (Gr11 \& Gr16) & 3050--11000 & 1.0 & 400 \& 600\\
2014-10-22 & 56952.47 & 26.5 & LCO+FLOYDS & 3200--10000 & 1.6 & 400--700 \\
2014-10-24 & 56954.21 & 28.2 & ESO-NTT + SOFI & 9000--25000 & 1.0 & 600--2200 \\
2014-10-27 & 56957.47 & 31.5 & ANU+WiFeS & 3050--11000 & - & 3000 \\
2014-11-14 & 56975.16 & 49.2 & ESO-NTT + EFOSC (Gr11 \& Gr16) & 3050--11000 & 1.0 & 400 \& 600 \\
2014-11-15 & 56976.19 & 50.2 & ESO-NTT + SOFI & 9000--25000 & 1.0 & 600--2200 \\
2014-12-16 & 57007.51 & 81.5 & LCO+FLOYDS & 3200--10000 & 1.6 & 400--700 \\
2014-12-22 & 57013.08 & 87.1 & ESO-NTT + SOFI & 9000--25000 & 1.0 & 600--2200 \\
2015-01-13 & 57035.06 & 109.1 & ESO-NTT + EFOSC (Gr11 \& Gr16) & 3050--11000 & 1.0 & 400 \& 600 \\
2015-01-20 & 57042.06 & 116.1 & ESO-NTT + SOFI & 9000--25000 & 1.0 & 600--2200 \\
2015-01-27 & 57049.04 & 123.0 & ESO-NTT + EFOSC (Gr13) & 3050--11000 & 1.0 & 350 \\
2015-08-19 & 57253.26 & 327.3 & ESO-NTT + EFOSC (Gr11 \& Gr16) & 3050--11000 & 1.0 & 400 \& 600\\
\hline
\end{tabular}

%% file: main.bbl
\begin{thebibliography}{}
\expandafter\ifx\csname natexlab\endcsname\relax\def\natexlab#1{#1}\fi
\providecommand{\url}[1]{\href{#1}{#1}}
\providecommand{\dodoi}[1]{doi:~\href{http://doi.org/#1}{\nolinkurl{#1}}}
\providecommand{\doeprint}[1]{\href{http://ascl.net/#1}{\nolinkurl{http://ascl.net/#1}}}
\providecommand{\doarXiv}[1]{\href{https://arxiv.org/abs/#1}{\nolinkurl{https://arxiv.org/abs/#1}}}

\bibitem[{{Agnoletto}(2009)}]{2006gy_agnoletto}
{Agnoletto}, I. 2009, in American Institute of Physics Conference Series, Vol. 1111, American Institute of Physics Conference Series, ed. G.~{Giobbi}, A.~{Tornambe}, G.~{Raimondo}, M.~{Limongi}, L.~A. {Antonelli}, N.~{Menci}, \& E.~{Brocato}, 430--434, \dodoi{10.1063/1.3141587}

\bibitem[{{Bak Nielsen} {et~al.}(2018){Bak Nielsen}, {Hjorth}, \& {Gall}}]{2005ip_bak_neilsen_earlydust}
{Bak Nielsen}, A.-S., {Hjorth}, J., \& {Gall}, C. 2018, \aap, 611, A67, \dodoi{10.1051/0004-6361/201629904}

\bibitem[{{Balberg} \& {Loeb}(2011)}]{balberg_loeb_2011}
{Balberg}, S., \& {Loeb}, A. 2011, \mnras, 414, 1715, \dodoi{10.1111/j.1365-2966.2011.18505.x}

\bibitem[{{Becker}(2015)}]{hotpants}
{Becker}, A. 2015, {HOTPANTS: High Order Transform of PSF ANd Template Subtraction}, Astrophysics Source Code Library, record ascl:1504.004.
\newblock \doeprint{1504.004}

\bibitem[{{Bilinski} {et~al.}(2024){Bilinski}, {Smith}, {Williams}, {Smith}, {Leonard}, {Hoffman}, {Andrews}, \& {Milne}}]{IIn_specpol_bilinski}
{Bilinski}, C., {Smith}, N., {Williams}, G.~G., {et~al.} 2024, \mnras, 529, 1104, \dodoi{10.1093/mnras/stae380}

\bibitem[{{Bilinski} {et~al.}(2020){Bilinski}, {Smith}, {Williams}, {Smith}, {Andrews}, {Clubb}, {Zheng}, {Filippenko}, {Fox}, {Hosseinzadeh}, {Howell}, {Kelly}, {Milne}, {Sand}, {Hoffman}, {Leonard}, {Cargill}, {Casper}, {Halevy}, {Kim}, {Kumar}, {Pina}, \& {Yuk}}]{2014ab_bilinksi}
---. 2020, \mnras, 498, 3835, \dodoi{10.1093/mnras/staa2617}

\bibitem[{{Breeveld} {et~al.}(2011){Breeveld}, {Landsman}, {Holland}, {Roming}, {Kuin}, \& {Page}}]{sousa_zp}
{Breeveld}, A.~A., {Landsman}, W., {Holland}, S.~T., {et~al.} 2011, in American Institute of Physics Conference Series, Vol. 1358, Gamma Ray Bursts 2010, ed. J.~E. {McEnery}, J.~L. {Racusin}, \& N.~{Gehrels}, 373--376, \dodoi{10.1063/1.3621807}

\bibitem[{{Brimacombe} {et~al.}(2014){Brimacombe}, {Holoien}, {Stanek}, {Kochanek}, {Davis}, {Simonian}, {Basu}, {Beacom}, {Shappee}, {Prieto}, {Bersier}, {Szczygiel}, {Pojmanski}, {Conseil}, {Cruz}, {Hissong}, {Kiyota}, {Monard}, {Nicholls}, {Nicolas}, \& {Wiethoff}}]{14il_discovery}
{Brimacombe}, J., {Holoien}, T.~W.~S., {Stanek}, K.~Z., {et~al.} 2014, The Astronomer's Telegram, 6525, 1

\bibitem[{{Brown} {et~al.}(2014){Brown}, {Breeveld}, {Holland}, {Kuin}, \& {Pritchard}}]{sousa_brown_2014}
{Brown}, P.~J., {Breeveld}, A.~A., {Holland}, S., {Kuin}, P., \& {Pritchard}, T. 2014, \apss, 354, 89, \dodoi{10.1007/s10509-014-2059-8}

\bibitem[{{Brown} {et~al.}(2009){Brown}, {Holland}, {Immler}, {Milne}, {Roming}, {Gehrels}, {Nousek}, {Panagia}, {Still}, \& {Vanden Berk}}]{sousa_reduction_brown_2009}
{Brown}, P.~J., {Holland}, S.~T., {Immler}, S., {et~al.} 2009, \aj, 137, 4517, \dodoi{10.1088/0004-6256/137/5/4517}

\bibitem[{{Brown} {et~al.}(2013){Brown}, {Baliber}, {Bianco}, {Bowman}, {Burleson}, {Conway}, {Crellin}, {Depagne}, {De Vera}, {Dilday}, {Dragomir}, {Dubberley}, {Eastman}, {Elphick}, {Falarski}, {Foale}, {Ford}, {Fulton}, {Garza}, {Gomez}, {Graham}, {Greene}, {Haldeman}, {Hawkins}, {Haworth}, {Haynes}, {Hidas}, {Hjelstrom}, {Howell}, {Hygelund}, {Lister}, {Lobdill}, {Martinez}, {Mullins}, {Norbury}, {Parrent}, {Paulson}, {Petry}, {Pickles}, {Posner}, {Rosing}, {Ross}, {Sand}, {Saunders}, {Shobbrook}, {Shporer}, {Street}, {Thomas}, {Tsapras}, {Tufts}, {Valenti}, {Vander Horst}, {Walker}, {White}, \& {Willis}}]{LCO_telescopes_instruments}
{Brown}, T.~M., {Baliber}, N., {Bianco}, F.~B., {et~al.} 2013, \pasp, 125, 1031, \dodoi{10.1086/673168}

\bibitem[{{Chatzopoulos} {et~al.}(2012){Chatzopoulos}, {Wheeler}, \& {Vinko}}]{CW12}
{Chatzopoulos}, E., {Wheeler}, J.~C., \& {Vinko}, J. 2012, \apj, 746, 121, \dodoi{10.1088/0004-637X/746/2/121}

\bibitem[{{Chatzopoulos} {et~al.}(2013){Chatzopoulos}, {Wheeler}, {Vinko}, {Horvath}, \& {Nagy}}]{CW13}
{Chatzopoulos}, E., {Wheeler}, J.~C., {Vinko}, J., {Horvath}, Z.~L., \& {Nagy}, A. 2013, \apj, 773, 76, \dodoi{10.1088/0004-637X/773/1/76}

\bibitem[{{Chevalier}(1982)}]{Chevalier_1982_self-similar}
{Chevalier}, R.~A. 1982, \apj, 258, 790, \dodoi{10.1086/160126}

\bibitem[{{Chevalier}(1998)}]{Chevalier_1998}
---. 1998, \apj, 499, 810, \dodoi{10.1086/305676}

\bibitem[{{Childress} {et~al.}(2014){Childress}, {Scalzo}, {Yuan}, {Zhang}, {Ruiter}, {Seitenzahl}, {Schmidt}, \& {Tucker}}]{14il_classification}
{Childress}, M., {Scalzo}, R., {Yuan}, F., {et~al.} 2014, The Astronomer's Telegram, 6536, 1

\bibitem[{{Childress} {et~al.}(2016){Childress}, {Tucker}, {Yuan}, {Scalzo}, {Ruiter}, {Seitenzahl}, {Zhang}, {Schmidt}, {Anguiano}, {Aniyan}, {Bayliss}, {Bento}, {Bessell}, {Bian}, {Davies}, {Dopita}, {Fogarty}, {Fraser-McKelvie}, {Freeman}, {Kuruwita}, {Medling}, {Murphy}, {Murphy}, {Owers}, {Panther}, {Sweet}, {Thomas}, \& {Zhou}}]{awsnap_2016}
{Childress}, M.~J., {Tucker}, B.~E., {Yuan}, F., {et~al.} 2016, \pasa, 33, e055, \dodoi{10.1017/pasa.2016.47}

\bibitem[{{Chugai}(2001)}]{Chugai_2001_1998S}
{Chugai}, N.~N. 2001, \mnras, 326, 1448, \dodoi{10.1111/j.1365-2966.2001.04717.x}

\bibitem[{{Chugai} \& {Danziger}(1994)}]{chugai_danziger_1994}
{Chugai}, N.~N., \& {Danziger}, I.~J. 1994, \mnras, 268, 173, \dodoi{10.1093/mnras/268.1.173}

\bibitem[{{Chugai} {et~al.}(2004){Chugai}, {Blinnikov}, {Cumming}, {Lundqvist}, {Bragaglia}, {Filippenko}, {Leonard}, {Matheson}, \& {Sollerman}}]{Chugai_1994w_2004}
{Chugai}, N.~N., {Blinnikov}, S.~I., {Cumming}, R.~J., {et~al.} 2004, \mnras, 352, 1213, \dodoi{10.1111/j.1365-2966.2004.08011.x}

\bibitem[{{de Jager}(1998)}]{ysg_de_jager}
{de Jager}, C. 1998, \aapr, 8, 145, \dodoi{10.1007/s001590050009}

\bibitem[{{de Jager} {et~al.}(1988){de Jager}, {Nieuwenhuijzen}, \& {van der Hucht}}]{mass_loss_rsg_de_jager}
{de Jager}, C., {Nieuwenhuijzen}, H., \& {van der Hucht}, K.~A. 1988, \aaps, 72, 259

\bibitem[{{Dessart} {et~al.}(2015){Dessart}, {Audit}, \& {Hillier}}]{Dessart_IIn_simulation}
{Dessart}, L., {Audit}, E., \& {Hillier}, D.~J. 2015, \mnras, 449, 4304, \dodoi{10.1093/mnras/stv609}

\bibitem[{{Dickinson} {et~al.}(2023){Dickinson}, {Smith}, {Andrews}, {Milne}, {Kilpatrick}, \& {Milisavljevic}}]{15ua_dickinson}
{Dickinson}, D., {Smith}, N., {Andrews}, J.~E., {et~al.} 2023, arXiv e-prints, arXiv:2302.04958, \dodoi{10.48550/arXiv.2302.04958}

\bibitem[{Dickinson(2021)}]{14il_dickinson_thesis}
Dickinson, D.~A. 2021, AN OPTICAL STUDY OF THE SUPERLUMINOUS TYPE IIN SUPERNOVA ASASSN-14IL.
\newblock \url{http://hdl.handle.net/10150/666597}

\bibitem[{{Filippenko}(1997)}]{filippenko97_review}
{Filippenko}, A.~V. 1997, \araa, 35, 309, \dodoi{10.1146/annurev.astro.35.1.309}

\bibitem[{{Fox} {et~al.}(2009){Fox}, {Skrutskie}, {Chevalier}, {Kanneganti}, {Park}, {Wilson}, {Nelson}, {Amirhadji}, {Crump}, {Hoeft}, {Provence}, {Sargeant}, {Sop}, {Tea}, {Thomas}, \& {Woolard}}]{2005ip_orifox_dust}
{Fox}, O., {Skrutskie}, M.~F., {Chevalier}, R.~A., {et~al.} 2009, \apj, 691, 650, \dodoi{10.1088/0004-637X/691/1/650}

\bibitem[{{Fransson} {et~al.}(2002){Fransson}, {Chevalier}, {Filippenko}, {Leibundgut}, {Barth}, {Fesen}, {Kirshner}, {Leonard}, {Li}, {Lundqvist}, {Sollerman}, \& {Van Dyk}}]{1995N_fransson}
{Fransson}, C., {Chevalier}, R.~A., {Filippenko}, A.~V., {et~al.} 2002, \apj, 572, 350, \dodoi{10.1086/340295}

\bibitem[{{Fransson} {et~al.}(2014){Fransson}, {Ergon}, {Challis}, {Chevalier}, {France}, {Kirshner}, {Marion}, {Milisavljevic}, {Smith}, {Bufano}, {Friedman}, {Kangas}, {Larsson}, {Mattila}, {Benetti}, {Chornock}, {Czekala}, {Soderberg}, \& {Sollerman}}]{2010jl_fransson}
{Fransson}, C., {Ergon}, M., {Challis}, P.~J., {et~al.} 2014, \apj, 797, 118, \dodoi{10.1088/0004-637X/797/2/118}

\bibitem[{{Gal-Yam}(2012)}]{slsn_gal-yam_2012}
{Gal-Yam}, A. 2012, Science, 337, 927, \dodoi{10.1126/science.1203601}

\bibitem[{{Gal-Yam}(2019)}]{slsn_gal-yam_2019}
---. 2019, \araa, 57, 305, \dodoi{10.1146/annurev-astro-081817-051819}

\bibitem[{{Gal-Yam} \& {Leonard}(2009)}]{2005gl_gal-yam}
{Gal-Yam}, A., \& {Leonard}, D.~C. 2009, \nat, 458, 865, \dodoi{10.1038/nature07934}

\bibitem[{{Gall} {et~al.}(2014){Gall}, {Hjorth}, {Watson}, {Dwek}, {Maund}, {Fox}, {Leloudas}, {Malesani}, \& {Day-Jones}}]{gall2014}
{Gall}, C., {Hjorth}, J., {Watson}, D., {et~al.} 2014, \nat, 511, 326, \dodoi{10.1038/nature13558}

\bibitem[{{Gangopadhyay} {et~al.}(2020){Gangopadhyay}, {Turatto}, {Benetti}, {Misra}, {Kumar}, {Cappellaro}, {Pastorello}, {Tomasella}, {Vanni}, {Fiore}, {Morales-Garoffolo}, {Elias-Rosa}, {Singh}, {Dastidar}, {Ochner}, {Tartaglia}, {Kumar}, \& {Pandey}}]{2012ab_gangopadhyay}
{Gangopadhyay}, A., {Turatto}, M., {Benetti}, S., {et~al.} 2020, \mnras, 499, 129, \dodoi{10.1093/mnras/staa2606}

\bibitem[{{Gehrels} {et~al.}(2004){Gehrels}, {Chincarini}, {Giommi}, {Mason}, {Nousek}, {Wells}, {White}, {Barthelmy}, {Burrows}, {Cominsky}, {Hurley}, {Marshall}, {M{\'e}sz{\'a}ros}, {Roming}, {Angelini}, {Barbier}, {Belloni}, {Campana}, {Caraveo}, {Chester}, {Citterio}, {Cline}, {Cropper}, {Cummings}, {Dean}, {Feigelson}, {Fenimore}, {Frail}, {Fruchter}, {Garmire}, {Gendreau}, {Ghisellini}, {Greiner}, {Hill}, {Hunsberger}, {Krimm}, {Kulkarni}, {Kumar}, {Lebrun}, {Lloyd-Ronning}, {Markwardt}, {Mattson}, {Mushotzky}, {Norris}, {Osborne}, {Paczynski}, {Palmer}, {Park}, {Parsons}, {Paul}, {Rees}, {Reynolds}, {Rhoads}, {Sasseen}, {Schaefer}, {Short}, {Smale}, {Smith}, {Stella}, {Tagliaferri}, {Takahashi}, {Tashiro}, {Townsley}, {Tueller}, {Turner}, {Vietri}, {Voges}, {Ward}, {Willingale}, {Zerbi}, \& {Zhang}}]{swift}
{Gehrels}, N., {Chincarini}, G., {Giommi}, P., {et~al.} 2004, \apj, 611, 1005, \dodoi{10.1086/422091}

\bibitem[{{Graham} {et~al.}(2014){Graham}, {Sand}, {Valenti}, {Howell}, {Parrent}, {Halford}, {Zaritsky}, {Bianco}, {Rest}, \& {Dilday}}]{2009ip_graham}
{Graham}, M.~L., {Sand}, D.~J., {Valenti}, S., {et~al.} 2014, \apj, 787, 163, \dodoi{10.1088/0004-637X/787/2/163}

\bibitem[{{Groh} {et~al.}(2009){Groh}, {Hillier}, {Damineli}, {Whitelock}, {Marang}, \& {Rossi}}]{groh_hillier_lbv_2009}
{Groh}, J.~H., {Hillier}, D.~J., {Damineli}, A., {et~al.} 2009, \apj, 698, 1698, \dodoi{10.1088/0004-637X/698/2/1698}

\bibitem[{{Guillochon} {et~al.}(2018){Guillochon}, {Nicholl}, {Villar}, {Mockler}, {Narayan}, {Mandel}, {Berger}, \& {Williams}}]{mosfit}
{Guillochon}, J., {Nicholl}, M., {Villar}, V.~A., {et~al.} 2018, \apjs, 236, 6, \dodoi{10.3847/1538-4365/aab761}

\bibitem[{{Henden} {et~al.}(2015){Henden}, {Levine}, {Terrell}, \& {Welch}}]{apass_dr9}
{Henden}, A.~A., {Levine}, S., {Terrell}, D., \& {Welch}, D.~L. 2015, in American Astronomical Society Meeting Abstracts, Vol. 225, American Astronomical Society Meeting Abstracts \#225, 336.16

\bibitem[{{Hillier} {et~al.}(2001){Hillier}, {Davidson}, {Ishibashi}, \& {Gull}}]{hillier_2001_eta_carinae}
{Hillier}, D.~J., {Davidson}, K., {Ishibashi}, K., \& {Gull}, T. 2001, \apj, 553, 837, \dodoi{10.1086/320948}

\bibitem[{{Hosseinzadeh} {et~al.}(2022){Hosseinzadeh}, {Berger}, {Metzger}, {Gomez}, {Nicholl}, \& {Blanchard}}]{Bumpy_lk_typeI_slsn_Hosseinzadeh}
{Hosseinzadeh}, G., {Berger}, E., {Metzger}, B.~D., {et~al.} 2022, \apj, 933, 14, \dodoi{10.3847/1538-4357/ac67dd}

\bibitem[{Hosseinzadeh \& Gomez(2022)}]{lk_fitting}
Hosseinzadeh, G., \& Gomez, S. 2022, Light Curve Fitting, v0.6.0,  Zenodo, \dodoi{10.5281/zenodo.6519623}

\bibitem[{{Ishii} {et~al.}(2024){Ishii}, {Takei}, {Tsuna}, {Shigeyama}, \& {Takahashi}}]{Ishii_2024_narrow_lines}
{Ishii}, A.~T., {Takei}, Y., {Tsuna}, D., {Shigeyama}, T., \& {Takahashi}, K. 2024, \apj, 961, 47, \dodoi{10.3847/1538-4357/ad072b}

\bibitem[{{Jencson} {et~al.}(2016){Jencson}, {Prieto}, {Kochanek}, {Shappee}, {Stanek}, \& {Pogge}}]{2010jl_jencson}
{Jencson}, J.~E., {Prieto}, J.~L., {Kochanek}, C.~S., {et~al.} 2016, \mnras, 456, 2622, \dodoi{10.1093/mnras/stv2795}

\bibitem[{{Jiang} {et~al.}(2020){Jiang}, {Jiang}, \& {Ashley Villar}}]{Jiang_IIn_analytical_mosfit}
{Jiang}, B., {Jiang}, S., \& {Ashley Villar}, V. 2020, Research Notes of the American Astronomical Society, 4, 16, \dodoi{10.3847/2515-5172/ab7128}

\bibitem[{{Jones} {et~al.}(2004){Jones}, {Saunders}, {Colless}, {Read}, {Parker}, {Watson}, {Campbell}, {Burkey}, {Mauch}, {Moore}, {Hartley}, {Cass}, {James}, {Russell}, {Fiegert}, {Dawe}, {Huchra}, {Jarrett}, {Lahav}, {Lucey}, {Mamon}, {Proust}, {Sadler}, \& {Wakamatsu}}]{6dF2004}
{Jones}, D.~H., {Saunders}, W., {Colless}, M., {et~al.} 2004, \mnras, 355, 747, \dodoi{10.1111/j.1365-2966.2004.08353.x}

\bibitem[{{Jones} {et~al.}(2009){Jones}, {Read}, {Saunders}, {Colless}, {Jarrett}, {Parker}, {Fairall}, {Mauch}, {Sadler}, {Watson}, {Burton}, {Campbell}, {Cass}, {Croom}, {Dawe}, {Fiegert}, {Frankcombe}, {Hartley}, {Huchra}, {James}, {Kirby}, {Lahav}, {Lucey}, {Mamon}, {Moore}, {Peterson}, {Prior}, {Proust}, {Russell}, {Safouris}, {Wakamatsu}, {Westra}, \& {Williams}}]{host_galaxy_z}
{Jones}, D.~H., {Read}, M.~A., {Saunders}, W., {et~al.} 2009, \mnras, 399, 683, \dodoi{10.1111/j.1365-2966.2009.15338.x}

\bibitem[{{Kankare} {et~al.}(2012){Kankare}, {Ergon}, {Bufano}, {Spyromilio}, {Mattila}, {Chugai}, {Lundqvist}, {Pastorello}, {Kotak}, {Benetti}, {Botticella}, {Cumming}, {Fransson}, {Fraser}, {Leloudas}, {Miluzio}, {Sollerman}, {Stritzinger}, {Turatto}, \& {Valenti}}]{2009kn_kankare}
{Kankare}, E., {Ergon}, M., {Bufano}, F., {et~al.} 2012, \mnras, 424, 855, \dodoi{10.1111/j.1365-2966.2012.21224.x}

\bibitem[{{Katsuda} {et~al.}(2016){Katsuda}, {Maeda}, {Bamba}, {Terada}, {Fukazawa}, {Kawabata}, {Ohno}, {Sugawara}, {Tsuboi}, \& {Immler}}]{2010jl_xray_asymmetry_katsuda}
{Katsuda}, S., {Maeda}, K., {Bamba}, A., {et~al.} 2016, \apj, 832, 194, \dodoi{10.3847/0004-637X/832/2/194}

\bibitem[{{Kiewe} {et~al.}(2012){Kiewe}, {Gal-Yam}, {Arcavi}, {Leonard}, {Emilio Enriquez}, {Cenko}, {Fox}, {Moon}, {Sand}, {Soderberg}, \& {CCCP}}]{kiewe_iin_sample_2012}
{Kiewe}, M., {Gal-Yam}, A., {Arcavi}, I., {et~al.} 2012, \apj, 744, 10, \dodoi{10.1088/0004-637X/744/1/10}

\bibitem[{{Kochanek} {et~al.}(2011){Kochanek}, {Szczygiel}, \& {Stanek}}]{1961V_progenitor}
{Kochanek}, C.~S., {Szczygiel}, D.~M., \& {Stanek}, K.~Z. 2011, \apj, 737, 76, \dodoi{10.1088/0004-637X/737/2/76}

\bibitem[{{Li} {et~al.}(2011){Li}, {Leaman}, {Chornock}, {Filippenko}, {Poznanski}, {Ganeshalingam}, {Wang}, {Modjaz}, {Jha}, {Foley}, \& {Smith}}]{li_SNe_rate}
{Li}, W., {Leaman}, J., {Chornock}, R., {et~al.} 2011, \mnras, 412, 1441, \dodoi{10.1111/j.1365-2966.2011.18160.x}

\bibitem[{{Maeda} {et~al.}(2013){Maeda}, {Nozawa}, {Sahu}, {Minowa}, {Motohara}, {Ueno}, {Folatelli}, {Pyo}, {Kitagawa}, {Kawabata}, {Anupama}, {Kozasa}, {Moriya}, {Yamanaka}, {Nomoto}, {Bersten}, {Quimby}, \& {Iye}}]{2010jl_maeda}
{Maeda}, K., {Nozawa}, T., {Sahu}, D.~K., {et~al.} 2013, \apj, 776, 5, \dodoi{10.1088/0004-637X/776/1/5}

\bibitem[{{Martin} {et~al.}(2015){Martin}, {Hambsch}, {Margutti}, {Tan}, {Curtis}, \& {Soderberg}}]{martin_2009ip}
{Martin}, J.~C., {Hambsch}, F.~J., {Margutti}, R., {et~al.} 2015, \aj, 149, 9, \dodoi{10.1088/0004-6256/149/1/9}

\bibitem[{{Mattila} {et~al.}(2008){Mattila}, {Meikle}, {Lundqvist}, {Pastorello}, {Kotak}, {Eldridge}, {Smartt}, {Adamson}, {Gerardy}, {Rizzi}, {Stephens}, \& {van Dyk}}]{mattila_2006jc_dust}
{Mattila}, S., {Meikle}, W.~P.~S., {Lundqvist}, P., {et~al.} 2008, \mnras, 389, 141, \dodoi{10.1111/j.1365-2966.2008.13516.x}

\bibitem[{{Matzner} \& {McKee}(1999)}]{Matzner_Mckee_1999}
{Matzner}, C.~D., \& {McKee}, C.~F. 1999, \apj, 510, 379, \dodoi{10.1086/306571}

\bibitem[{{Mauerhan} {et~al.}(2013){Mauerhan}, {Smith}, {Silverman}, {Filippenko}, {Morgan}, {Cenko}, {Ganeshalingam}, {Clubb}, {Bloom}, {Matheson}, \& {Milne}}]{2011ht_mauerhan}
{Mauerhan}, J.~C., {Smith}, N., {Silverman}, J.~M., {et~al.} 2013, \mnras, 431, 2599, \dodoi{10.1093/mnras/stt360}

\bibitem[{{McCully} {et~al.}(2018){McCully}, {Volgenau}, {Harbeck}, {Lister}, {Saunders}, {Turner}, {Siiverd}, \& {Bowman}}]{Banzai}
{McCully}, C., {Volgenau}, N.~H., {Harbeck}, D.-R., {et~al.} 2018, in Society of Photo-Optical Instrumentation Engineers (SPIE) Conference Series, Vol. 10707, Software and Cyberinfrastructure for Astronomy V, ed. J.~C. {Guzman} \& J.~{Ibsen}, 107070K, \dodoi{10.1117/12.2314340}

\bibitem[{{Moriya} {et~al.}(2013){Moriya}, {Maeda}, {Taddia}, {Sollerman}, {Blinnikov}, \& {Sorokina}}]{takashi_IIn_analytical}
{Moriya}, T.~J., {Maeda}, K., {Taddia}, F., {et~al.} 2013, \mnras, 435, 1520, \dodoi{10.1093/mnras/stt1392}

\bibitem[{{Moriya} {et~al.}(2014){Moriya}, {Maeda}, {Taddia}, {Sollerman}, {Blinnikov}, \& {Sorokina}}]{Mass_loss_IIn_moriya_2014}
---. 2014, \mnras, 439, 2917, \dodoi{10.1093/mnras/stu163}

\bibitem[{{Nicholl}(2018)}]{2018Nicholl}
{Nicholl}, M. 2018, Research Notes of the American Astronomical Society, 2, 230, \dodoi{10.3847/2515-5172/aaf799}

\bibitem[{{Nicholl} {et~al.}(2020){Nicholl}, {Blanchard}, {Berger}, {Chornock}, {Margutti}, {Gomez}, {Lunnan}, {Miller}, {Fong}, {Terreran}, {Vigna-G{\'o}mez}, {Bhirombhakdi}, {Bieryla}, {Challis}, {Laher}, {Masci}, \& {Paterson}}]{2016aps_nicholl}
{Nicholl}, M., {Blanchard}, P.~K., {Berger}, E., {et~al.} 2020, Nature Astronomy, 4, 893, \dodoi{10.1038/s41550-020-1066-7}

\bibitem[{{Nyholm} {et~al.}(2017){Nyholm}, {Sollerman}, {Taddia}, {Fremling}, {Moriya}, {Ofek}, {Gal-Yam}, {De Cia}, {Roy}, {Kasliwal}, {Cao}, {Nugent}, \& {Masci}}]{iptf13z_nyholm}
{Nyholm}, A., {Sollerman}, J., {Taddia}, F., {et~al.} 2017, \aap, 605, A6, \dodoi{10.1051/0004-6361/201629906}

\bibitem[{{Nyholm} {et~al.}(2020){Nyholm}, {Sollerman}, {Tartaglia}, {Taddia}, {Fremling}, {Blagorodnova}, {Filippenko}, {Gal-Yam}, {Howell}, {Karamehmetoglu}, {Kulkarni}, {Laher}, {Leloudas}, {Masci}, {Kasliwal}, {Mor{\r{a}}}, {Moriya}, {Ofek}, {Papadogiannakis}, {Quimby}, {Rebbapragada}, \& {Schulze}}]{nyholm_IIn_survey_2020}
{Nyholm}, A., {Sollerman}, J., {Tartaglia}, L., {et~al.} 2020, \aap, 637, A73, \dodoi{10.1051/0004-6361/201936097}

\bibitem[{{Ofek} {et~al.}(2014){Ofek}, {Zoglauer}, {Boggs}, {Barri{\'e}re}, {Reynolds}, {Fryer}, {Harrison}, {Cenko}, {Kulkarni}, {Gal-Yam}, {Arcavi}, {Bellm}, {Bloom}, {Christensen}, {Craig}, {Even}, {Filippenko}, {Grefenstette}, {Hailey}, {Laher}, {Madsen}, {Nakar}, {Nugent}, {Stern}, {Sullivan}, {Surace}, \& {Zhang}}]{Ofek_2010jl_modelling}
{Ofek}, E.~O., {Zoglauer}, A., {Boggs}, S.~E., {et~al.} 2014, \apj, 781, 42, \dodoi{10.1088/0004-637X/781/1/42}

\bibitem[{{Poznanski} {et~al.}(2012){Poznanski}, {Prochaska}, \& {Bloom}}]{poznanski2012}
{Poznanski}, D., {Prochaska}, J.~X., \& {Bloom}, J.~S. 2012, \mnras, 426, 1465, \dodoi{10.1111/j.1365-2966.2012.21796.x}

\bibitem[{{Pozzo} {et~al.}(2004){Pozzo}, {Meikle}, {Fassia}, {Geballe}, {Lundqvist}, {Chugai}, \& {Sollerman}}]{pozzo_1998s_dust}
{Pozzo}, M., {Meikle}, W.~P.~S., {Fassia}, A., {et~al.} 2004, \mnras, 352, 457, \dodoi{10.1111/j.1365-2966.2004.07951.x}

\bibitem[{{Roming} {et~al.}(2005){Roming}, {Kennedy}, {Mason}, {Nousek}, {Ahr}, {Bingham}, {Broos}, {Carter}, {Hancock}, {Huckle}, {Hunsberger}, {Kawakami}, {Killough}, {Koch}, {McLelland}, {Smith}, {Smith}, {Soto}, {Boyd}, {Breeveld}, {Holland}, {Ivanushkina}, {Pryzby}, {Still}, \& {Stock}}]{uvot}
{Roming}, P. W.~A., {Kennedy}, T.~E., {Mason}, K.~O., {et~al.} 2005, \ssr, 120, 95, \dodoi{10.1007/s11214-005-5095-4}

\bibitem[{{Schlafly} \& {Finkbeiner}(2011)}]{milkyway_reddening}
{Schlafly}, E.~F., \& {Finkbeiner}, D.~P. 2011, \apj, 737, 103, \dodoi{10.1088/0004-637X/737/2/103}

\bibitem[{{Schlegel}(1990)}]{schlegel1990_IIn}
{Schlegel}, E.~M. 1990, \mnras, 244, 269

\bibitem[{{Smartt} {et~al.}(2015){Smartt}, {Valenti}, {Fraser}, {Inserra}, {Young}, {Sullivan}, {Pastorello}, {Benetti}, {Gal-Yam}, {Knapic}, {Molinaro}, {Smareglia}, {Smith}, {Taubenberger}, {Yaron}, {Anderson}, {Ashall}, {Balland}, {Baltay}, {Barbarino}, {Bauer}, {Baumont}, {Bersier}, {Blagorodnova}, {Bongard}, {Botticella}, {Bufano}, {Bulla}, {Cappellaro}, {Campbell}, {Cellier-Holzem}, {Chen}, {Childress}, {Clocchiatti}, {Contreras}, {Dall'Ora}, {Danziger}, {de Jaeger}, {De Cia}, {Della Valle}, {Dennefeld}, {Elias-Rosa}, {Elman}, {Feindt}, {Fleury}, {Gall}, {Gonzalez-Gaitan}, {Galbany}, {Morales Garoffolo}, {Greggio}, {Guillou}, {Hachinger}, {Hadjiyska}, {Hage}, {Hillebrandt}, {Hodgkin}, {Hsiao}, {James}, {Jerkstrand}, {Kangas}, {Kankare}, {Kotak}, {Kromer}, {Kuncarayakti}, {Leloudas}, {Lundqvist}, {Lyman}, {Hook}, {Maguire}, {Manulis}, {Margheim}, {Mattila}, {Maund}, {Mazzali}, {McCrum}, {McKinnon}, {Moreno-Raya}, {Nicholl}, {Nugent}, {Pain}, {Pignata}, {Phillips}, {Polshaw}, {Pumo}, {Rabinowitz},
  {Reilly}, {Romero-Ca{\~n}izales}, {Scalzo}, {Schmidt}, {Schulze}, {Sim}, {Sollerman}, {Taddia}, {Tartaglia}, {Terreran}, {Tomasella}, {Turatto}, {Walker}, {Walton}, {Wyrzykowski}, {Yuan}, \& {Zampieri}}]{pessto_smartt_2015}
{Smartt}, S.~J., {Valenti}, S., {Fraser}, M., {et~al.} 2015, \aap, 579, A40, \dodoi{10.1051/0004-6361/201425237}

\bibitem[{{Smith}(2014)}]{Smith_mass_loss_2014}
{Smith}, N. 2014, \araa, 52, 487, \dodoi{10.1146/annurev-astro-081913-040025}

\bibitem[{{Smith}(2017)}]{smith_IIn_review}
---. 2017, in Handbook of Supernovae, ed. A.~W. {Alsabti} \& P.~{Murdin} (Springer), 403, \dodoi{10.1007/978-3-319-21846-5_38}

\bibitem[{{Smith} \& {Andrews}(2020)}]{2017hcc_smith_andrews}
{Smith}, N., \& {Andrews}, J.~E. 2020, \mnras, 499, 3544, \dodoi{10.1093/mnras/staa3047}

\bibitem[{{Smith} {et~al.}(2023){Smith}, {Andrews}, {Milne}, {Filippenko}, {Brink}, {Kelly}, {Yuk}, \& {Jencson}}]{2015dalate_nathansmith_2023}
{Smith}, N., {Andrews}, J.~E., {Milne}, P., {et~al.} 2023, arXiv e-prints, arXiv:2312.00253, \dodoi{10.48550/arXiv.2312.00253}

\bibitem[{{Smith} {et~al.}(2008){Smith}, {Chornock}, {Li}, {Ganeshalingam}, {Silverman}, {Foley}, {Filippenko}, \& {Barth}}]{2006tf_smith}
{Smith}, N., {Chornock}, R., {Li}, W., {et~al.} 2008, \apj, 686, 467, \dodoi{10.1086/591021}

\bibitem[{{Smith} {et~al.}(2009{\natexlab{a}}){Smith}, {Hinkle}, \& {Ryde}}]{smith_2009_vy_canis_majoris}
{Smith}, N., {Hinkle}, K.~H., \& {Ryde}, N. 2009{\natexlab{a}}, \aj, 137, 3558, \dodoi{10.1088/0004-6256/137/3/3558}

\bibitem[{{Smith} \& {McCray}(2007)}]{2006gy_smith_mccray_shellshockeddiffusion}
{Smith}, N., \& {McCray}, R. 2007, \apjl, 671, L17, \dodoi{10.1086/524681}

\bibitem[{{Smith} \& {Owocki}(2006)}]{smith_owocki_linedrivenwinds}
{Smith}, N., \& {Owocki}, S.~P. 2006, \apjl, 645, L45, \dodoi{10.1086/506523}

\bibitem[{{Smith} {et~al.}(2012){Smith}, {Silverman}, {Filippenko}, {Cooper}, {Matheson}, {Bian}, {Weiner}, \& {Comerford}}]{2010jl_smith_dust}
{Smith}, N., {Silverman}, J.~M., {Filippenko}, A.~V., {et~al.} 2012, \aj, 143, 17, \dodoi{10.1088/0004-6256/143/1/17}

\bibitem[{{Smith} {et~al.}(2009{\natexlab{b}}){Smith}, {Silverman}, {Chornock}, {Filippenko}, {Wang}, {Li}, {Ganeshalingam}, {Foley}, {Rex}, \& {Steele}}]{2005ip_smith}
{Smith}, N., {Silverman}, J.~M., {Chornock}, R., {et~al.} 2009{\natexlab{b}}, \apj, 695, 1334, \dodoi{10.1088/0004-637X/695/2/1334}

\bibitem[{{Smith} {et~al.}(2010){Smith}, {Miller}, {Li}, {Filippenko}, {Silverman}, {Howard}, {Marcy}, {Bloom}, {Nugent}, {Ghez}, {Lu}, {Yelda}, {Bernstein}, \& {Colucci}}]{2009ip_precursor_smith_2010}
{Smith}, N., {Miller}, A., {Li}, W., {et~al.} 2010, in American Astronomical Society Meeting Abstracts, Vol. 215, American Astronomical Society Meeting Abstracts \#215, 430.16

\bibitem[{{Smith} {et~al.}(2011){Smith}, {Li}, {Miller}, {Silverman}, {Filippenko}, {Cuillandre}, {Cooper}, {Matheson}, \& {Van Dyk}}]{2010jl_smith}
{Smith}, N., {Li}, W., {Miller}, A.~A., {et~al.} 2011, \apj, 732, 63, \dodoi{10.1088/0004-637X/732/2/63}

\bibitem[{{Speagle}(2020)}]{dynesty}
{Speagle}, J.~S. 2020, \mnras, 493, 3132, \dodoi{10.1093/mnras/staa278}

\bibitem[{{Stritzinger} {et~al.}(2012){Stritzinger}, {Taddia}, {Fransson}, {Fox}, {Morrell}, {Phillips}, {Sollerman}, {Anderson}, {Boldt}, {Brown}, {Campillay}, {Castellon}, {Contreras}, {Folatelli}, {Habergham}, {Hamuy}, {Hjorth}, {James}, {Krzeminski}, {Mattila}, {Persson}, \& {Roth}}]{2005ip_2006jd_stritzinger}
{Stritzinger}, M., {Taddia}, F., {Fransson}, C., {et~al.} 2012, \apj, 756, 173, \dodoi{10.1088/0004-637X/756/2/173}

\bibitem[{{Svirski} {et~al.}(2012){Svirski}, {Nakar}, \& {Sari}}]{Svirski_2012}
{Svirski}, G., {Nakar}, E., \& {Sari}, R. 2012, \apj, 759, 108, \dodoi{10.1088/0004-637X/759/2/108}

\bibitem[{{Taddia} {et~al.}(2013){Taddia}, {Stritzinger}, {Sollerman}, {Phillips}, {Anderson}, {Boldt}, {Campillay}, {Castell{\'o}n}, {Contreras}, {Folatelli}, {Hamuy}, {Heinrich-Josties}, {Krzeminski}, {Morrell}, {Burns}, {Freedman}, {Madore}, {Persson}, \& {Suntzeff}}]{taddia_IIn_sample_2013}
{Taddia}, F., {Stritzinger}, M.~D., {Sollerman}, J., {et~al.} 2013, \aap, 555, A10, \dodoi{10.1051/0004-6361/201321180}

\bibitem[{{Tartaglia} {et~al.}(2020){Tartaglia}, {Pastorello}, {Sollerman}, {Fransson}, {Mattila}, {Fraser}, {Taddia}, {Tomasella}, {Turatto}, {Morales-Garoffolo}, {Elias-Rosa}, {Lundqvist}, {Harmanen}, {Reynolds}, {Cappellaro}, {Barbarino}, {Nyholm}, {Kool}, {Ofek}, {Gao}, {Jin}, {Tan}, {Sand}, {Ciabattari}, {Wang}, {Zhang}, {Huang}, {Li}, {Mo}, {Rui}, {Xiang}, {Zhang}, {Hosseinzadeh}, {Howell}, {McCully}, {Valenti}, {Benetti}, {Callis}, {Carracedo}, {Fremling}, {Kangas}, {Rubin}, {Somero}, \& {Terreran}}]{2015da_tartaglia}
{Tartaglia}, L., {Pastorello}, A., {Sollerman}, J., {et~al.} 2020, \aap, 635, A39, \dodoi{10.1051/0004-6361/201936553}

\bibitem[{{Valenti} {et~al.}(2014){Valenti}, {Sand}, {Pastorello}, {Graham}, {Howell}, {Parrent}, {Tomasella}, {Ochner}, {Fraser}, {Benetti}, {Yuan}, {Smartt}, {Maund}, {Arcavi}, {Gal-Yam}, {Inserra}, \& {Young}}]{Valenti_floyds}
{Valenti}, S., {Sand}, D., {Pastorello}, A., {et~al.} 2014, \mnras, 438, L101, \dodoi{10.1093/mnrasl/slt171}

\bibitem[{{Valenti} {et~al.}(2016){Valenti}, {Howell}, {Stritzinger}, {Graham}, {Hosseinzadeh}, {Arcavi}, {Bildsten}, {Jerkstrand}, {McCully}, {Pastorello}, {Piro}, {Sand}, {Smartt}, {Terreran}, {Baltay}, {Benetti}, {Brown}, {Filippenko}, {Fraser}, {Rabinowitz}, {Sullivan}, \& {Yuan}}]{valenti_lcogtsnpipe}
{Valenti}, S., {Howell}, D.~A., {Stritzinger}, M.~D., {et~al.} 2016, \mnras, 459, 3939, \dodoi{10.1093/mnras/stw870}

\bibitem[{{van Loon} {et~al.}(2005){van Loon}, {Cioni}, {Zijlstra}, \& {Loup}}]{mass_loss_rsg_van_loon}
{van Loon}, J.~T., {Cioni}, M. R.~L., {Zijlstra}, A.~A., \& {Loup}, C. 2005, \aap, 438, 273, \dodoi{10.1051/0004-6361:20042555}

\bibitem[{{Vink}(2018)}]{Vink2018}
{Vink}, J.~S. 2018, \aap, 619, A54, \dodoi{10.1051/0004-6361/201833352}

\bibitem[{{Vink} \& {de Koter}(2002)}]{LBV_winds_Vink_deKoter}
{Vink}, J.~S., \& {de Koter}, A. 2002, \aap, 393, 543, \dodoi{10.1051/0004-6361:20021009}

\bibitem[{Virtanen {et~al.}(2020)Virtanen, Gommers, Oliphant, Haberland, Reddy, Cournapeau, Burovski, Peterson, Weckesser, Bright, {van der Walt}, Brett, Wilson, Millman, Mayorov, Nelson, Jones, Kern, Larson, Carey, Polat, Feng, Moore, {VanderPlas}, Laxalde, Perktold, Cimrman, Henriksen, Quintero, Harris, Archibald, Ribeiro, Pedregosa, {van Mulbregt}, \& {SciPy 1.0 Contributors}}]{2020SciPy-NMeth}
Virtanen, P., Gommers, R., Oliphant, T.~E., {et~al.} 2020, Nature Methods, 17, 261, \dodoi{10.1038/s41592-019-0686-2}

\bibitem[{{Wang} {et~al.}(2022){Wang}, {Liu}, {Lin}, {Wang}, {Dai}, {Li}, \& {Song}}]{iPTF14hls_Wang22}
{Wang}, L.-J., {Liu}, L.-D., {Lin}, W.-L., {et~al.} 2022, \apj, 933, 102, \dodoi{10.3847/1538-4357/ac7564}

\bibitem[{{Woosley} {et~al.}(2007){Woosley}, {Blinnikov}, \& {Heger}}]{2006gy_woosley}
{Woosley}, S.~E., {Blinnikov}, S., \& {Heger}, A. 2007, \nat, 450, 390, \dodoi{10.1038/nature06333}

\bibitem[{{Yaron} \& {Gal-Yam}(2012)}]{wiserep}
{Yaron}, O., \& {Gal-Yam}, A. 2012, \pasp, 124, 668, \dodoi{10.1086/666656}

\bibitem[{{Zhang} {et~al.}(2012){Zhang}, {Wang}, {Wu}, {Chen}, {Chen}, {Liu}, {Huang}, {Liang}, {Zhao}, {Lin}, {Wang}, {Dennefeld}, {Zhang}, {Zhai}, {Wu}, {Fan}, {Zou}, {Zhou}, \& {Ma}}]{2010jl_zhang}
{Zhang}, T., {Wang}, X., {Wu}, C., {et~al.} 2012, \aj, 144, 131, \dodoi{10.1088/0004-6256/144/5/131}

\end{thebibliography}
